\documentclass{emulateapj}
\usepackage{lscape}
\newcommand{\km}{${\rm km\,s}^{-1}$}
\newcommand{\fuse}{{\em FUSE}}
\newcommand{\hst}{{\em HST}}

\newcommand{\stis}{{STIS}}
\newcommand{\calstis}{{CALSTIS}}
\newcommand{\calfuse}{{CALFUSE}}

\newcommand{\hei}{He$\;${\small\rm I}\relax}
\newcommand{\hii}{H$\;${\small\rm II}\relax}

\newcommand{\cii}{C$\;${\small\rm II}\relax}
\newcommand{\cli}{Cl$\;${\small\rm I}\relax}

\newcommand{\civ}{C$\;${\small\rm IV}\relax}

\newcommand{\nv}{N$\;${\small\rm V}\relax}

\newcommand{\ovi}{O$\;${\small\rm VI}\relax}

\newcommand{\pii}{P$\;${\small\rm II}\relax}
\newcommand{\mgii}{Mg$\;${\small\rm II}\relax}

\newcommand{\sii}{S$\;${\small\rm II}\relax}
\newcommand{\siii}{Si$\;${\small\rm II}\relax}

\newcommand{\aliii}{Al$\;${\small\rm III}\relax}

\newcommand{\siiv}{Si$\;${\small\rm IV}\relax}

\newcommand{\feii}{Fe$\;${\small\rm II}\relax}
\newcommand{\heii}{He$\;${\small\rm II}\relax}
\newcommand{\feiii}{Fe$\;${\small\rm III}\relax}
\newcommand{\Niii}{Ni$\;${\small\rm II}\relax}

\newcommand{\ratio}{N(\mbox{\civ})/N(\mbox{\siiv})}
\newcommand{\ratioa}{N(\mbox{\civ})/N(\mbox{\ovi})}
\newcommand{\ratiob}{N(\mbox{\nv})/N(\mbox{\ovi})}

\slugcomment{Accepted for publication in the ApJ}
\shortauthors{Lehner et al.}
\shorttitle{Highly-Ionized Plasmas in the Milky Way}
\begin{document}

\title{Fundamental Properties of the Highly Ionized Plasmas in the Milky Way\altaffilmark{1}}
\author{N.\ Lehner, W.\ F. Zech, J.\ C. Howk}
\affil{Department of Physics, University of Notre Dame, 225 Nieuwland Science Hall, Notre Dame, IN 46556}

\and

\author{B. \ D. Savage}
\affil{Department of Astronomy, University of Wisconsin, 475 North Charter Street, Madison, WI 53706}

\altaffiltext{1}{Based on observations made with the NASA/ESA Hubble Space Telescope, obtained at the Space Telescope Science Institute, which is operated by the Association of Universities for Research in Astronomy, Inc. under NASA contract No. NAS5-26555. Based also on observations made with the NASA-CNES-CSiA Far Ultraviolet Spectroscopic Explorer. FUSE is operated for NASA by the Johns Hopkins University under NASA contract NAS5-32985.}

\begin{abstract}
The cooling transition temperature gas in the interstellar medium (ISM), traced by the high ions, \siiv, \civ, \nv\ and \ovi, helps to constrain the flow of energy from the hot ISM with $T >10^6$  K to the warm ISM with $T < 2\times 10^4$ K.  We investigate the properties of this gas along the lines of sight to 38 stars in the Milky Way disk using 1.5--2.7 \km\ resolution spectra of \siiv, \civ, and \nv\ absorption from the Space Telescope Imaging Spectrograph (STIS),  and 15 \km\  resolution spectra of \ovi\ absorption from the {\it Far Ultraviolet Spectroscopic Explorer (FUSE)}. The absorption by \siiv\ and \civ\ exhibits broad and narrow components while only broad components are seen in \nv\ and \ovi.  The narrow components imply gas with $T < 7 \times 10^4$ K and trace two distinct types of gas. The strong, saturated, and narrow \siiv\ and \civ\ components trace the gas associated with the vicinities of O-type stars and their supershells. The weaker narrow \siiv\ and \civ\ components trace gas in the general ISM that is photoionized by the EUV radiation from cooling hot gas or has radiatively cooled in a non-equilibrium manner from the transition temperature phase, but rarely the warm ionized medium (WIM) probed by \aliii. The broad \siiv, \civ, \nv, and \ovi\ components trace collisionally ionized gas that is very likely undergoing a cooling transition from the hot ISM to the warm ISM. The cooling process possibly provides the regulation mechanism that produces $ \langle N($\civ$)/N($\siiv$) \rangle = 3.9 \pm 1.9$.   The cooling process also produces absorption lines where the median and mean values of the line widths increase with the energy required to create the ion.
\end{abstract}

\section{Introduction}\label{s-intro}
The ions \ovi, \nv, \civ, and \siiv\ (a.k.a. high ions) are found in highly ionized plasmas throughout the Universe.  The detection of absorption by these ions in wide-ranging locations emphasizes the universal importance of such plasmas in interstellar medium and intergalactic medium.  Absorption lines from these high ions have been observed in the disk and halo of the Milky Way as well as in other nearby galaxies, in fast moving gas around galaxies, in the intergalactic medium, and even in primordial galaxies probed by the damped Ly$\alpha$ systems. \citep[e.g.,][]{jenkins78,cowie81,savage94,savage03,savage06,huang95,howk02, hoopes02,lehner07,sembach03,collins04,fox04,simcoe02,danforth08,tripp08,thom08,cooksey10,wolfe00,fox07,lehner08b}. 

For plasmas in which electron collisions with metal ions dictate the ionization state of the gas, the fractional abundance of these high ions peak at temperatures $T_p \sim (0.6$--$3) \times 10^5$ K depending on their ionizing energies (see below). Plasmas in this temperature regime can be produced through moderate velocity shocks ($v_s \ga 70$--$150$ \km) or through the production of hotter coronal temperature gas ($>10^6$ K) which subsequently cools through this temperature regime.  In the latter case, the quantity of high ions is related to the mass cooling rate of the coronal material. Furthermore, because the high ions likely probe matter that is rapidly cooling from coronal temperatures to temperatures more appropriate for the general diffuse ISM ($T \la 10^4$ K), the high ions may often trace matter in the interfaces between these two temperature regimes. Thus, the high ions are often used as indirect probes of energetic stellar feedback, the injection of energy into the interstellar medium (ISM) from massive stars through their stellar winds and supernovae, in galactic environments.  A significant fraction of this energy will be thermalized, producing the hot, highly ionized plasmas probed directly or indirectly by the high ions.

\begin{figure}[tbp]
\epsscale{1.1} 
\plotone{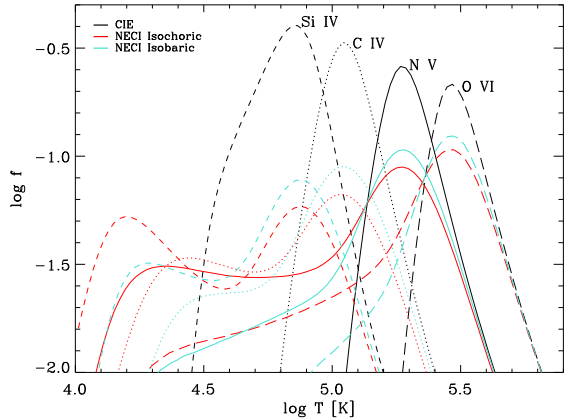}
\caption{Ionization fraction of \ovi\ (long-dashed lines), \nv\ (solid lines), \civ\ (dotted lines), and \siiv\ (dashed lines) in collisional ionization equilibrium and non-equilibrium collisional ionization according to the models with solar abundances of \citet{gnat07}. 
\label{f-ionicf}}
\end{figure}

There are complications, of course, in interpreting the relationship of the high ions to feedback.  Low density astrophysical plasmas with temperatures at which the high ions have their abundance peaks are significantly less stable than gas at higher or lower temperatures. Gas at $T \sim (0.6$--$3) \times 10^5$ K with near solar metallicities cools very rapidly since thermal electrons are able to excite the valence electrons into the upper states of the strong resonance transitions of these same high ions.  The subsequent rapid removal of energy from the gas through spontaneous photon emission cools the gas much more rapidly that it can typically recombine.  Thus, low density plasmas at these temperatures are not likely to be in a state of collisional ionization equilibrium (CIE).  In Fig.~\ref{f-ionicf}, we show the equilibrium and non-equilibrium ionization fractions of \siiv, \civ, \nv, and \ovi\ produced in a radiatively cooling gas at solar metallicity from the calculations of \citet{gnat07}.  While the high ions peak in abundance at $T_p$, in non-equilibrium ionization (NEI), a substantial fraction of the high ions is present at temperatures much lower than $T_p$ \citep{shapiro76,edgar86}. If such NEI conditions are indeed important, high ions in this cooler temperature regime should be readily observable if studied with appropriate spectral resolution. Yet, the importance of this cooler highly ionized phase remains largely unknown as we show below.

Several models for the production of high ions under NEI conditions have been presented in the literature, including conductive interfaces, where the gas is initially heated up (evaporative phase) and then cooling when radiative cooling dominates (condensing phase), turbulent mixing layers (TMLs) in which velocity shear across an interface becomes unstable, mixing hot and warm gas in a turbulent fashion, shock ionization, and several variants of these processes \citep[e.g., expanding superbubbles, supernova remnants, see, e.g.,][]{borkowski90,weaver77,begelman90,spitzer96,dopita96,indebetouw04a,balsara08}. It is important to note that all these collisional ionization models study mostly the evolution of the highly ionized gas that is cooling from the hot phase. Indeed, while heating the gas can take place in the evaporative phase of a conductive interface, this process is short-lived \citep[$<2\times10^6$ yr,][]{borkowski90}  and hence unlikely to be seen often in absorption \citep[but see][]{savage06}. Therefore the properties of the high ions give mostly clues on the physical mechanisms occurring in gas cooling from a hotter phase.

Radiation by massive hot stars or OB associations may also produce hard enough photons to ionize \civ\ and \siiv\ (ionizing energies $E_i>48$ and $34$ eV). While the addition of energetic radiation from hot plasmas in the winds of these stars can boost the amount of \civ\ relative to \siiv\ \citep[][]{black80,cowie81,knauth03}, photoionization still fails to produce significant amounts of \ovi\ ($E_i>114$ eV) or \nv\ ($E_i>78$ eV) \citep[e.g.,][]{knauth03}. Only in a Str\"omgren sphere produced by a pure hydrogen white dwarf can a substantial amount of \ovi\ and \nv\ be found \citep{dupree83}. However, the chance of intersecting \hii\ regions of radii $<10$ pc over 1--3 kpc path in order to produce a substantial column density is quite small \citep{dupree83}, suggesting that this process is unlikely important for most studies of these high ions. (In the present study we will show an absence of strong narrow \ovi\ or \nv\ absorption, demonstrating this source of high ions does not contribute significantly to our sample.)

A major shortcoming of the existing high ion surveys in the galaxy is that they were performed using moderate resolutions, with $\Delta v \approx 10$--$30$ \km\ (FWHM). If CIE conditions apply or the highly ionized gas is seen mostly at $T>10^5$ K, such resolution would be sufficient to derive the physical parameters (the column density and Doppler parameter or $b$-value) from the absorbing gas.  However, a few observations of the high ions obtained at resolutions $\Delta v \approx 1.5$--3 \km\ \citep{savage94,fitzpatrick97, sembach97,brandt99,savage01a,sterling02,fox03} have not only confirmed the presence of broad absorption with $b \ga 10$ \km\ (which could trace the $\ga 10^5$ K gas), but also revealed narrow absorbing components with $b \la 10$ \km, principally in \civ\ and \siiv, implying gas temperatures $\la (2$--$7) \times 10^4$ K. Only two measurements of \ovi\ have been made at comparable resolution by the Interstellar Medium Absorption Profile Spectrograph (IMAPS), and neither sight line shows narrow components like those seen in \civ\ and \siiv\ \citep[][E.B. Jenkins 2010 private communication]{jenkins98}.

The narrow components seen in these few studies could be evidence for the NEI plasmas predicted by collisional ionization models \citep[e.g.,][]{gnat10,kwak10,edgar86,shapiro76}, although the narrow \civ\ and \siiv\ components are cool enough that they could also trace photoionized gas (as was often assumed in these earlier studies).  The sample of Milky Way sight lines observed at sufficiently high resolution to study this narrow absorption has, however, been too small and never been combined to elucidate the importance of the warm highly ionized plasma probed by these narrow components. 

In this paper, we employ high resolution (1.5--2.6 \km) and relatively high signal-to-noise archival \hst/STIS E140H spectra of 38 early type stars in the Milky Way to study the highly ionized gas. We complement these data with \fuse\ observations (where \ovi\ is found) and \hst/STIS E230H observations (providing the \aliii\ doublet) when available. As \aliii\ absorption ($18\la E_i\la 28$ eV) is found solely in ionized gas with $T\la 7\times 10^4 $ K, and is believed to be principally present in the warm ionized medium (WIM) of the Milky Way \citep{howk99}, valuable information can be obtained by comparing, e.g., the \aliii\ and \siiv\ profiles; if those are similar, it would corroborate that these two ions are found in the same gas, but as we will show this is not typical, favoring a different origin for these ions. 

\begin{figure*}[tbp]
\epsscale{1} 
\plotone{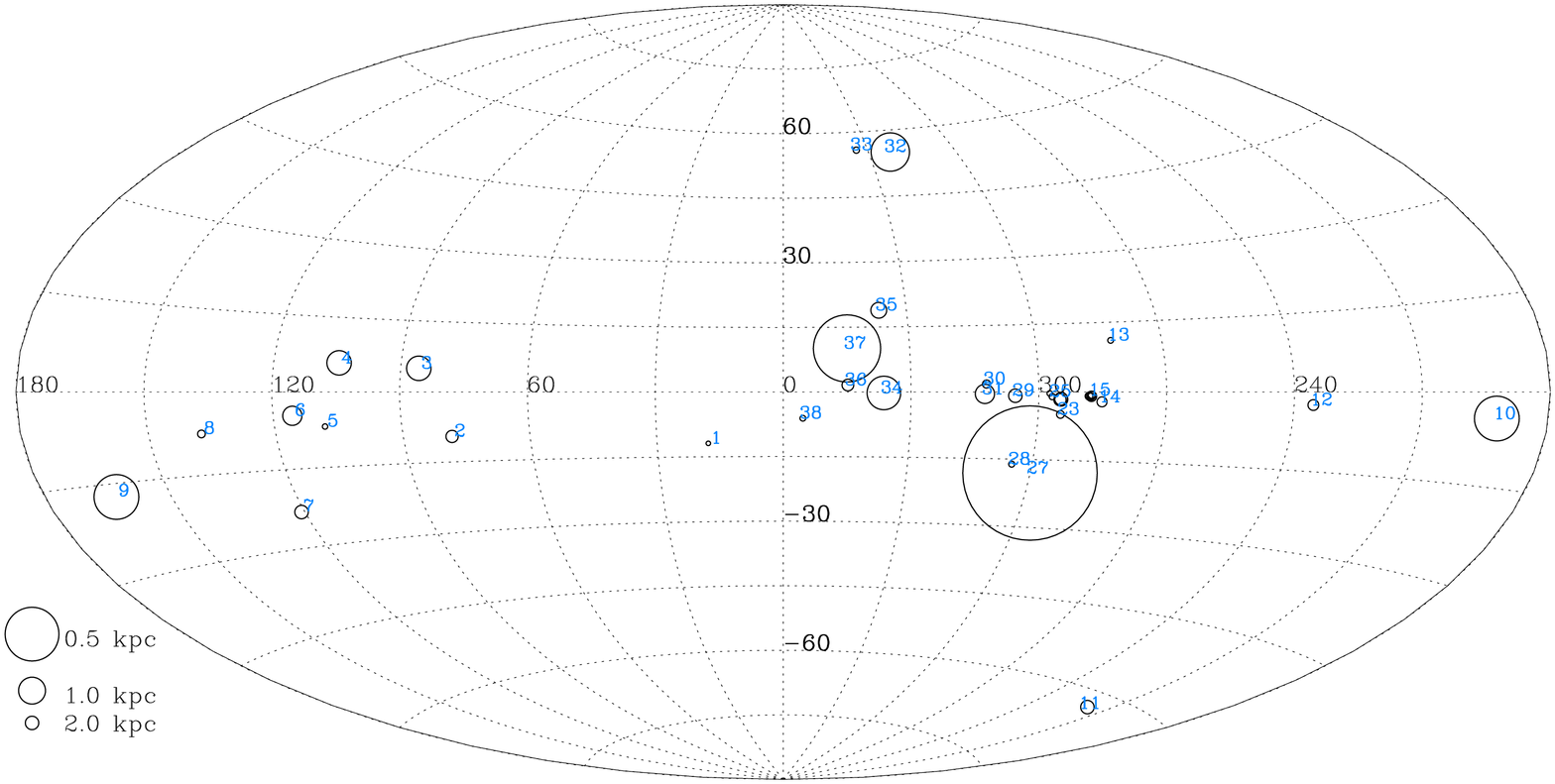}
\caption{Sky distribution of the lines of sight in an Aitoff projection. 
The sizes of circles are inversely proportional to the distance. The IDs of the stars are indicated. Near the Carina region 
(star 15) we did not mark the stars 16 to 19 and near 25, stars 24 and 26 are not marked because of the high density of 
stars in these two directions. 
\label{f-map1}}
\end{figure*}

The main aim of this paper is to characterize for the first time the properties of the transition temperature plasmas in the Galactic disk using all the accessible high ion absorption lines observed with the highest spectral resolution and a large sample of stars probing different physical regions.  In order to test the existing models and the possible origin(s) of ionizing mechanisms, we systematically relate the observed properties with the predictions inferred from collisional ionization and photoionization models, such as collisional ionization equilibrium and non-equilibrium collisional ionization \citep[CIE, NEI, e.g.,][]{sutherland93,gnat07} conductive interfaces \citep[CIs, e.g.,][]{boehringer87,borkowski90,gnat10}, shock ionization \citep[SI, e.g.,][]{dopita96,gnat09}, turbulent mixing layers \citep[TMLs, e.g.,][]{begelman90,slavin93,esquivel06,kwak10}, wind-blown bubbles or supernovae remnants \citep[SNRs, e.g.,][]{castor75,weaver77,slavin92,shelton98}, photoionization by early-type stars or a hot cooling plasma \citep[e.g.,][]{knauth03}.

Our manuscript is organized as follows. The main findings of our survey are presented in \S\ref{s-prop}, discussed in \S\ref{s-disc}, and summarized in \S\ref{s-sum}. In \S\ref{s-data} the description of our sample, the data reduction and analyses are presented. In one of the appendices we in particular summarize the estimates of the average densities of \aliii, \siiv, \civ, \nv, and \ovi\ in the Galactic disk based on the present survey.

\section{Sample, Data Reduction and Analysis}\label{s-data}
\subsection{Description of the Sample}\label{s-description}

In order to achieve our goals listed in \S\ref{s-intro}, we searched the multi-mission archive at Space Telescope (MAST) for all the OB-type stars that were observed with \hst/STIS using the E140H setting, which provides the highest resolution ($R>114,000$) and the wavelength coverage for \civ, \siiv, and \nv\ as well as other species. These stars are particularly suited to study the ISM on large distant path as they are bright and provide generally a stellar continuum suitable for interstellar study. OB-type stars with emission lines or late B-type stars with low projected rotational velocities are, however, inadequate for studying the absorption from the high ions and were rejected from the sample. Only Galactic OB-type stars were considered because the interstellar spectra of hot stars in the Small Magellanic Cloud (SMC) have the Galactic and SMC high-ion components blended. We also required that at least two  of the three high ions (i.e., \siiv, \civ, and/or \nv) were covered by the E140H observations and detected with sufficient signal-to-noise (S/N). These criteria resulted in the present sample of 38 stars. 

In Table~\ref{t-data}, the 38 stars are listed by alphabetical/numerical orders while in Table~\ref{t-datao} they are sorted by ascending galactic longitudes; we use the latter sorting to number the stars with a unique ID number. We will refer mostly to this ID number in the remaining paper (e.g., star 11 refers to star HD18100). For these 38 stars we also systematically searched for complementary STIS E230H data where the \aliii\ $\lambda$$\lambda$1854, 1862 lines are covered; 14 stars were found. Even if \ovi\ was not observed at high resolution, its relationship with the other high ions is important to describe and decipher. We therefore also search the \fuse\ (which provides the wavelength coverage of \ovi) archive at MAST, and except for stars 10 and 32, the stars in our sample were observed with \fuse. 

Table~\ref{t-data} is discussed further in the next section, and basic information about the spectral type, luminosity class, coordinate, magnitude ($V$), radial velocity ($v_\star$), distance, and $z$-height for each star is given in Table~\ref{t-datao}. For stars that were not in the sample of \citet{bowen08}, the distances were derived following the method described by these authors (see the Appendix of their paper for further information and Zech 2010). 

\begin{figure*}[tbp]
\epsscale{0.9} 
\plotone{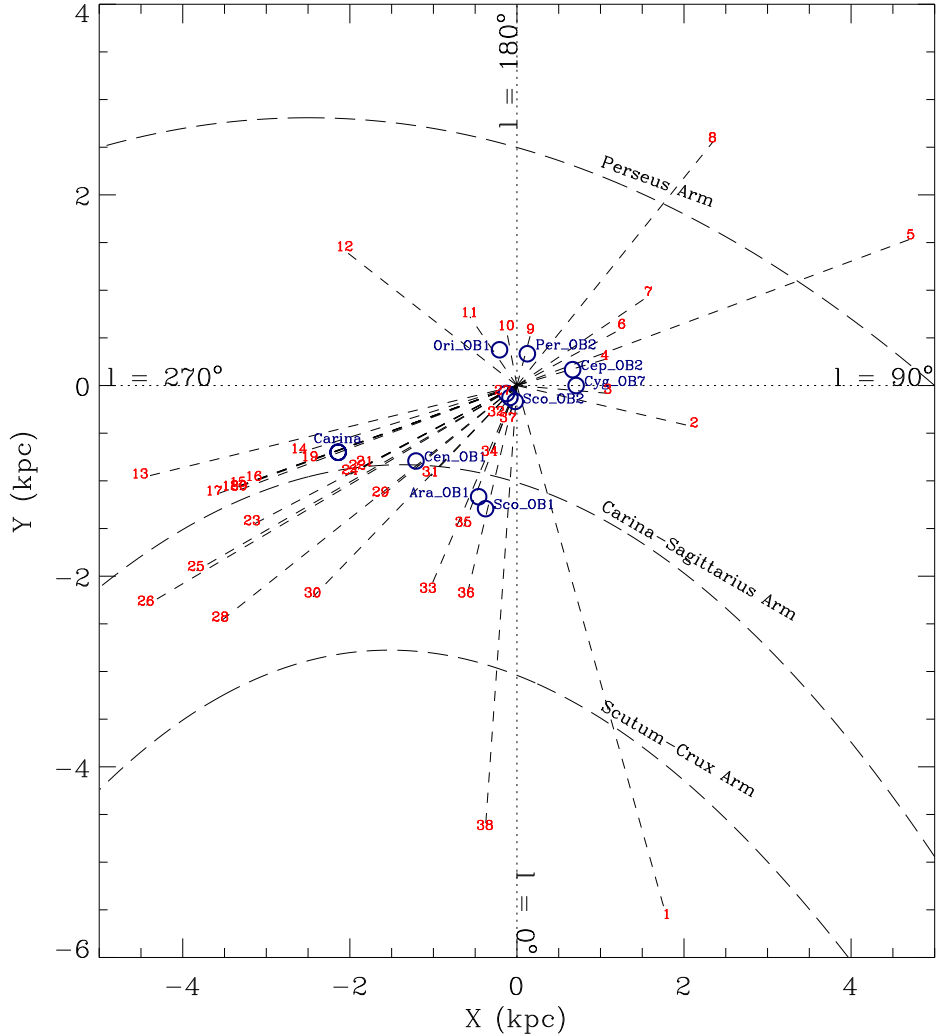}
\caption{Distribution of the lines of sight looking down into the Galactic plane. Approximate positions of the spiral arms 
are shown (from models of Vall\'ee 2002, adapted from Fig.~6 in Bowen et al. 2009). 
Carina and prominent OB associations (where $R>0$ along the line of sight) are indicated by the blue circles. 
\label{f-map2}}
\end{figure*}

The distribution of the stars in an aitoff projection is shown in Fig.~\ref{f-map1}, while in Fig.~\ref{f-map2} the distribution of the stars projected on the Galactic plane is displayed. The sight lines are distributed throughout the Galaxy, but most of them are located near the Galactic plane within $|b| < 10 \degr$ and height above the Galactic plane $|z|<0.7$ kpc. Only 4 stars are at $|z|> 1$ kpc. The distances of the stars in our sample range from 0.2 to 6 kpc, with an average distance of 2.7 kpc; about 30\% have $d<2$ kpc, 50\% $2<d<4$ kpc, and 20\% $4<d<6$ kpc. The last column of Table~\ref{t-datao} indicates some known regions that the sight lines cross. Except for stars 9 to 12, all the stars cross at least one of the radio Loops (I, II, III, IV). Several stars probe OB associations. All the stars probe the interarm region between the Perseus and Carina arms as a consequence of the position of the Earth in the Milky Way. Many also probe the interarm region between Carina and Scutum-Crux. Two stars pass the Perseus and Scutum-Crux Arms, respectively. The major grouping of stars occurs in and toward the Carina Nebula (stars 15 to 20). 

In summary, the assembled sample probes a variety of regions of the Galaxy with some being very energetically active (e.g., Carina Nebula) while others being more quiescent. A priori, the active regions may have an impact on the production of the high ions \citep[e.g.,][]{castor75,weaver77}, and especially \siiv\ and \civ.   Observationally, e.g., \citet{lehner07} directly show the impact of the stellar environments in the LMC on the high-ion absorption: supergiant shells, X-ray bright superbubbles, \hii\ regions produce strong, narrow saturated \siiv\ and \civ\ absorption in the LMC. As we will show, this is observed in the Galaxy as well. It is therefore useful to differentiate X-ray bright from darker regions. Following \citet{bowen08}, we use the parameter $R$, which scales with the excess of 0.25--1 keV X-ray excess emission in the region surrounding the star. By definition, a $R=0$ value is a region free of any X-ray emission while $R\ge 2$ is a target situated in a bright X-ray emission region, and $R=1$ is the intermediate case, generally a star near a diffuse enhancement or at the edge of a bright X-ray emission region. In Table~\ref{t-datao}, we summarize the ``$R$'' class. For the sight lines in common with \citet{bowen08}, we simply adopted their $R$-values.  As \citet{bowen08} discussed, the separation between active (``bubble'') and quieter (``non-bubble'') regions is sometimes subjective. We emphasize that regions with large $R$ identify highly disturbed environments where  the high ions may be produced, but not necessarily by processes directly related to the production of the X-rays. We also note that there is (unsurprisingly) some dependence between the type of the target star and the $R$-value along the sight line to the target star: 1) for types earlier than O7, 11 sight lines along these stars have $R>0$ and only 1 has $R=0$; 2) for types later than O9, the sample is not as bias with 17 $R=0$ sight lines  and 9 $R>0$ sight lines. 

\subsection{STIS E140 and E230H Data}\label{s-stisdata}

For this study only observations made with the E140H and E230H high-resolution \stis\ setup were selected. The observations were obtained with three different apertures: $0\farcs1 \times 0\farcs03$ (Jenkins slit), $0\farcs2 \times 0\farcs09$, and $0\farcs2 \times 0\farcs2$. These three apertures each have their own line-spread function (LSF), which we adopt from the \stis\ Instrument Handbook \citep{dressel07}. For E140H, they have a resolving power ($R \equiv \lambda/\Delta\lambda$) of $200,000$, $114,000$, and $114,000$, respectively, corresponding to a velocity resolution of 1.5 \km, 2.7 \km, and 2.7 \km\ (FWHM) (the resolution in terms of $b$-values -- FWHM\,$= 2 (\ln 2)^{0.5} b$ -- is 0.9 \km\ and 1.6 \km\ for the FWHM of 1.5 \km\ and 2.7 \km, respectively). The resolving power for E230H with the Jenkins slit is slightly less with $R \simeq 168,000$ but similar ($R\simeq 114,000$) for the $0\farcs2 \times 0\farcs09$ aperture.  For a full description of the design and construction of \stis\, see Woodgate et al. (1998), and a summary of the \stis\ on-orbit performance is given by \citet{kimble98}. We also refer the reader to \citet{jenkins01} who discuss E140H data in more detail.  

Table~\ref{t-data} summarizes the \stis\ data for each star, including the aperture used, and the wavelength coverage.  The data were reduced with the \calstis\ (ver. 2.22) pipeline in order to provide orbital Doppler shift adjustments, detector nonlinearity corrections, dark image subtraction, flat-field division, background subtraction, wavelength zero-point calibration, and to convert the wavelengths into the heliocentric reference frame. The individual exposures were weighted by their inverse variance. As \calstis\ does not coadd the various echelle-order, one of us (J.C. Howk) developed a series of IDL routines to coadd them in order to produce a single coadded spectrum. We then applied a velocity shift to each star in order to transform the heliocentric reference frame into the local standard of rest (LSR) frame (which assumes a solar motion of $+19.5$ \km\ in the direction $(l,b)\approx (56\degr,+23\degr)$).  Table~\ref{t-datao} gives the LSR correction ($v_c$) applied for each star. We emphasize that the absolute velocity uncertainty of the \stis\ E140H and E230H observations is excellent with an overall precision of $\sim$0.3 \km\ \citep{ayres08}. 

\subsection{\fuse\ Data}\label{s-fusedata}

The \fuse\ data exist for 36 of the 38 stars of the sample.   The spectral resolution is  far cruder (a factor $>5$) than the STIS E140H--E230H resolution, with $R\simeq 20,000$ (FWHM\,$\sim$$15$ \km). The data were reduced using the \calfuse\ (ver. 3.2) pipeline.  The \calfuse\ processing is described in \citet{dixon07}. In short, the detector and scattered-light background were scaled and subtracted by \calfuse. The extracted spectra associated with the separate exposures were aligned by cross-correlating the positions of absorption lines, and then co-added. In order to maximize the S/N, we coadded the various segments where applicable noting that we did not find any evidence for a reduction in the spectral resolution of the \ovi\ absorption line.  We also ensured before coadding the various segments that none was affected by fixed-pattern noise. 

As the \fuse\ wavelength calibration remains uncertain, the zero point in the final {\fuse}\ wavelength scale was established by  shifting the average {\fuse}\ velocity to the STIS 140M velocity of the same species or probing similar  types of gas. For example, we compare the H$_2$ absorption near the \ovi\ $\lambda$1031 line to \cli\ $\lambda\lambda$1347, 1379 or  \mgii\ $\lambda\lambda$1239, 1240, \siii\ $\lambda\lambda$1304,1526, and \feii\ $\lambda\lambda$1608, 1611 to \siii\ $\lambda$1020 and \feii\ $\lambda$1055. We note that the \fuse\ relative wavelength calibration accuracy is generally better than $\pm 5$ \km, but in a few cases larger shifts (up to about $\pm 10$ \km) were found between the same species observed at different wavelengths. Although the \calfuse\ version used by \citet{bowen08} is different, we refer to this paper for a more detailed description of the \fuse\ observations. 

\subsection{Analysis of the Interstellar Absorption}\label{s-anal}

We made use of two methods to analyze the data, namely the apparent optical depth (AOD) method described by Savage \& Sembach (1991), and a component fitting method using the code described by Fitzpatrick \& Spitzer (1997). While the former method does not provide the Doppler parameters (or $b$-values), velocities, and column densities in the individual components (assuming a Maxwellian distribution), the AOD method allows us to estimate directly the average velocities and total column densities of a given species from its absorption profile. The AOD method also allows us to produce a pixel-by-pixel analysis of the absorption profiles with no prior assumption about the velocity distribution, which is valuable, e.g., for comparing the variation of the column density ratio of two different species  as a function of velocities along the line of sight. These two methods are therefore complementary. To derive the column densities and measure the velocities, we used the  updated atomic parameters compiled by \citet{morton03}. 

\subsubsection{Continuum Placement and Possible Line Contamination}\label{s-cont}

The first step for analyzing an absorption profile is to define the stellar continuum level near the profile. Specifically, the continua were modeled near \siiv\ $\lambda$$\lambda$1393, 1402, \civ\ $\lambda$$\lambda$1548, 1550, \nv\ $\lambda$$\lambda$1238, 1242, \aliii\ $\lambda$$\lambda$1854, 1862, and \ovi\ $\lambda$1031. For comparison with tracers of cooler or less ionized gas, we also used systematically \siii\ $\lambda$1526 and \mgii\ $\lambda$1239 (if \mgii\ was not available, we instead used \sii\ $\lambda$1250 or \pii\ $\lambda$1532). \siii\ $\lambda$1526 is a relatively strong transition and its profile is generally saturated, allowing us to decipher the velocity interval over which the neutral and weakly ionized gas is present. The \mgii\ $\lambda$1239 (or \sii\ $\lambda$1250 or \pii\ $\lambda$1301) absorption profiles are generally not saturated, showing the velocities where most of the neutral gas is present and allowing us to have detailed information about the component structures to compare with the high-ion profiles. 

Each continuum was fitted by a Legendre polynomial of a degree $d$ within $200$--$400$ \km\ from the interstellar absorption line of interest following the methodology described in \citet{sembach92}. The degree  ($1 \le d\le 8$) depended on the complexity of the stellar continuum near the absorption lines.  The left panel on Fig.~\ref{f-sum} for a given sight line in the Appendix~\ref{s-fig} shows the adopted continua for the \nv, \civ, \siiv, and \aliii\ lines. This figure shows that i) in many cases the continuum was simple enough to be modeled by a low order Legendre polynomial ($\leq5$), and ii) in cases where the continuum is complicated, one transition of the doublet often has a simpler continuum. In complicated cases, where the absorption was weak, i.e., no saturation was present (${\rm max}(\tau_a(v))\la 1$),  we used an iterative approach to fit the continuum where we transformed the absorption profile into an apparent column density ($N_a(v)$) profile (see below and Fig.~\ref{f-sum}) and then check that the $N_a(v)$ profiles for each transition of the doublet match each other. In cases where they did not, we adjusted the continuum in the most complex profile of the two transitions. In cases, where the profiles are saturated or ${\rm max}(\tau_a(v))> 1$, we relied on part of the profiles where $\tau_a(v)\la 1$. We emphasize that this iterative procedure represents a minority of cases ($\approx 20\%$ of the sample). 

We also stress that for any absorption the continuum is critical for accurate results but it is even more crucial near weak transitions, such as the \nv\ doublet, where a small alteration in the continuum may have a dramatic effect on the line parameters themselves (line width, equivalent width, column density).  Hence we systematically modeled two stellar continua near the \nv\ transitions that appear adequate, and if the differences in the $N$ and/or $b$ values were larger than about 40\%, those are regarded uncertain.

For the \ovi\ absorption we refer the reader to \citet{lehner01} and \citet{bowen08} regarding specific issues for analyzing this absorption line, including the modeling of the continuum placement and the removal of the HD contamination. We, however, note that our independent continuum placements resulted in measurements generally overlapping within 1$\sigma$ with those produced  by \citet{bowen08} (see \S\ref{s-aod} for more details). 

We also do not show the continuum placement near the low ions, but note that generally the modeling was in most cases straightforward.   

We noted above that \ovi\ can be contaminated by an HD line at $1031.909$ \AA. The \ovi\ $\lambda$1031 is also surrounded by two H$_2$ lines at 1031.557 and  1032.354 \AA. The latter can be seen in 3 of our targets at about 120 \km\ (star 16, 17, and 38), but it does not directly contaminate the \ovi\ absorption, and we therefore did not attempt to model it. The only other line that shows some possible contamination is \siiv\ $\lambda$1393, where at about $-90$ \km, some weak \Niii\ $\lambda$1393.324 can be present. This line can be often seen in the \siiv\ $\lambda$1393 panels in Fig.~\ref{f-sum} of the Appendix~\ref{s-fig} and is for the majority of the sightlines not an issue because the \siiv\ absorption profile does not reach these high negative velocities. However, this is not the case for stars 15, 16, 17, 18, 19, and 34. For these stars, we first checked the strength of \Niii\ $\lambda$1454.852, which is a factor about 3.3 stronger than \Niii\ $\lambda$1393. For star 34, we concluded \Niii\ $\lambda$1393 would be too weak to be of any concern, and indeed the profile fit (see \S\ref{s-pf}) to the \siiv\ doublet is excellent (see Fig.~\ref{f-sum}). For the other stars, we artificially increased the error of the normalized flux vector in the appropriate velocity region (that we defined using the absorption profile of \Niii\ $\lambda$1454) of \siiv\ $\lambda$1393 so that the profile fit is heavily weighted by  \siiv\ $\lambda$1402 in that region of the spectrum. We prefer this method over attempting to model \Niii\ $\lambda$1393 using other \Niii\ lines because there is no other \Niii\ transition with similar strength. We finally note that the correction is anyway very small (see Fig.~\ref{f-sum}) as non-weighted fits gave very similar results. 

\subsubsection{AOD Measurements}\label{s-aod}
In the AOD method the absorption profiles are converted into apparent column densities per unit velocity $N_a(v) = 3.768\times 10^{14} \ln[F_c(v)/F_{\rm obs}(v)]/(f\lambda)$ cm$^{-2}$\,(\km)$^{-1}$, where $F_c(v)$ and $F_{\rm obs}(v)$ are the continuum and observed fluxes as a function of velocity, respectively, $f$ is the oscillator strength of the absorption and $\lambda$ is in \AA.  The total column density was obtained by integrating over the absorption profile  $N_a = \int_{v_1}^{v_2} N_a(v)dv$. The values of the velocity, $v_a$, is obtained from $v_a = \int_{v_1}^{v_2} v N_a(v)dv / N_a $ \citep[see][]{sembach92}. The errors for the individual transitions include both statistical and continuum errors \citep[see][]{sembach92}. When no detection was observed, we estimated a 3$\sigma$ upper limit following the method described by \citet{lehner08}. 

In Table~\ref{t-aod}, the average velocities and total apparent column densities are summarized for \siiv\ $\lambda$$\lambda$1393, 1402, \civ\ $\lambda$$\lambda$1548, 1550, \nv\ $\lambda$$\lambda$1238, 1242, \aliii\ $\lambda$$\lambda$1854, 1862 and \ovi\ $\lambda$1031 when available. This table also gives the velocity interval over which the profiles were integrated. We try to use, as much as possible, the same velocity interval for all the ions as the overall absorption ion profiles were observed over similar velocities (despite differences in the profile structure). However, in two cases (stars 16 and 17), the \ovi\ is partially contaminated by an H$_2$ line, and hence a smaller velocity interval was used in the integration. The affected targets are marked by a $^*$ symbol in Table~\ref{t-aod}, and the reader can refer to Fig.~\ref{f-sum} to see the contaminated velocities. We noted above while \siiv\ $\lambda$1393 can be contaminated by \Niii\ $\lambda$1393, for the stars where this contamination occurs, the \siiv\ absorption is saturated; therefore this weak contamination has not effect on the total column density and we did not correct for it. 

From the integration of the AOD profiles, we derived a column density for each ion. In  cases where the absorption profiles are not saturated, an average of $N_a$ of each line doublet was made. In cases where the strong line of the doublet is saturated, $N_a$ of the weaker line was adopted \citep[although we emphasize that a small saturation correction may be needed in these cases, see][]{savage91}. In cases where both transitions of the doublet are saturated, $N_a$ for the weaker transition was adopted as a lower limit. When none of the transitions of a given doublet was detected, we adopted the most stringent limit. Near \nv, the continuum placement remains very uncertain for a few stars, and for those we list colons instead of formal errors in Table~\ref{t-aod}.

For the \ovi\ absorption we can compare our results for the 21 stars in common with \citet{bowen08}. For 17 of them, our independent measurements are consistent within $\la 0.05$--0.1 dex. For 4 of them (sight lines 2, 3, 24, 37), the agreement is within 0.2--0.4 dex. For those stars, the continuum placement is extremely complex and is the main reason for the difference. We also note that for the cases where the continua are very complicated, Bowen et al. errors are often larger than ours. This is because we did not explore the continuum placements over the same interval of possibilities. In particular, for the 4 discrepant cases, their values overlap with ours within their 1--2$\sigma$ errors. Another 5 stars ($b>10\degr$) are common with the Galactic halo sample from \citet{zsargo03}, and in all these cases, our and their \ovi\ column densities are consistent within 1$\sigma$. 

In Table~\ref{t-aodf}, we also give the average density ($n \equiv N/d$) along the sight line for each ion based on the total apparent column densities. We discuss further the results from the average densities in the Appendix~\ref{s-vol}.

From the total column densities and the AOD profiles, we can derive the ionic ratios, which are useful for comparison with collisional ionization and photoionization models (see \S\ref{s-ionratios}). Table~\ref{t-aodr} lists the ionic ratios where, for a given sight line, the values in the first row are derived from the integrated column densities (from Table~\ref{t-aodf}) while the ionic ratios given in the second row are the mean and standard deviation from the pixel-by-pixel comparison  of the AOD profiles (see Fig.~\ref{f-sum}). In the latter, we transformed the STIS spectra to the \fuse\ resolution and estimated the ionic ratios in each pixel (1 pixel size being about 15 \km, the \fuse\ resolution) where the absorption was detected at better than 2$\sigma$. In general the values between the two methods are in good agreement (but not always -- the most extreme case being star 36 -- owing to some possible saturation in the \siiv\ $\lambda$1402 profile), but the standard deviation values are generally larger than the formal errors derived from the propagation error in the ratios of the column densities, giving a better sense of the extent of the variability in the ionic ratios across the velocity profiles. The ionic ratios will be discussed in more detail in \S\ref{s-ionratios}. 

\subsubsection{Profile Fitting}\label{s-pf}
While the AOD method provides useful information about the absorption profiles, in order to separate an absorption profile into individual components, we used the method of component fitting which models the absorption profile as individual Maxwellian components (also loosely referred as ``clouds").  We used a modified version of the software described in \citet{fitzpatrick97}.  The best-fit values describing the gas were determined by comparing the model profiles convolved with an instrumental line-spread function (LSF) with the data.  The LSFs are not purely gaussian, and we adopted the LSFs from the \stis\ Instrument Handbook \citep{dressel07}.  The three parameters $N_i$, $b_i$, and $v_i$ for each component, $i$, are input as initial guesses and were subsequently varied to minimize $\chi^2$.  The fitting process enables us to find the best fit of the component structure using the data from one or more transitions of the same ionic species simultaneously.

We applied this component-fitting procedure to the \civ, \siiv, \nv, and \aliii\ doublets. All the ions are fitted independently  (i.e., we did not assume a common component structure for all the ions a priori), but both lines of the doublet are simultaneously fitted.  We always started each fit with the smallest number of components that reasonably modeled the profiles, and added more components as necessary.  In this manner, if the addition of a particular component did not improve the $\chi^2$ goodness of fit parameter, we removed that component.  We allowed the software to determine the components freely (i.e., we did not fix any of the input parameters).  This procedure was repeated for each profile until the best fit was achieved. In  most complicated cases, we also tested the profile fitting by significantly changing the input $N_i$, $b_i$, and $v_i$ to test the uniqueness of the fit. In general, the new results converged to the former fit. However, in the most complicated cases (especially sight lines probing the Carina region), it is possible to find another solution with very similar reduced-$\chi^2$.  In those cases, we adopted the fit that we believed most accurately described the profiles. However, we stress that these sight lines have many components (see Table~\ref{t-fit} and Fig.~\ref{f-sum}), and the larger resulting uncertainty generally affects solely one or two components, not the majority of the components. We finally bear in mind that even though the $\chi^2$ goodness of fit may be good, in saturated regions of the profiles (where the flux reaches zero), it is impossible to accurately model these regions because there is no way to determine the correct number of components. The resulting fit usually has one (or very few) component(s) that fits the saturated region, and the column densities and $b$-values are unreliable. However, we emphasize that $b$ cannot be larger in the saturated components, and therefore the estimate is a robust upper limit (but $N$ directly depends on $b$ in the saturated regime, and hence $N$ cannot be considered as lower limit -- in that case the AOD method gives a more robust lower limit). 

The results from component fitting for \aliii, \siiv, \civ, and \nv\ are given in  Table~\ref{t-fit}, and the component models can be seen in Figures~\ref{f-sum} where both the individual components and global fits are shown. In total, the number of fitted components are 56, 138, 130, and 24 for \aliii, \siiv, \civ, and \nv, respectively. The lower number of fitted components for \aliii\ is the result simply of a smaller sample of sight line (14 compared to 38 for \civ\ or \siiv). \nv\ is not always detected, but also its absorption is generally well fitted with one (or two) component(s). If the saturated components are discarded, the numbers of components are 53, 127, and 129 for \aliii, \siiv, and \civ, respectively. Evidently from these numbers, \siiv\ is the most affected ion with saturation. 

Broad components, with $b\ga 20$ \km, may sometimes be less reliable, especially when they are blended with other components or the S/N level is not high and  $\tau$ is small. For example, when the profiles are complex, a very broad component may be fitted principally to reduce the $\chi^2$  while several narrower components could be more adequate. We note that while each species was fitted independently, some of the components seen at similar velocities in the various species may trace similar gas (see \S~\ref{s-match}). Therefore, a posteriori, we can compare the results between the high ions to further support (or not) the results from the profile fits. In the cases where there is some doubt, we did not revisit the fits because in order to do so we would need to alter our rule of minimum number of fitted components with no statistical improvement in the $\chi^2$ (and we would actually have little constraints to define any additional components). In the Appendix~\ref{s-abroad} we review some possible troublesome cases for each fitted ion. 

We finally emphasize that a limitation of the profile fitting  (especially in limited S/N) and despite the use of the highest resolution available is that a broad component can be either a single broad component or a superposition of many narrower components with smaller $N$, different $b$, and dispersed over a certain range of velocities. This complicates the interpretation of the nature of the broad components, and as we will see later, the interpretation is model-dependent. We bear in mind this issue throughout the text, but as we will show the broad (blended or not) components have in any case different properties than the narrow (unblended) components. 

\begin{figure}[tbp]
\epsscale{1} 
\plotone{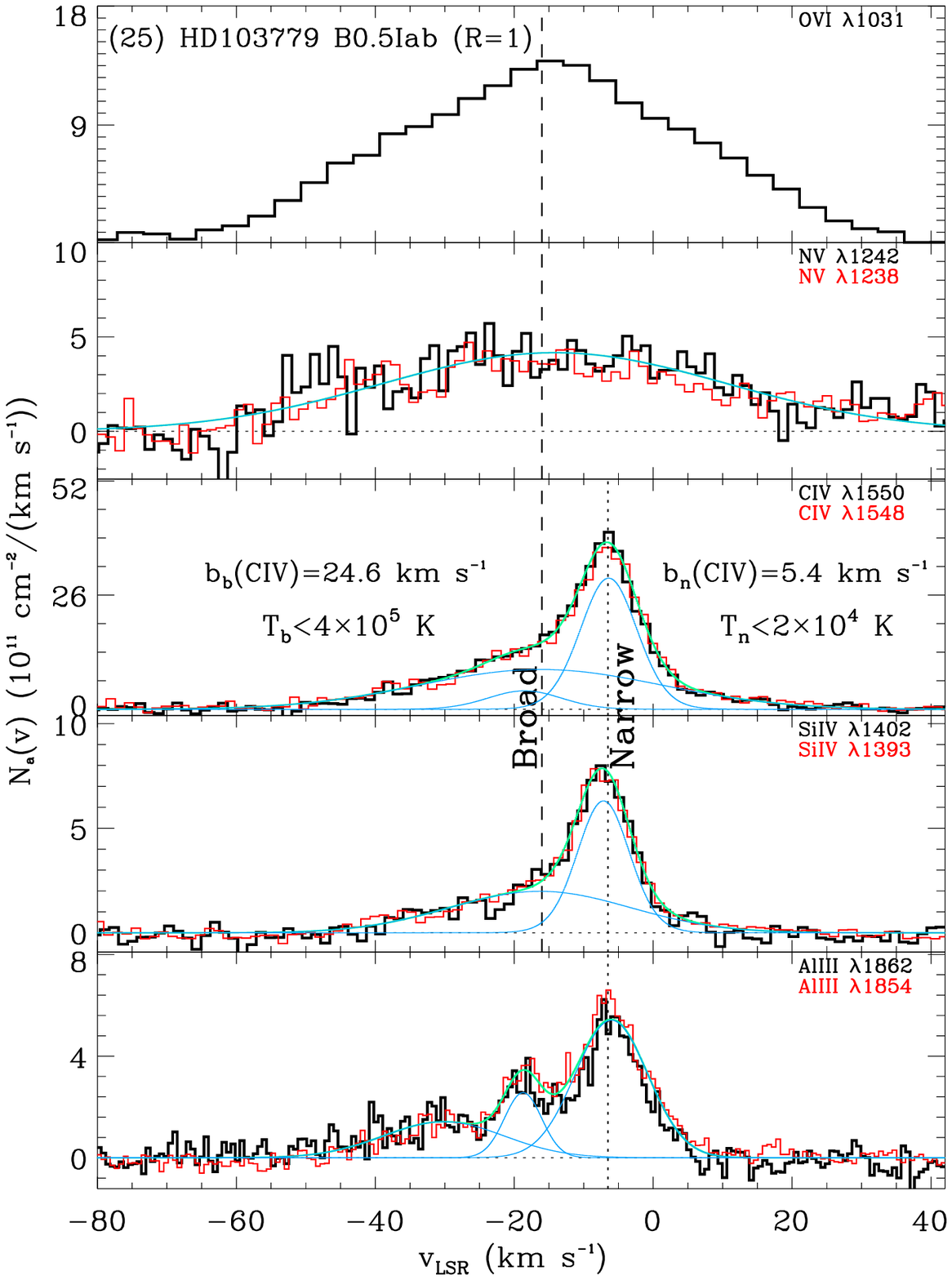}
\caption{Example of broad and narrow \civ\ and \siiv\ components. The AOD profiles are shown in black and red. The fitted profiles were transformed in AOD profiles and are shown in blue (individual components) and green (full modeled profiles).  The narrow \civ\ and \siiv\ components align with  one component of \aliii\ (but as we emphasize in the text, this is a rare occurence), and the broad \civ\ and \siiv\ component aligns well with that of \ovi\ and \nv. But note that no narrow component is detected in the \ovi\ or \nv\ profile, and no broad component is seen in the \aliii\ profile. We also indicate the temperature implied from the $b$-values of the main broad and narrow \civ\ components. 
\label{f-aodex1}}
\end{figure}

\section{General Properties of the Highly Ionized Gas}\label{s-prop}
\subsection{Broad Inferences from the Velocity Profiles}
Some examples of high-ion profiles are highlighted in Figs.~\ref{f-aodex1} and \ref{f-aodex2}, and we refer the reader to Fig.~\ref{f-sum} in the Appendix~\ref{s-fig} for all the profiles. In Fig.~\ref{f-aodex1}, we show a relatively simple example where both narrow and broad components are seen in the \civ\ and \siiv\ profiles. In this case, the narrow component aligns with one of the \aliii\ components  while the broad component aligns with that of \ovi\ and \nv. No narrow component is detected in the \ovi\ or \nv\ profile, and no broad component is seen in the \aliii\ profile. However, not all the profiles are that simple, and in Fig.~\ref{f-aodex2}, we show three examples of somewhat more complicated profiles, highlighting the diversity in the high-ion profiles and their differences as a function of the $R$-value, i.e., the amount of X-ray emission along the line of sight tracing the gas in prominent OB associations (see \S\ref{s-description}). An inspection of the AOD profiles allows us to outline some general properties of the high ions, but before proceeding further, we need to be more explicit in our definition of narrow and broad components.

\begin{figure*}[tbp]
\epsscale{1} 
\plotone{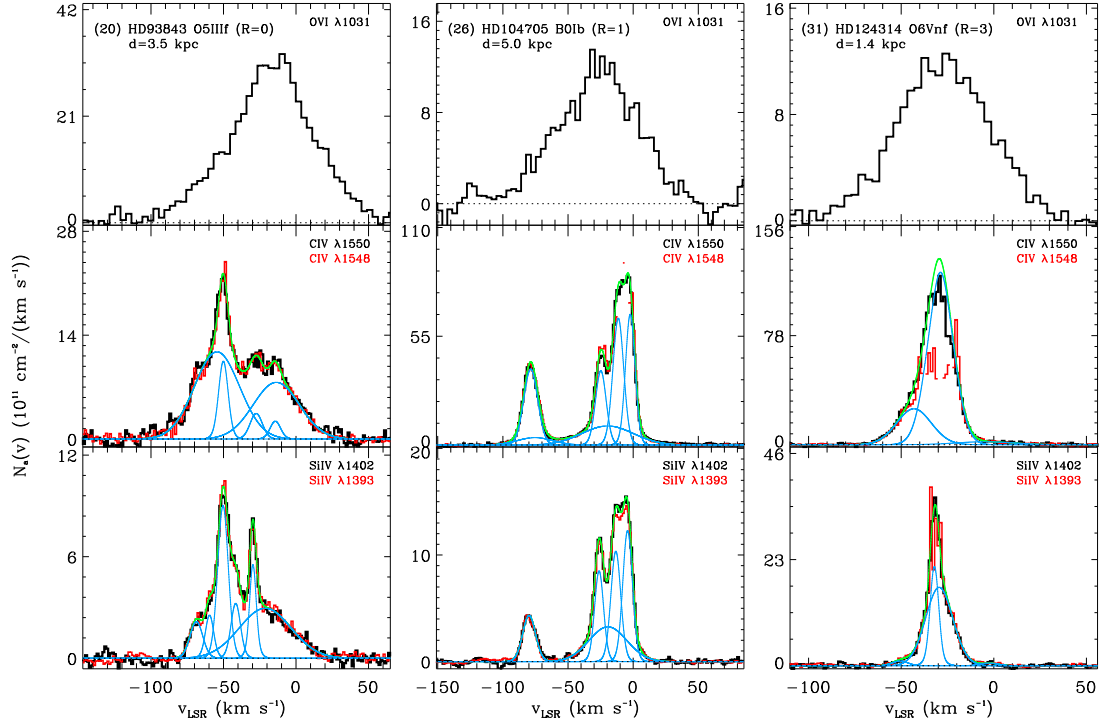}
\caption{Examples showing the diversity of component behavior in the \civ\ and \siiv\ profiles. Three examples of high ion profiles along different $R$-value sight line and type of stars. The AOD profiles are shown in black and red. The fitted profiles were transformed in AOD profiles and are shown in blue (individual components) and green (full modeled profiles). While the strength of \ovi\ changes, the shape of the \ovi\ profile is not too different between these sight lines. On the other hand, the \civ\ and \siiv\ AOD profile are different in strength and shape. Toward the hot star and $R=3$ region (right panel), \civ\ and \siiv\ are both saturated, while they are not toward the hot star and $R=0$ region (left panel) or cooler star and $R=1$ region (middle panel). Also note how some of the \siiv\ and \civ\ components align well, and for some of the components, they are only seen in one of the ions. 
\label{f-aodex2}}
\end{figure*}

The highly ionized gas can have temperatures that are warm (a few times $10^4$ K) or in the transition regime ($10^5$--$10^6$ K). Observationally, the temperature is not directly measured or estimated, but a limit can be inferred from the $b$-values derived from fitting the absorption profiles since  $b$ is a function of both the thermal and nonthermal motions of the gas. Assuming a Maxwellian distribution, the broadening of an interstellar absorption line can be written:
\begin{equation}
\label{e-b}
b^2=b^2_{\rm th} + b^2_{\rm nt}= \frac{2 k T}{A}+b^2_{\rm nt},
\end{equation} 
where $A$ is the atomic weight, $k$ is Boltzmann's constant, $T$ is the temperature,  and $b_{\rm nt}$ is the nonthermal contribution to the broadening. It is therefore judicious to define a $b$ cutoff,  $b_c$, between broad and narrow components where the narrow components probe gas temperature of $7 \times 10^4$ K in the absence of nonthermal broadening ($5 \times10^4$ K if we assume roughly equal thermal and nonthermal contributions to the line width). This value, of course, varies for each species:  $b_c= 6.5$ \km\ for \siiv\ and \aliii, 10 \km\ for \civ, 9 \km\ for \nv, and 8.5 \km\ for \ovi.  With the help of this definition, and using the above figures and more generally all the sample shown in  Fig.~\ref{f-sum} in the Appendix, we can draw some apparent properties from the AOD and fitted profiles:

1) Narrow and broad components are widespread in the \siiv\ and \civ\ profiles, but narrow components are rarely found in the profiles of \nv\ or \ovi\ absorption. 

2) Most of the narrow \civ\ and \siiv\ components align with each other, implying they probe the same gas.  Similarly, many of the broad components seen in \siiv, \civ, \nv, and \ovi\ align with each other suggesting a common origin. 

3)  The aspect of the \nv\ and \ovi\ AOD profiles is often similar, being broad with little structure. This implies the lower resolution \fuse\ observations of \ovi\ are mostly adequate to trace the  true component structure of the \ovi\ absorption.  This is important because there are more detections of  \ovi\ than \nv\ so  the \ovi\ absorption can be reliably compared to the \siiv\ and  \civ\ absorption. 

4)   \siiv\ and \civ\ have very different component behavior than \aliii\ in many cases. Therefore, in these cases \siiv\ and \aliii\ do not arise in the same gas.   

5) In most cases, the velocity profiles of high, intermediate, and low ions are observed over similar velocity-intervals. However, there is not a direct correspondence between the component structure of the low ions (e.g., \mgii\ or \sii) and \aliii\ or the high ions. 

6) Saturated components are absent in the \nv\ and \ovi\ profiles, but can be present in the other high ions. All the saturated  \siiv\ components are narrow and only appear toward $R>0$ regions and O7 to O3 stars (with one exception). These elements suggest that these components are associated with the environments of the O-type stars, especially since the velocities of these components are systematically blue-shifted relative to the star velocities.  

7) The ratios of the AOD profiles for the high ions usually exhibit substantial variations as a function of velocity.  This implies that several processes affect the origins of the high ions along most lines of sight.

All these characteristics show that while the highly ionized absorption is complicated and the sight lines probe different regions of the Milky Way, some general properties can be recovered from these high resolution spectra. In particular, it seems apparent that \civ\ and \siiv\ trace both the warm and transition temperature gas, while \ovi\ and \nv\ seem to only probe the latter type of gas. It also appears that extremely strong components of \siiv\ and \civ\ arise in the stellar environment of the hottest stars. In the following sections, we explore more quantitatively these emerging properties.

\subsection{Correlation between the Ions}\label{s-match}
\subsubsection{High Ions and \aliii}

The inspection of the absorption profiles shows that the various high ions and \aliii\  sometimes share the same components. The main aim of this section is to try to quantify this further, i.e., we wish to know what are the fractions of matching components between \aliii, \siiv, \civ, \nv, and \ovi. The result from this analysis informs us if these ions trace the same gas or not.  We use the profile-fit results summarized in Table~\ref{t-fit}, comparing \aliii\ ($18.8\le E_i\le 28.5$ eV) with \siiv\ ($33.5\le E_i\le 45.1$ eV), \siiv\ with \civ\ ($47.9\le E_i\le 64.5$ eV), and \civ\ with \nv\ ($77.5\le E_i\le 97.8$ eV). Because \ovi\ is so important for understanding the highly ionized plasma since it has the highest available $E_i$ ($113.9 \le E_i\le 134$ eV), we also fitted the \ovi\ for sight lines where \nv\ is detected in order to compare these two ions \citep[we note our results compare to those of][but we prefer to use the velocity calibration of our data for this velocity comparison]{bowen08}. Pairing adjacent ions on the ionization energy scale should in principle allow us to compare ions that are more likely to be found in similar gas phases. 

Our first test is to make a velocity comparison using the central velocities of the components derived from the profile fits. In order to do this, we compared the central velocities of the respective ions, and paired the components if the velocity separation between them is less than 5 \km\ (and often it is actually less than 3 \km). In some cases a given ion may have a broad and narrow components at a similar velocity but the compared ion has only a narrow or broad component at the same central velocity. In this case, we paired the components of the same kind (i.e., narrow with narrow components and broad with broad components, see below). We find that for \aliii\ and \siiv, 51\% of the components have similar velocities, for \siiv\ and \civ, this is 69\%; for \civ\ and \nv, 67\%; and for \nv\ and \ovi, 82\%. The respective high ions correlate much better than \aliii\ with \siiv. However, this approach is not particularly robust as the central velocities could match each other coincidentally. 

Another condition for testing if the ions are present in the same gas makes use of the $b$-values of the above matching components. Assuming the same $T$ and $b_{\rm nt}$ for two ions, a solution to Eqn.~\ref{e-b} provides support that these two ions exist in the same gas (at a single temperature). While the assumption of same $T$ and $b_{\rm nt}$  is quite standard, we see below that it may break down and may not be warranted, at least in the highly ionized plasma studied here. With these assumptions, if there is no solution to Eqn.~\ref{e-b}, either the compared ions are at different gas-temperatures and/or have different nonthermal velocities, or the components are not associated with each other and just happen to share a similar central velocity by chance.  Another possibility is that the profiles are not correctly modeled (either because a Maxwellian distribution is not adequate or because the numbers of components is incorrect, which may be in particular an issue for the broad or saturated components).

\begin{figure*}[tbp]
\epsscale{1} 
\plotone{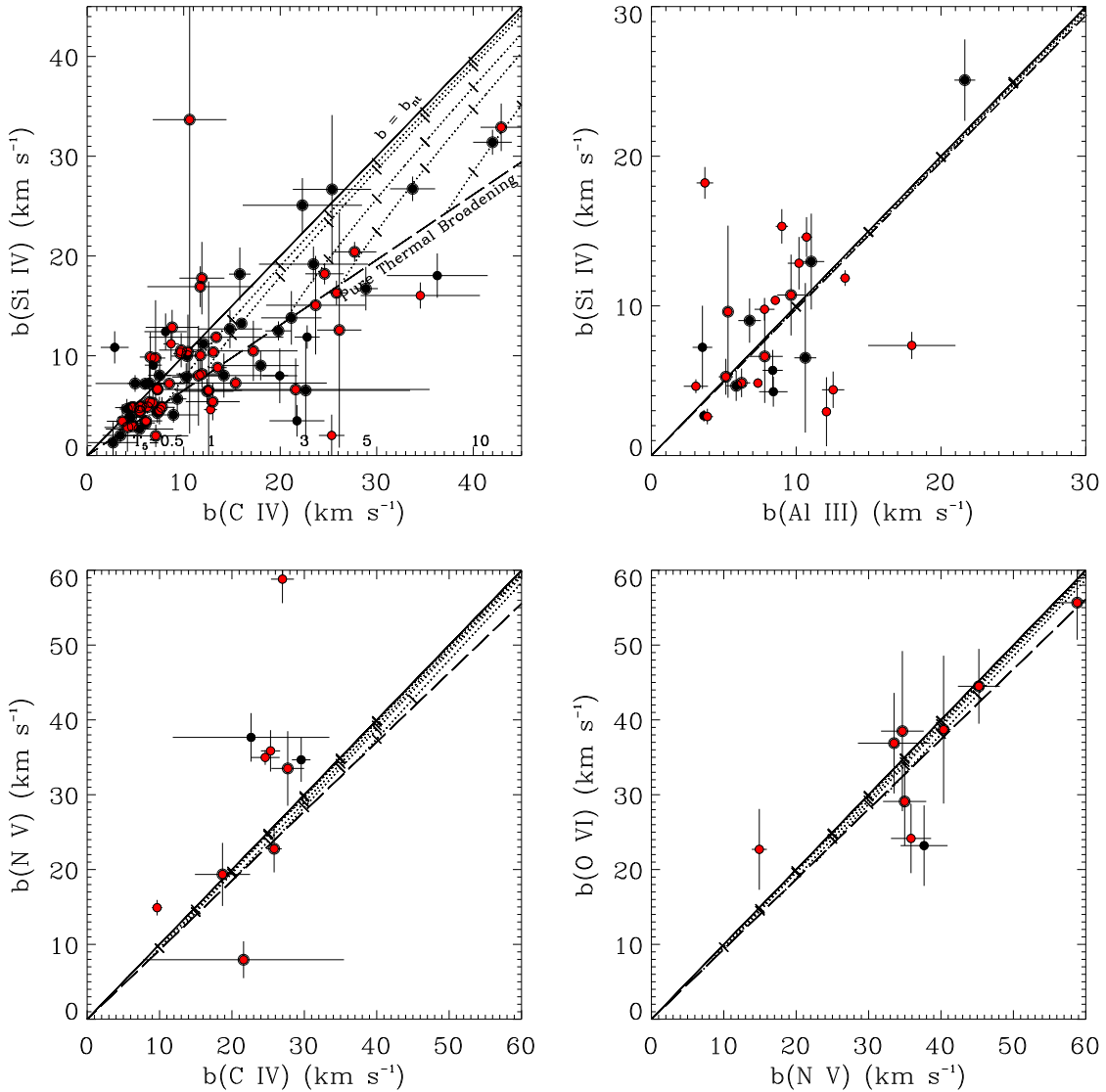}
\caption{Comparison of the Doppler parameters for the velocity matching components. {\it Note that here the data are shown with 2$\sigma$ errors}. The solid line in each panel is the broadening due to purely nonthermal motions while the long-dashed line is the broadening dominated by thermal motions. The thicker and larger circles represent data that are less than 2$\sigma$ away from these lines. The dotted lines (that can be seen best in the \siiv--\civ\ panel) represent the loci of constant temperature where the temperature is given in units of $10^5$ K  and  the tick marks on each dotted line indicate the values of $b_{\rm nt}$, are spaced by $5$ \km, and where the first tick mark is $b_{\rm nt} = 10$, $15$, $25$, and $30$ \km\ for $T_5 = 0.5$, $1$, $3$, and $5$ respectively. 
\label{f-compcs}}
\end{figure*}

In Fig.~\ref{f-compcs}, we show the $b$-value comparison between the ions.  Components for which Eqn.~\ref{e-b} is valid for a single, common temperature should lie between the purely thermal and nonthermal lines, but this figure also shows that several data points do not. Keeping only those components consistent with Eqn.~\ref{e-b} at a single temperature at the $2\sigma$ level, the fractions of matching components drop to  19\% for \aliii\ and \siiv, 61\% for \siiv\ and \civ,  27\% for \civ\ and \nv, and 55\% for \nv\ and \ovi. Because the $b$-values of saturated components are uncertain, these were not included in the sample. This affects mostly \siiv\ and \aliii\ where  most of the saturated components are found. In most cases, the saturated \siiv\ components generally match with corresponding \civ\ components, and hence the 61\% fraction is a lower limit for these two ions. This also applies to \aliii\ and \siiv\ saturated components in most cases.\footnote{Star 17 could appear as an exception when only the velocities are compared. However, the profile of \siiv\ is so strongly saturated that its single fitted component is unlikely to be adequate in this case. It seems that the \siiv\ profiles in the velocity-regions where they are saturated could be decomposed in two components, one aligning with \civ\ and the other one aligning with \aliii.} These saturated components are found exclusively along $R>0$ sight lines and all the stars but one (star 19 of type O9.7\,Ib) are O7 to O3 stars. For all these stars but star 19, the strong saturated components are blueshifted relative to the stellar velocity by $-20$ to $-55$ \km. These elements provide strong evidence that the saturated components directly trace the photoionization from the O-type stars and their environments (see \S\ref{s-disc}). From  Fig.~\ref{f-compcs} and the high frequency of matched components, there is little doubt that on the one hand \nv\ and \ovi\ often co-exist in similar gas phase, and on the other hand \civ\ and \siiv\ are also found frequently in the same gas phase (which may or may not be the same phase as traced by \ovi--\nv\ as we discuss below). 

The $b$-values of \nv\ and \ovi\ shown in Fig.~\ref{f-compcs} are in a majority of cases similar, but as we will see in \S\ref{s-bdist}, on average $\langle b($\ovi$)\rangle >\langle b($\nv$)\rangle $. For these two ions, $20 \la b \la 60$ \km\ would imply temperatures up to $3\times 10^6$ K if the broadening is purely thermal, temperature at which little \ovi\ and even less \nv\ are expected to be observed (see Fig.~\ref{f-ionicf}). This, in turn, signals that nonthermal motions may be important and/or multiple hidden (blended) velocity components could be often present. 

 The larger difference in the atomic weight between Si and C allows us to better separate the turbulent and thermal contributions to the $b$-values than the other paired ions. For \civ\ and \siiv, there is a cluster of data lying at $b($\civ$) \la 10$ \km, implying $T< 7 \times 10^4$ K, i.e., a temperature where the highly ionized gas is not expected to be in CIE (see Fig.~\ref{f-ionicf}). There are several components with $10 \la b($\civ$) \la 20$ \km\ that probe both temperature regimes, $T< 7 \times 10^4$ K and $T \la (1$--$2)\times 10^5$ K. In these cases, several data lie on the pure thermal broadening line, but many are also located in regions that imply that nonthermal motions are as important as thermal motions (or even dominant). Finally, for $b($\civ$) \ga 20$ \km, we can distinguish several data with $b_{\rm nt} \ga b_{\rm th}$. Some $b$-values of \civ\ and \siiv\ also seem to imply $T\sim (3$--$10)\times 10^5$ K, although within 2$\sigma$, many are consistent as well with cooler gas ($T\sim (1$--$2)\times 10^5$ K). According to simple CIE and NEI models (see Fig.~\ref{f-ionicf}), we would not expect to see a large amount of \civ\ and even less \siiv\ at $T> 3\times10^5$ K. This could indicate that multiple velocity components simulate a single Gaussian in these profiles. We, however, stress that the individual $b$ in that case must still be larger than $b_c$ in view of the different properties of the ionic ratios between the narrow and broad components (see \S\ref{s-ionratios} for more details). We finally note that  for the broad \civ\ components ($b>10$ \km), data points that depart from the solutions allowed by Eqn.~\ref{e-b} do so systematically such that $b($\civ$)>b($\siiv$)$.  This systematic offset suggests a physical origin for this trend. Indeed while in the above we make the standard assumption of a single temperature (and turbulence velocity), in many models of collisional ionization, the high ions need not be found at the same temperature.  In such models, the temperature increases with their ionization energies $E_i$.  As shown in Fig.~\ref{f-ionicf}, \siiv, \civ, and \nv\ peak in abundance at different temperatures ($T_{\rm p}$). For example at $T_{\rm p}$ of \civ, there is much less \nv\ in CIE or NEI conditions than at $T_{\rm p}($\nv$)$. In the conductive interface models of \citet{borkowski90} (see their Fig.~7), at $t = 10^6$ yr, $T({\mbox \siiv}) \sim 0.8 \times 10^5$ K, $T({\mbox \civ}) \sim 1.5 \times 10^5$ K, $T({\mbox \nv}) \sim 2.5 \times 10^5$ K.  Assuming $b_{\rm nt} \ll b_{\rm th}$, the models imply that $b($\civ$)>b($\siiv$)$ as it is observed in several cases. The increase of $E_i$ with $b$ could, of course, also be due to an increase in the nonthermal motions and we discuss this further in \S\ref{s-disc}.

For the \civ--\nv\ pairs, we found a comparable fraction of aligned components to that of \civ--\siiv\ pairs but a substantially smaller fraction of matching components based on the $b$-values.  We note that the \civ\ components considered  are quite broad, however, suggesting the components in these two ions have a common origin.  We saw above that $b($\civ$)\ge b($\siiv$)$ for the broad components and similarly here $b($\nv$)>b($\civ$)$. Although this trend is not observed for \ovi\ and \nv\ (possibly because the sample is small; in \S\ref{s-bdist} we will see that  $\langle b($\ovi$)\rangle >\langle b($\nv$)\rangle$). Hence, despite the failure of Eqn.~\ref{e-b} in numerous cases for the \civ--\nv\ pairs, it is not unreasonable to infer that {\it broad}\ \civ\ and \nv\ components may often trace the same gas. As \ovi\ and \nv\ trace the same gas, in turn, broad \civ\ and \ovi\ components must often trace the same highly ionized plasma. 

For the \aliii--\siiv\ pairs, the fractions based on a comparison of $v$ and $b$ are substantially smaller relative to the other ions. However, for these ions, collisional ionization models predict they should exist at overlapping temperatures and photoionization would produce a single temperature for both. Most of the paired components have $6.5 \la b($\aliii$) \la 15$ \km, which would imply $T>10^5$ K if $b_{\rm th} \gg b_{\rm nt}$. As no models predict large quantities of these ions at  $T>10^5$ K, nonthermal motions must be important and/or the fitted profiles must be composed of multiple unidentified velocity components. The data appear scattered around the pure-thermal/nonthermal broadening lines, i.e., there is no trend in the offset as opposed to what is observed for the \civ--\siiv\  or \civ--\nv\ pairs. The poor correlation between these two ions suggests therefore that the source of ionization of \aliii\ must be different from that of \siiv. As \aliii\ is a good tracer of the WIM \citep{savage90,howk99}, \siiv\ is not. The difference in $E_i$ is about a factor 2 between \aliii\ and \siiv, and more importantly the difference in $E_i$ between these two ions corresponds to the \hei\ absorption edge ($E_i=24.6$ eV) in stellar atmospheres. This may be the primary reason why \aliii\ is a good tracer of the WIM while \siiv\ is not as \aliii\ can arise in ionized gas by photons with $E_i<25.6$ eV, but not \siiv. 

\subsubsection{High and Low Ions}

It is beyond the scope of our goals to study the detailed properties of the low ions (e.g., \siii, \mgii, see Fig.~\ref{f-sum}), but from a comparison of the normalized profiles of the low, intermediate, and high ions, we observe that all these species are often observed over similar velocity intervals. However, there are few cases in which the strong \siii\ $\lambda$1526 may show some absorption without corresponding high-ion absorption or vice-versa: a good example is star 32 (see Fig.~\ref{f-sum}) where at $+30$ \km\ a \siii\ component is observed without a highly ionized counterpart component, while at $v_{\rm LSR} \le -30$ \km\ no \siii\ is observed, but high ion absorption can be observed from $-30$ to $-70$ \km. We also did not find a single trend in the relationship between the high, intermediate, and low ion components. In some cases, all the ions show components somewhat aligned, while in other cases there is no apparent correspondence in the profiles (except that they generally cover a similar full range of velocities). In other sight lines a few of the high and low ion components are aligned, but no relationship is apparent for other components.  For example, for stars 4 and 25, while some velocity shifts are seen between \civ, \siiv, \aliii, and \mgii, the overall shapes of the profiles in the main components are not too dissimilar. On the other hand, little correspondence can be found in the \civ, \siiv, \aliii, \mgii, and \siii\ absorption profiles in the spectra of stars 6, 8 and 34. 

The absence of a global trend between the low and high ions is not unexpected as, indicated in Table~\ref{t-datao}, the sight lines probe different physical regions in the Milky Way, where many different physical conditions and ionization mechanisms may occur. For example in the case of photoionization or a cooling flow, one may not necessarily expect  a close velocity correspondence between the highly ionized gas and the neutral gas. On the other hand, if the highly ionized gas is formed in an interface between the hot and neutral or weakly ionized gas, a closer velocity relationship between the neutral species and high ions may be expected, albeit with possible velocity shift between the high and low ion profiles \citep[e.g.,][]{boehringer87,savage06}. 
 
We finally note that there are also several  high-velocity clouds (HVCs), i.e., gas moving at $|v_{\rm LSR}| \ga 80$ \km, in these spectra. These HVCs are particularly interesting because they presumably arise from shocks in the interstellar gas. The Carina region is evidently extremely disturbed and has many high-velocity components. But these  HVCs are also observed in apparently far more quiescent region toward $R=0$ (stars 1, 2, 8, 13, 30) or $R=1$ sightlines (stars 26 and 36).  A previous study of this type of HVCs toward 3 nearby stars showed these HVCs have a complex ionization structure, e.g., where the gas is in a post-shock, overionized and cooling with only absorption from the singly and doubly ionized species \citep{trapero96}. A full study would require measuring the velocities and column densities of the neutral species, low ions (as well as excited states, such as \cii*) and relating those to the high ion velocity structure, but it is interesting to note that some of the properties discussed in Trapero et al. are readily seen in our profiles. For example,  at these velocities $\ga 80$ \km, \civ\ and \siiv\ can be easily created in a shock, and yet toward stars 1 and 13, no HVC high ion absorption is observed suggesting that the physical processes may be similar to those occuring toward 2 of the 3 stars in the Trapero et al. sample. A future detailed analysis of these high-velocity components should provide interesting results regarding the shock-ionization conditions of the ISM in different physical regions of the Milky Way. 

\subsection{$b$--$N$ distribution}\label{s-fndist}
Before we can quantify the $b$ distributions, it is important to understand the coupled $b$--$N$ distribution, as those two quantities are intertwined from both measurement and physical perspectives. In the physics of radiatively cooling or conductively heated gas, a relationship between the observed Doppler parameter and the column density is expected \citep{heckman02}. Following Heckman et al. \citep[but see also, e.g., ][]{edgar86}, the column density in the cooling gas (i.e., where the high-ion absorption occurs) is $N = \dot{N} t_{\rm cool}$, where $\dot{N} = v_{\rm cool} /n$ is the cooling rate per unit area and $t_{\rm cool } $ is the cooling time. The column density in the cooling plasma can be expressed as \citep[e.g.,][]{gnat07}:
\begin{equation}\label{e-ncool}
N = 4.34 k_{\rm B}   (\frac{3}{2}+s)  T  Z  v_{\rm cool}  \frac{f_{i}}{\Lambda} \,,
\end{equation}
where $Z$ is the abundance of the ion under consideration, $f_{i}$ its ionic fraction, $\Lambda$ the cooling function, and $s = 0$ is the isochoric cooling case while $s = 1 $ is the isobaric cooling case. Following \citet{heckman02}, the characteristic velocity $v_{\rm cool}$ can be identified as $b_{\rm nt}$ in Equation~\ref{e-b}, i.e., $b \simeq (0.0166 T/A + v^2_{\rm cool})^{0.5}$ (in \km). Using the \ovi\ results from various sources (galaxies, galactic haloes, high-velocity clouds, intergalactic absorbers), \citet{heckman02} argued that the apparent correlation between $N$ and $b$ is naturally explained by the physics of the cooling or heated gas and depends mostly on a characteristic velocity (cooling velocity, post-schock velocity, or isothermal sound speed depending on the physical scenario). The scatter in the correlation in that case can be explained by various temperatures and velocities in the cooling/heated gas. A similar relationship was found in the survey of \ovi\ in the Galactic disk \citep{bowen08}.

\begin{figure}[tbp]
\epsscale{1} 
\plotone{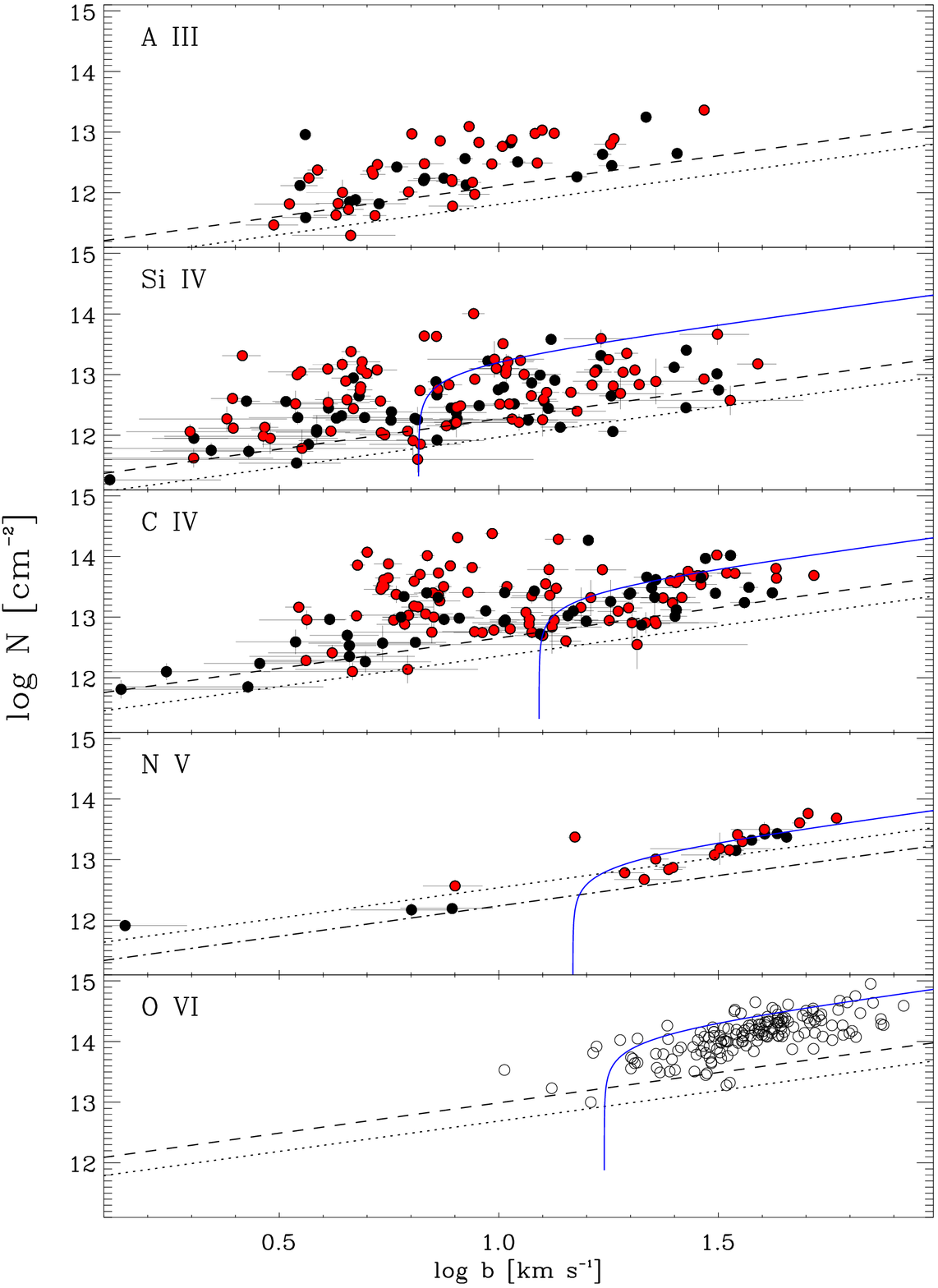}
\caption{The column density versus the Doppler parameter for \aliii, \siiv, \civ, \nv, and \ovi. The \ovi\ results are from \citet{bowen08} and were derived from the \fuse\ data that have a resolution of 15 \km\ (a factor $>$5 coarser than that of the other ions).  We do not show the error bars for \ovi\ for clarity \citep[the errors are shown in][and are not larger than the observed scatter]{bowen08}.  For the STIS data, the black filled circles are sight lines with $R=0$ and the red  filled circles are for $R>0$. For the \ovi\ sample, the reader should refer to Fig. 19 in \citet{bowen08} where the $R=0$ and $R>0$ samples are shown (no trend was observed).  The dashed, dotted, and dot-dashed lines show $N$ vs $b$ for a Gaussian line shape with peak optical depths of 0.2, 0.1, and 0.05, respectively. The blue line in the \siiv, \civ, \nv, and \ovi\ panels show the expected relationship between $N$ and $b$ in a cooling plasma assuming at  the peak abundance temperature, $T = (0.7,1.1, 1.9,2.9) \times 10^5$ K, respectively.  
\label{f-bn}}
\end{figure}

In Fig.~\ref{f-bn} we show the relation between $\log N$ and $\log b$ for the ions that we fitted as well as for \ovi\ from the results of \citet{bowen08} to directly compare them to the other high ions. As the S/N is limited, there is an observational bias present in these plots.  For a given $N$, components with larger $b$-values are more difficult to detect. The behavior of this bias is shown by the dashed, dotted, dot-dashed lines in this figure, which show the $N$--$b$ relationship for components with peak optical depths of 0.05, 0.1, and 0.2, respectively.  These can be interpreted as rough detection limits for data of various quality.  We show these three cases since the S/N ratio of our data is not uniform. Few components are found below these lines;  hence part of the general increase of $N$ with $b$ is indeed due to  the limited S/N of the data. In the \siiv, \civ, \nv, and \ovi\ panels we also plot the expected $N(b)$  from Eqn.~\ref{e-ncool} with assuming $T=T_p$ for each ion and determining $f$ and $\Lambda$ assuming CIE ion fractions from the calculations of \citet{gnat07}.  

From  Fig.~\ref{f-bn}, several inferences can be made: 

1) For \aliii, \siiv, \civ, there is no evidence for a simple relationship between $N$ and $b$ (but see below for the $R$ dependence) and there is significant scatter in $N$ for a given $b$ (limited by the detection limits on the low end). For example, for \siiv\ we find the points are randomly distributed over the range $N_{\rm min}<N \la 10^{13.8}$ cm$^{-2}$, where $N_{\rm min}$ is the minimum column density detectable at a given $b$. We note that the maximum value could be larger than given here depending on the true column density of the saturated components in our dataset.  The large scatter (especially at $b\le b_c$) suggests that photoionization by stars and other sources could play an important role, as a relationship between $N$ and $b$ is not expected in that case. 

2) For \nv\ and \ovi, there is a tight relation between $N$ and $b$, which seems to be reasonably well fitted at $b>15$ \km\ by the blue curve representing  Eqn.~\ref{e-ncool} with idealized assumptions. The Spearman correlation test gives $\rho = 0.90$ for \nv\ and 0.73 for \ovi, supporting the visual correlation. This correlation and fit suggest that collisional processes are the dominant ionizing source for the creation of \ovi\ and \nv. 

3) There is a general shift to higher $b$-values from \aliii, \siiv, \civ, \nv, to \ovi, i.e., $b$ increases with the ionization energy $E_i$. Because \nv\ absorption is predicted to be weak in NEI conditions (hence it is difficult to detect narrow \nv\ components) and \ovi\ is observed at low resolution, one may argue this shift is an artifact. From Fig.~\ref{f-ionicf}, the NEI models imply that \nv\ can be found at temperatures as low as $1.5$ to $6\times 10^4$ K, but the ionization fraction is a factor 3 (5) smaller in this temperature regime than in its peak temperature abundance in the isochoric (isobaric) models of \citet{gnat07}. The differences are even more dramatic for \ovi. This implies that it is indeed difficult to detect narrow \nv\ with the present S/N and \ovi\ in overall broad absorption. On the other hand, the {\it observed}\ maximum values of $N$ appear somewhat constant with $b$ for \civ\ and \siiv, and yet, narrow components are not seen in \nv\ and \ovi\ at any $b$, strongly suggesting that the higher ionization-energy ions are not affected in the same manner. 

4) For all the high ions, many data points are found at $b$-values that  imply temperatures much larger than $T_p$ (indicated by the vertical part of the blue curve in Fig.~\ref{f-bn}), where the  ionic fractions drop rapidly making the expected column densities of these ions very low and difficult to detect. If radiative cooling models as depicted in this figure (or other models that predict the observed column densities) are appropriate, nonthermal motions could be important and dominant in many cases.  Alternatively, low column-density clouds having smaller $b$ but with a dispersion in central velocities could mimic a single broad Gaussian line (but with individual $b$ greater than $b_c$, see our earlier remark and \S\ref{s-ionratios}).  This could happen if the sight lines intersect many interfaces.   We note that CI and TML models predict typically smaller interface column densities than found in our typical components \citep[e.g.,][]{kwak10,gnat10}.

5) The $N$--$b$ distributions do not seem to depend on $R$, except possibly for \civ\ where the dispersion in $N$ at $b < 10$ \km\ is larger for $R>0$ than for $R=0$. For \civ, the Spearman correlation test gives $\rho = 0.72$ for $R=0$ and $\rho = 0.21$ for $R>0$. In contrast, for \siiv, these values are $\rho = 0.56$ for $R=0$ and $\rho = 0.32$ for $R>0$. This suggests that \civ\ with $R = 0$ may be dominantly ionized by collisions (when a correlation between $N$ and $b$ is expected), but for $R>0$ sight lines photoionization becomes comparably important. This is not unexpected as in $R>0$, sources of ionizing radiation can be important and strong enough to produce \civ\ and \siiv\ absorption.  As discussed below, such radiation may include photons originating in stellar photospheres or in cooling hot gas in the environment of OB associations.

\begin{figure*}[tbp]
\epsscale{1} 
\plotone{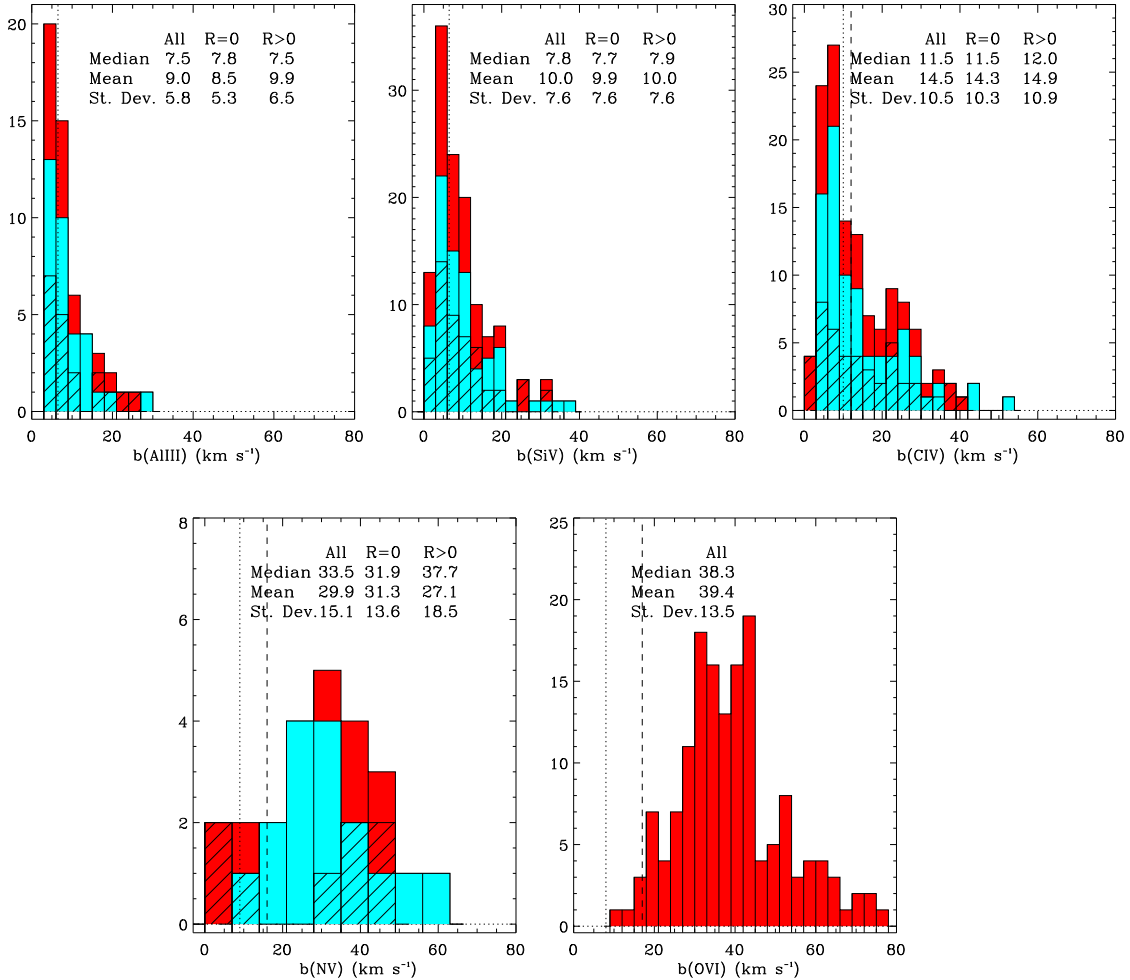}
\caption{Distribution of the Doppler parameter of \aliii, \siiv, \civ, \nv, and \ovi. The \ovi\ results are from \citet{bowen08} and were derived from the \fuse\ data that have a resolution of 15 \km\ (a factor $>$5 coarser than that of the other ions). In each panel the red histogram is for the whole sample {\it(All)}. In the panels showing the STIS data, the cyan histogram is for the sight lines with $R=0$, and the diagonal-crossed histogram is for the sight lines with $R> 0$. For each sample, the median, mean, and standard deviation of the $b$-values are given. The vertical dotted line shows the value of $b$ which would imply $T\sim 7\times 10^4$ K if the broadening is purely thermal, while the vertical dashed  lines show the expected $b$ value at  the peak abundance temperature in CIE from \citet{gnat07}. A bin size of 3 \km\ was adopted, except for \nv\ where it is 7 \km\ because the sample is smaller. 
\label{f-bhist}}
\end{figure*}

\subsection{Doppler Parameter Distribution}\label{s-bdist}
In this section we further quantify the $b$ distributions for each ion. The histograms showing the $b$-value distribution in our sample of components are displayed in Fig.~\ref{f-bhist}. Several sub-samples (all the data, $R=0$ and $R>0 $ sight lines) are considered, and the mean, median, and standard deviation for each of these samples are summarized in this figure. In the last panel we also show the $b$ distribution for \ovi\ based on the results of \citet{bowen08}. We tabulate statistics only for the whole sample (as discussed in Bowen et al., there is no apparent relationship with $R$). 

From  Fig.~\ref{f-bhist}, several immediate inferences can be made:

1) For \aliii, \siiv, and \civ, the distributions are skewed with a tail at large $b$, suggesting a lognormal distribution (see below). For \nv,  while the sample is small (we used a coarser bin size in Fig.~\ref{f-bhist}), there is no evidence of a tail and the distribution seems normally distributed. For the \ovi, there also seems to be a tail at large $b$. However $b($\ovi$)$ does not appear lognormal distributed as the \civ\ or \siiv. We also note that many of the very broad \ovi\ components ($b($\ovi$)\ga70$ \km) are often found in complicated, multiple-component absorption profiles \citep[see Fig. 24 in][]{bowen08}, so that the tail could be an artifact.

2) For all the ions, many components have $b>b_p$ and even $b\gg b_p$,  where $b_p$ is the $b$-value for purely thermal broadening with $T=T_p$, implying that nonthermal motions may dominate the broadening mechanism and/or there are multiple unidentified components in the broader profiles (in the latter case, the individual $b$ must greater than $b_c$; see our earlier remark and \S\ref{s-ionratios}).

\begin{figure}[tbp]
\epsscale{1} 
\plotone{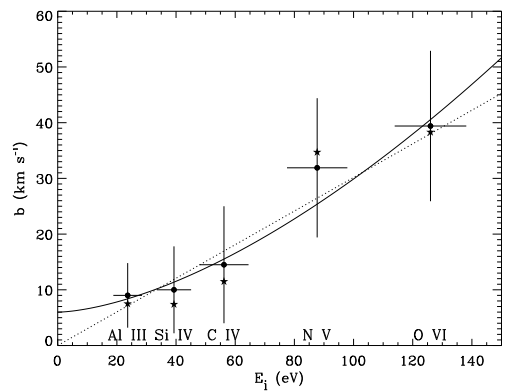}
\caption{The Doppler parameter vs. ionization energy. The circles and stars are the mean and median, respectively, of the $b$-values of the individual components from the whole sample. The vertical error bars are the dispersion around the mean and the horizontal error bars indicate the energy ranges for each ions specified at the bottom of the plot. The solid line is a geometric model while the dotted line is a linear fit to the data (see \S\ref{s-bdist}).
\label{f-bei}}
\end{figure}

\begin{figure*}[tbp]
\epsscale{1} 
\plotone{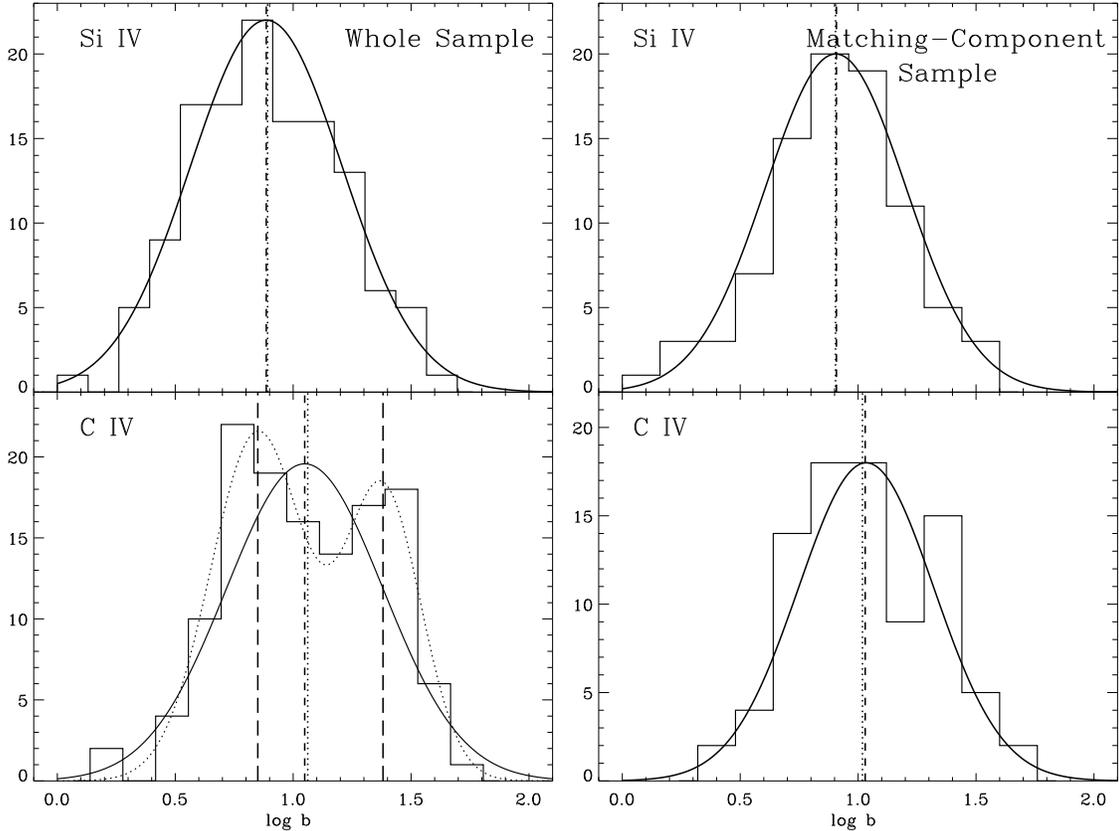}
\caption{Histogram distributions of $\log b$ for \siiv\ and \civ\ for the entire sample (left) and 
the sample where the components of \civ\ and \siiv\ are paired (right). The vertical dotted and dashed 
lines are the geometric mean and median of  $\log b$. The solid line is a Gaussian model using the geometric mean and multiplicative standard deviation values of $\log b$. For, \civ\ in the bottom left panel, the dotted line is a two-component Gaussian fit with the long-dashed vertical lines the centroids of each component.  
\label{f-blog}}
\end{figure*}

3) The importance of the tail as well as the range,  median, and mean of $b$-values increase with increasing $E_i$, and the shape of the distribution evolves with $E_i$.  In Fig.~\ref{f-bei} we show the increase in mean and median $b$ with $E_i$.  There appears to be a good correlation between $b$ and $E_i$.  The dotted line is a linear fit to the data that yields $\langle b\rangle [\mbox{\km}] = 0.3 E_i [{\rm} eV]$ while the solid line show a geometric dependence in the form of  $(\langle b\rangle-6) [\mbox{\km}] = 0.015 (E_i[{\rm} eV])^{1.6} $. This remarkable relationship between the observed $b$ and $E_i$ was noted by \citet{sembach94} with two sight lines, although their lower resolution data yielded larger $b$ for \aliii, \siiv, and \civ\ (but not \nv). We discuss further this correlation in \S\ref{s-disc}.

In order to characterize further the $b$ distribution, we use a Kolmogorov-Smirnov (KS) test to assess the shapes of the $b$-value distributions.  We find the distributions of $b$ for \aliii, \siiv, and \civ\ are unlikely to be consistent with a normal distribution, but they are consistent with a lognormal distribution. In that case, the geometric mean and multiplicative standard deviation of $b$ are more appropriate for characterizing these distributions. For \aliii, \siiv, and \civ, those are  $\mu_g =7.6 \pm 1.8$, $7.7 \pm 2.1$, $11.2 \pm 2.1$ \km.  In Fig.~\ref{f-blog}, the distributions of $\log b$ for \civ\ and \siiv\ are shown for the whole sample and the sample where the components of \siiv\ and \civ\ are matched, confirming visually the lognormal distribution.  In this figure, the thick solid lines show a Gaussian where the mean and standard deviation are derived from $\log b$ of each sample. There is some suggestion that the distribution of $\log b$ for the full sample of \civ\ may have two peaks (although this could be due in part to binning the sample), and  we show in this figure a two-component Gaussian fit to the distribution with central values and dispersions $\mu_g = 7.1\pm 1.6,24.0 \pm 1.5$ \km. 

A KS-test on the \nv\ sample shows that $b$ is consistent with a normal distribution, but is inconclusive for \ovi. However, as we discuss above, there might be larger uncertainties for $b($\ovi$)\ga70$ \km.  If we truncate the distribution at 70 \km, a KS-test implies that $b($\ovi$)$ is normally distributed. So as $E_i$ increases, not only do the mean, median, and range of $b$ increase, but the underlying distribution of $b$ changes as well.  

The above inferences apply to all the considered samples, and the $R$-values have little impact on the $b$ distribution. Although we note that for the $R=0$ sample, $b($\civ$)$ is consistent with a lognormal distribution, while for the $R>0$ sample, a KS test implies that $b($\civ$)$ is unlikely to be lognormal or normal distributed ($P< 0.03$).

\subsection{Column Density Distributions}\label{s-coldist}

\begin{figure*}[tbp]
\epsscale{1.1} 
\plotone{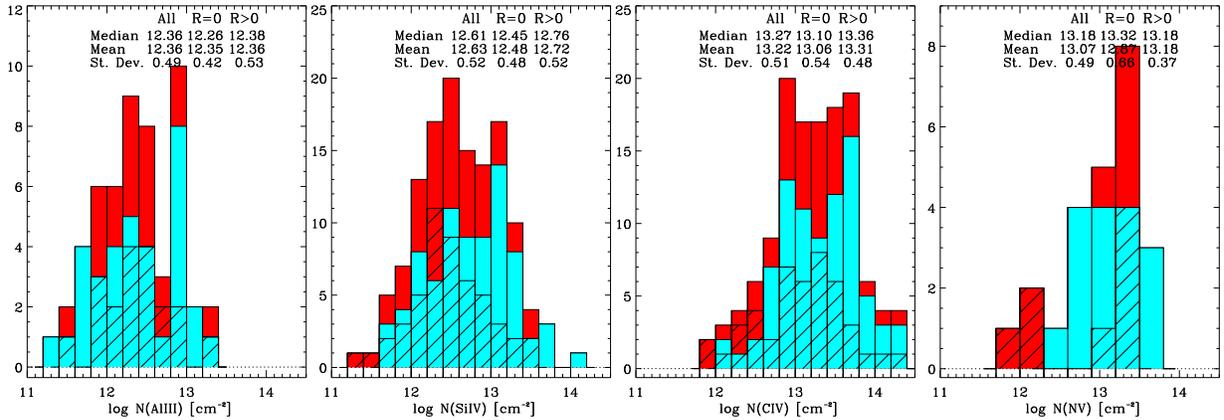}
\caption{Distribution of the logarithm of the column density of \aliii, \siiv, \civ, and \nv. In each panel 
the red histogram is for the whole sample {\it(All)}, the cyan histogram is for the sight lines with $R=0$, and the diagonal-crossed histogram is the sight lines with $R>0$.  For each sample, the median, mean, and standard deviation of $\log N$ are given. 
\label{f-nhist}}
\end{figure*}

Models for the production of high ions make different predictions regarding their column densities.  Since this study is the first where the column densities of the individual components were derived for a large sample, it is useful to discuss the column density distributions of the individual components of  \aliii, \siiv, \civ, and \nv.  We show the frequency distribution of logarithmic columns in Fig.~\ref{f-nhist}. As for the $b$-distribution, we consider the whole sample as well as sub-samples in $R$.  The geometric means, medians, and multiplicative standard deviations of these samples are summarized in this figure. We note that since the $N$ distributions are consistent with a lognormal distribution, we give the geometric mean and multiplicative standard deviation. 

For each species the dispersion in the column density distribution is large. Although the $N$ distributions are difficult to interpret (principally owing to limited S/N and its effect on the low column density end of the distribution), it is the case that $N$ is observationally most frequent in the intervals 12.1 to 13.3 dex for \siiv, 12.8 to 13.8 dex for \civ, and 13.0 to 13.5 dex for \nv. The upper (non-detected) and lower (saturated components) limits are not summarized in this figure. \civ\ is not affected by upper and lower limits (only one of each in this sample). Inclusion of lower limits in \siiv\ would change somewhat the distribution of the whole and $R>0$ samples, as several saturated and narrow components may have $\log N($\siiv$)>13.8$. On the other hand, \nv\ is often not seen, and several of the upper limits could be as low as  $\log N($\nv$)<12.2$--12.5, a factor 5--10 below the average. While it is unlikely that narrow components may be further decomposed, we cannot rule out that broad components may have multiple blended components mimicking a single one.  In this case the unseen individual components would have smaller column densities.

For \aliii\ and \nv, there does not appear to be any difference between the various $R$ samples. However, the mean and median of $N$ for \civ\ and \siiv\ are a 0.2--0.4 dex (1.5--2.5 times) larger in the $R>0$ sample than in the $R=0$ sample.  In order to better discern the origin of these differences, we summarize in Table~\ref{t-statn} the mean, dispersion, and median of the column density for the narrow and broad components for the whole, $R=0$, and $R>0$ samples. For all the ions there is a marginal (0.05--0.14 dex), but systematic, increase in the column densities of the broad components going from the $R=0$ to $R>0$ sight lines.  However, the narrow components show major differences in the column densities (except for \aliii, reinforcing our earlier conclusion that \aliii\ probes in most cases a different type of gas than \siiv) between these samples.  For \siiv, the column density in the narrow components is on average 0.25 dex (1.8 times) larger in the $R>0$ sample than in the $R=0$ (and this, without including the saturated components where $N($\siiv$)$ is likely to be even larger). For \civ, the difference is even more extreme, with a 0.63 dex (a factor 4) increase. This is certainly because \civ\ is not affected by saturation as much as \siiv\ is, allowing us to derive more reliable column density at the high end. However, this also means that the narrow \civ\ component sample is directly affected by O7 and earlier type stars. The difference between the $R=0$ and $R>0$ narrow components implies that the $R>0$ sight lines must probe regions of intense activity with enough high energy photons to photoionize significant quantities of \siiv\ and \civ.  However, the marginal increase in broad component column densities  along $R=0$ and $R>0$ sight lines shows these are minimally affected by the X-ray bright regions.

From Table~\ref{t-statn}, for all the ions the mean or median column density of the broad components is larger on average by 0.5 dex (a factor 3) than that of narrow components. The exception is \civ\ in the $R>0$ sample, for which the mean and median are similar for the narrow and broad components. 

The observed intervals of column densities in the individual components of \civ\ and \nv\ somewhat overlap with those predicted by collisional ionizing models \cite[see, e.g., the summary table in][and references therein]{indebetouw04a}.  The exception is the TML models of \citet{slavin93}, which predict much smaller high ion column densities, in fact much lower than is detectable with these data. However, the more recent TML models with NEI calculations by \citet{kwak10} predict larger \civ, \nv, and \ovi\ column densities than the CIE calculations, somewhat relaxing the needed number of TMLs to match the observed column densities. For \siiv, most of the models (except the white dwarf and halo SNR models) under-predict the amount of \siiv\ per component, likely because photoionization is generally not included in these models.

\subsection{Fraction of Broad and Narrow Components}\label{s-broadnarrow}

In the previous sections we inferred that both narrow and broad components appear important for \civ\ and \siiv, but we have not yet quantified their relative importance.  Considering the whole sample we find the percentage of broad  components (those with $b>b_c$) is $(58 \pm 11)\%$ for \aliii, $(59 \pm 7)\%$ for \siiv, $(54 \pm 6)\%$ for \civ, and $(83 \pm 17)\%$ for \nv. For \ovi, we find $(100 \pm 8)\%$, but we emphasize that $b_c$ is similar to the instrumental $b$-value in that case. It is interesting to note that even though \aliii\ is not considered a high ion, the fraction of broad/narrow components for \aliii\ is similar to the fraction in \siiv\ and \civ. The column of  \aliii\ in $T>10^5$ K gas is expected to be extremely small.  Thus, the broad components in \aliii\ are likely blends of multiple narrow components, or, their broadening is dominated by nonthermal effects. We examined if the physical regions probed by the sight lines (Galactic thin and thick disk, or Carina nebula, or sight lines with various $R$ values) have any influence on these values; we found no significant difference between various samples \citep[see][]{zech10}. 

In the previous section (and see Table~\ref{t-statn}), we showed that the fraction of the total column density in narrow components is on average $N_{\rm narrow}/N_{\rm tot} \sim 0.2$ and 0.3 for \civ\ and \siiv, respectively.  For $R>0$ sight lines, the ratio for \civ\ is somewhat higher with $N_{\rm narrow}/N_{\rm tot} \sim 0.5$. Combining these results reveals that the narrow components of \aliii, \siiv, and \civ\ make up an appreciable amount of the total number of components and of the total column density along Galactic sight lines. We stress that these fractions are based on average values,  the dispersions are large, and the narrow saturated components (especially in \siiv) that potentially have very large columns are not included.   In sight lines piercing the Carina region, for example, we find $N_{\rm narrow}/N_{\rm tot} \ga 0.5$ for \siiv.

\subsection{Component Density}\label{s-compkpc}
The component density (number of components per kpc), $\delta$, serves as a measure of the number of gaseous interfaces or cloud-like structures in the interstellar gas along a give sight line. This may provide additional constraints on theoretical models, as some of those predict the expected number of interfaces along a given sight line.  The component densities were calculated by summing the average number of broad or narrow components in each sight line, and dividing the result by the distance to the target star.   The average numbers of broad and narrow component per kpc are, respectively: 

\noindent
-- for \aliii,  $  \delta_b = 0.98  \pm	  0.18  $ kpc$^{-1}$ and $   \delta_n = 0.94  \pm       0.20 $ kpc$^{-1}$; \\
-- for \siiv,   $  \delta_b = 1.09  \pm    0.12 $ kpc$^{-1}$ and $   \delta_n =  1.12  \pm      0.14 $ kpc$^{-1}$; \\
-- for \civ,    $  \delta_b = 1.03  \pm    0.12 $ kpc$^{-1}$ and $   \delta_n = 0.87  \pm       0.11 $ kpc$^{-1}$; \\
-- for \nv,     $  \delta_b = 0.39  \pm    0.09 $ kpc$^{-1}$,    
\\
where the errors were estimated assuming Poisson statistics. We emphasize that narrow components are unlikely to be further decomposed, except possibly for the saturated components. However, this may not be  the case for the broad components. Therefore, the average values for the broad component sample should be considered as upper limits.  

As for the fraction of broad/narrow components, we do not find variations in these values that are correlated with the $R=0$ or $R>0$ subsamples.  The values for sight lines probing the general thin disk are similar to those seen in the Carina region, although the large number of saturated components for Carina sight lines, particularly for \siiv\ and \aliii, makes this comparison difficult.  Sight lines that pierce the thick disk show a smaller component density, with the number of components per kpc roughly half that found in the thin disk for \civ\ and \siiv.

\subsection{Ionic Ratios}\label{s-ionratios}

Theoretical models of the ionization mechanisms that give rise to \siiv, \civ, \nv, and \ovi\ make different predictions for the ratios of these ions \citep[e.g.,][]{spitzer96}, and column density ratios have often been used in attempts to understand the ionization and physical conditions within the highly ionized gas traced by these ions. In practice the ionic ratios do not usually remove the ambiguities regarding the ionization conditions \citep[e.g., see the attempt in][]{indebetouw04} for several reasons: (1) the absorption profiles are usually derived from data with coarser resolution than used here, blending unrelated components with one another;
 (2) several processes take place simultaneously which, when blended, produce ionic ratios that are composites of those processes;  and (3) the physics and hypotheses adopted in the models are vastly oversimplified. The use of high resolution data should improve the situation, at least for points (1) and (2). We therefore investigate below the ionic ratios both from our sample of matching components and from a pixel-by-pixel analysis of the velocity profiles.

\subsubsection{The $\ratio$ ratio}\label{s-civsiiv}

$\ratio$ is not the first choice of ionic ratio to understand collisional ionizing processes since photoionization may be quite important, especially for \siiv\ (but see below).  However, the large strengths of \siiv\ and \civ\ make the detection of these ions easy, and many studies have therefore reported $\ratio$ derived from data obtained at various spectral resolutions. Despite the wide variety of environments and of possible ionization mechanisms probed by the previous observations, $\ratio$  is found to be relatively constant \citep{pettini82,sembach92}.  Most recently, \citet{savage09} find $ \langle \ratio \rangle_{\rm g} = 3.6 \pm 1.6$ in the Galactic disk (where the subscript $g$ refers to a geometric mean and a multiplicative standard deviation). A similar value is also found in the Galactic halo \citep{savage09,sembach92}. However, \citet{fox03} and \citet{savage09} noted, based on a small sample of high resolution data decomposed into individual components (as in this work), that the situation could be more complicated.  They found a much larger dispersion in $\ratio$, albeit based on a very limited number of sight lines.

\begin{figure}[tbp]
\epsscale{1} 
\plotone{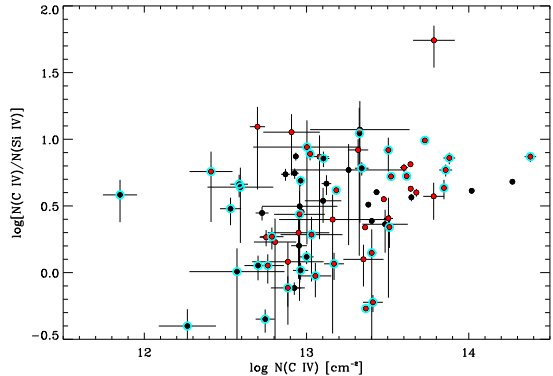}
\caption{Ratio of column densities, $\ratio$, for matching components against $N($\civ$)$.  The black filled circles are sight lines with $R=0$ and the red  filled circles are for $R>0$.  The cyan circles highlight components with $b<10$ \km. 
\label{f-ration}}
\end{figure}

\begin{figure*}[tbp]
\epsscale{1} 
\plotone{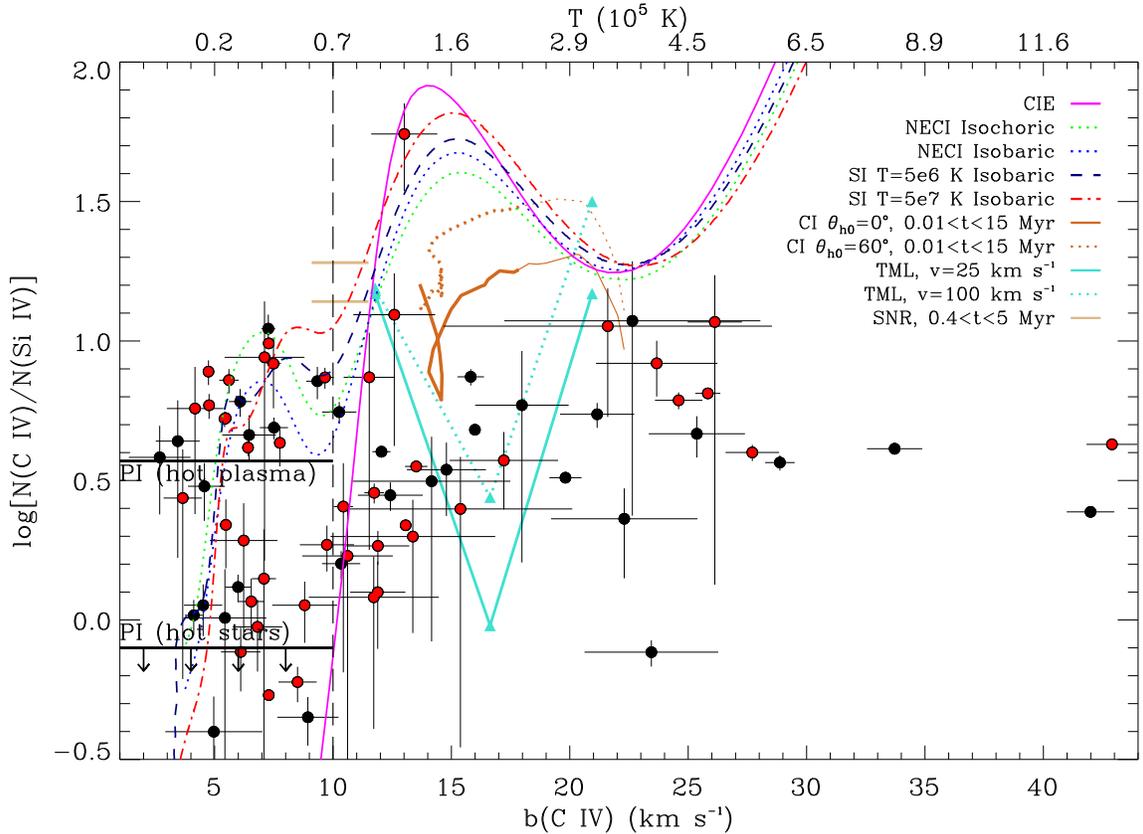}
\caption{Ratio of column densities, $\ratio$, for matching components against $b($\civ$)$.  The black filled circles are sight lines with $R=0$ and the red  filled circles are for $R>0$.   Several models are shown and are plotted relative to the gas temperature shown on the top $x$-axis. If thermal broadening dominates, the top and bottom $x$-axes have a 1:1 relationship.  Several models are shown and are overplotted. For the CI models, the thicker part of the curve is for $t>0.1$ Myr while the thinner part of the curve is for $0.01<t<0.1$ Myr. 
\label{f-compmod}}
\end{figure*}

In Figs.~\ref{f-ration} and \ref{f-compmod}, we show $\ratio$ as a function of $\log N($\civ$)$ and  $b($\civ$)$, respectively, for the individual matching components. In both figures, we differentiate between $R=0$ and $R>0$ sight lines. In Fig.~\ref{f-ration}, we further separate the broad and narrow \civ\ components. In Fig.~\ref{f-compmod}, we also show the predictions of several ionization models including CIE and NECI \citep{gnat07}, CIs \citep{borkowski90}, TMLs \citep{slavin93}, SNR \citep{slavin92}, SI \citep{gnat09}, as well as expected ratios from photoionization from stars and hot plasmas \citep{knauth03}.\footnote{Where the models used older or different estimates of the solar abundance, we have adjusted the results to current solar abundance estimates adopted from \citet{asplund09}.  It would, however, be preferable to recalculate the models with updated atomic parameters and solar abundances as the abundance correction may not be simply linear. However, we also note that dust depletion effects may also be important and those are often not taken into account in the models \citep[except in the TMLs model of][]{slavin93}.} Note that the ionic ratios derived from photoionization from a hot plasma may take lower or higher values depending on the assumed ionization parameter and density \citep{black80,cowie81,knauth03}, but the ratio from photoionization by stars in this figure is an upper limit. The $N($\civ$)/N($\siiv$)$ model-ratios are plotted against the temperature. As the observed $b$ provides only an upper limit on $T$, the data points relative to the top $x$-axis should be considered upper limits, i.e., the models could be shifted to higher $b$-values  in the presence of nonthermal motions or multiple thermally broadened Gaussian components.

Fig.~\ref{f-ration} shows a larger scatter in $\ratio$ at $\log N($\civ$)\la 13.5$ than at larger column density. The scatter is on average about 1 dex at $\log N($\civ$)\la 13.5$. Fig.~\ref{f-compmod} shows that $\ratio$ at $b($\civ$)<10$--$12$ \km\ is more variable on average than at larger $b$-values. We find  $\langle \ratio \rangle_{\rm g} = 3.3 \pm 2.5$ for the whole sample.  A KS test shows that the distribution of $\ratio$ is consistent with a lognormal distribution. The mean ratio is similar to values derived using lower resolution data, but the dispersion is notably larger.  The larger dispersion could arise owing to the impact of $R>0$ regions tracing gas in pronounced OB associations or the inclusion of narrow components in the sample. The sample used in \citet{savage09} is essentially comparable to our $R=0$ sample since they excluded any sight lines with prominent \hii\ regions, sight lines through SNRs, and toward stars with strong foreground X-ray emission. We find  $ \langle \ratio \rangle_{\rm g} = 3.0 \pm 2.4$ and $3.5 \pm 2.6$ in the $R=0$ and $R>0$ samples. The dispersions are large in both samples. The larger dispersion and different mean in the $R=0$ sample than  that derived in the low resolution data implies that the information on the narrow components is essentially lost in the instrumental broadened profiles.  This is not entirely suprising as in the $R=0$ sample the impact of the narrow component columns are on average relatively small  ($N_{\rm narrow}\sim 0.2$--$0.3 N_{\rm total}$). We also emphasize that  saturated (and narrow) components are not included in the above average values. The ratio of these components  imply $\ratio \ll 0.1$, and hence  $\langle \ratio \rangle$ would decrease and the deviation would increase if these components were included in the sample. These are found in the Carina and $R>0$ regions, and the low $\ratio$ values are likely due to the dense \hii\ regions present along these lines of sight where \siiv\ is preferentially ionized relative to \civ. 

If we consider the sample with $b($\civ$)> 10$ \km\ (i.e., the broad component sample), $\ratio$ is lognormal distributed with $ \langle \ratio \rangle_{\rm g} = 3.9 \pm 1.9$ (where the high and low extrema are removed from the sample).  This mean and dispersion are similar to earlier results using lower resolution data and integrated column densities. On the other hand, the  $b($\civ$)\le 10$ \km\ sample is not lognormal distributed. Its geometric mean is $ \langle \ratio \rangle_{\rm g} = 2.7 \pm 2.7$ and its median 4.1 (the mean and median would decrease if saturated components were considered).

This discussion suggests there is a canonical value and a lognormal distribution for $\ratio$  in the broad component sample but not for the narrow components.  The high dispersion at low $b$ could be caused by several factors.  At $T<7\times10^4$ K, a larger range of processes could contribute to the ionization, including photoionization and non-equilibrium cooling. Near $10^5$ K, the gas cools rapidly (faster than the recombination rate), ``freezing in'' the existing \siiv\ and \civ\ ions \citep{kafatos73,shapiro76}. The NEI models suggest large variations in ionic ratios over small temperature ranges in this regime (see Fig.~\ref{f-compmod}), which could also contribute to the dispersion. If photoionization is important, the radiation responsible for ionizing these species is unlikely to arise in photospheres of OB stars, as Fig.~\ref{f-ration} shows such ionization is only able to match the properties of a small number of components (although to those should be added the components where \siiv\ is saturated). Cooling hot gas could be a significant source of high energy (X-ray and EUV) radiation able to ionize these high ions \citep{black80,cowie81,knauth03}. Thus, the ionized gas in the narrow components is either photoionized by high energy photons, or is the remains of a once hot collisionally ionized gas that has rapidly cooled in a nonequilibrium manner. However, we cannot tell from the observations if one process dominates.

In the gas probed by broad components ($b($\civ$)\ge 10$ \km), some regulating mechanism must take place, possibly via turbulent motions occuring during the cooling processes, to keep $\ratio$ fairly constant.\footnote{At first glance, the low dispersion of $\ratio$  in the broad component sample could be thought as an observational artifact in view of the observational limitation of the $N$ range as $b$ increases. However, using a sample with $\log N($\siiv$)\ge 12.3$ and  $\log N($\civ$)\ge 12.8$ (i.e., where $N_{\rm min}$ and $N_{\rm max}$ are about the same for any $b$, see Fig.~\ref{f-bn}), the distribution of $\ratio$ does not change, implying that the low dispersion in that ionic ratio at $b($\civ$)>b_p($\civ$)$ is a real effect.} We discussed above that nonthermal motions may dominate the broadening, or multiple velocity components may simulate a broad component in many cases. In view of the remarkably smaller dispersion in the broad component sample, if multiple velocity components exist in the broad components, the individual components cannot be smaller than $b_c$, otherwise we should see a similarly large dispersion in $\ratio$ as for the components with $b<b_c$.  Allowing for large nonthermal motions or line blending  would seem to favor the TMLs and possibly CIE. However, collisional ionization models generally assume that \siiv\ is only produced by collisional ionization, while it can be more easily photoionized than \civ; i.e., an unknown amount of the observed \siiv\ may be produced by photoionization by stars or a hot plasma.  As we discuss now in \S\ref{s-civovi}, $\ratioa$ and $\ratiob$ favor all the collisional ionization models except CIE.

\begin{figure*}[tbp]
\epsscale{1} 
\plotone{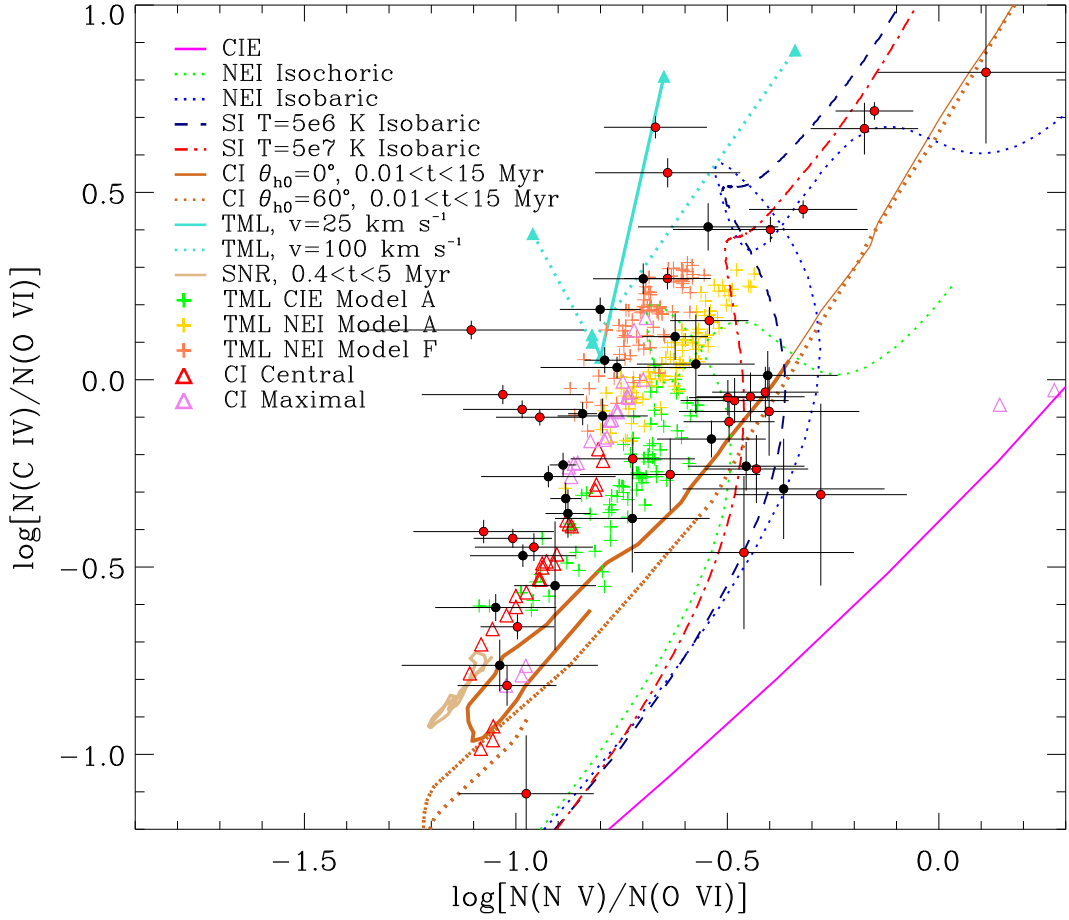}
\caption{Ratio of column densities, $\ratioa$ vs. $\ratiob$, in 15 \km\ wide pixels from the pixel-by-pixel analysis for 14 stars with detected \nv. Four pixels are typically displayed for each star with circles and attached error bars. The black  circles are sight lines with $R=0$ and the red   circles are for $R>0$. Several models are shown and are overplotted. The brown lines are the CI models from \citet{borkowski90}, while the open triangle symbols are the CI models from \citet{gnat10} (from their Table~3).
For the CI models from \citet{borkowski90} , the thicker part of the curve is for $t>0.1$ Myr while the thinner part of the curve is for $0.01<t<0.1$ Myr. The cyan lines are the TML models from \citet{slavin93}, while the plus symbols represent the recent model TML models from \citet{kwak10}.
 \label{f-ratioallp}}
\end{figure*}

\subsubsection{$\ratioa$ vs. $\ratiob$}\label{s-civovi}
The  ratios of $\ratioa$ and $\ratiob$ can a priori be used to better test collisional ionization processes as \ovi\ and \nv\ cannot be produced in any significant quantities by photoionization\footnote{Except in the Str\"omgren spheres produced by pure hydrogen white dwarves \citep{dupree83}. But as these authors noted, it is unlikely to produce the bulk of the observed column densities because of their small cross-sections. Furthermore, if that source were important, it would produce strong narrow \ovi\ and \nv\ components that are not seen.} in view of their large ionization energies, while \civ\ is less easily photoionized than \siiv\ in Galactic conditions.  As \ovi\ has only been observed with a resolution of about 15 \km, we used the AOD profiles to determine the ionic ratios. Rather than using the integrated column densities, we derived the ionic ratios in each \fuse\ resolution element ($\sim$15 \km, i.e., the STIS spectra were degraded to the \fuse\ resolution) where 2$\sigma$ absorption is detected following our description in \S\ref{s-aod} (see also the AOD ratio panels in Fig.~\ref{f-sum}). In  Fig.~\ref{f-ratioallp}, we show the ratios derived using this "pixel-by-pixel" analysis for 14 stars with detected \nv.  Two to six pixels are  displayed for each star with circles and attached error bars. We noted that \civ\  often has narrow components. As those are not found in \nv\ and \ovi, we checked if excluding the pixels where a narrow \civ\ component is observed would change the results and found no qualitative change. 

In this figure, we also plot the $\ratioa$ vs. $\ratiob$ tracks  predicted by various models (see previous section, but we also include the recent conductive interface model by Gnat et al. 2010 and TML model by Kwak \& Shelton 2010).\footnote{While boxes have often been used in this type of presentation for showing the predicted ratios, they encompass a much wider range of values that is actually allowed by the models.} There is a good overall agreement between the data and the predicted column-density ratios. The observed scatter and correlation of $\ratioa$ vs. $\ratiob$ follow well that of the models. While the correlation is likely affected by the presence of $1/N($\ovi$)$ on both axes (which applies to both the models and observations), other ionic ratios (e.g., $\ratio$ vs. $N($\civ$)/N($\nv$)$, see below) do not show the same agreement or correlation.  The distribution of the observed ratios overlaps many of the models, and no model is favored.  This is not surprising, as the physical and ionization conditions change from sight line to sight line and with velocity (location) along a single sight line.  However, the present data are not consistent with the predictions made by the CIE models.

For both ionic ratios, the scatter is smaller in the $R=0$ sample than in the $R>0$ sample: in the $R=0$ sample, $-0.7\la \log [N($\civ$)/N($\ovi$)] \la + 0.5$ and $-1.0\la \log [N($\nv$)/N($\ovi$)] \la -0.3$ while in the $R>0$ sample $-1.1\la \log [N($\civ$)/N($\ovi$)] \la + 0.8$ and $-1.1\la \log [N($\nv$)/N($\ovi$)] \la + 0.1$. For the whole sample, we find that $\langle \log [N($\civ$)/N($\ovi$)] \rangle \simeq -0.1$, $\langle \log [N($\nv$)/N($\ovi$)] \rangle \simeq -0.6$. These intervals and averages are not different from the values found in the Galactic halo \citep[e.g.,][]{sembach92,savage97,indebetouw04}. The scatter in $\ratioa$ vs. $\ratiob$ may be larger in the halo according to Fig.~4 of \citet{indebetouw04}, but as no error bars are displayed by these authors, it is difficult to judge the effects of the errors on the observed scatter.

We also considered $\ratio$ vs. $N($\civ$)/N($\nv$)$ as all these ions were observed at high resolution, but the display of these data resulted in a scatter diagram.  We finally note that in most studies the integrated column densities are considered. Using these integrated ratios, we found that some data points are consistent with the model tracks, and some are not. Several of these ratios are limits that could possibly be consistent with the model, leaving more ambiguities than the pixel-by-pixel analysis presented above. 

\section{Discussion}\label{s-disc}
With this survey, the number of analyzed interstellar high ions (\siiv, \civ, \nv) using  high resolution STIS spectra of hot stars in the Galactic disk and lower halo has increased by a factor $\sim$10. We also systematically compared these high ions with  the \ovi\ absorption observed in lower resolution \fuse\ spectra of the same stars. The analysis and interpretation of these observations have revealed a wealth of new information on the properties of the highly ionized gas. Prior to this work, the comparison of observations and theories was mostly based on ionic ratios often derived from the integrated column densities. Theories for the origins of the highly ionized gas can now be tested beyond simply the ionic ratios. In fact, Fig.~\ref{f-ratioallp} shows that ionic ratios alone provide poor diagnostics for inferring the dominant physical processes along a given sight line (the degeneracy could be removed if other ions are included, e.g., \siiv, but as we discuss above, photoionization is generally not well treated in the models). One of the most important results of this study is the finding that \siiv\ and \civ\ exhibit both  narrow and broad absorption components while \nv\ and \ovi\ only exhibit broad absorption components.  A quantitative analysis of the properties of the gas in these different types of absorbers allows us to identify three types of highly ionized plasmas in the Galactic disk where the majority of our stars reside:
\\
a) warm photoionized gas tracing radiative feedback of mostly O-type stars;
\\
b) warm ($<7 \times 10^4$ K) gas tracing mechanical feedback either directly as  radiatively cooling gas after a shock or through interface physics, or indirectly as the result of photoionization powered by the radiation of hot plasmas; and,
\\
c) transition temperature ($\sim$$10^5$--$10^6$ K) gas tracing mechanical feedback. 

Types (a) and (b) are only observed in the  narrow components of \civ\ and \siiv\ absorption profiles. Type (c) is observed in all the high ions studied in this work, i.e., in the broad components of \siiv, \civ, \nv, \ovi. Types (a), (b), and (c) are observed in the Galactic disk, but only (b) and (c) are observed toward the four Galactic halo stars.

Our results show that (a) and (b) can be simply distinguished based on the presence or absence of saturation in the \siiv\ absorption profiles. Indeed,  the strong, saturated \siiv\ components are {\em solely}\ found  toward $R>0$ regions and all the target stars for $R>0$ sight lines  are O7 to O3 giant and main-sequence stars, with a single exception (star 19 of type O9.7\,Ib). The strong saturated components along these sight lines are blueshifted relative to the stellar velocity by $-20$ to $-55$ \km. There is a single star that have a type earlier than O7 and the sight line to the target star passes through a $R=0$ region (star 20). In that case, the \siiv\ profiles are not saturated, suggesting that indeed both the type of star and enhanced X-ray emission region along the path to the target star are required for producing the saturated \siiv\ components. All these elements provide strong evidence that the saturated components are the direct result of radiative feedback from the O-type stars and the expanding shells they power.  For these components, $\ratio$ is generally $\la 1$ (see the AOD ratios in Fig.~\ref{f-sum}), so radiation from the stars themselves must indeed be a dominant source of ionization (\S\ref{s-civsiiv}).\footnote{For the one exception, the sight line to star 19, the saturated component is redshifted relative to the stellar velocity, possibly indicating that the sight line may capture gas associated with the foreground Carina nebula.} 

The properties of the saturated \siiv\ and \civ\ components are very similar to those found in the LMC by \citet{lehner07}. These authors analyzed the UV spectra of 3 O-type and one WR LMC stars and found similarly strong, narrow \siiv\ components as well as associated strong \civ\ and \feiii. H\,$\alpha$ emission along these LMC sight lines had the same velocity as the strong \siiv\ and \civ\ absorption. Thus, the strong high ion absorption was associated with the H\,$\alpha$-emitting LMC \hii\ regions and supershells visible in H\,$\alpha$ images of these regions. Hence, the saturated \siiv\ components with accompanying strong \civ\ and other tracers of photoionized gas (\aliii, \feiii, H\,$\alpha$) are signatures of the gas in the vicinity of  extremely massive and hot stars. 

On the other hand, the weaker narrow \siiv\ components are found toward any type of target stars and with no preferred  $R$-value. The velocities are not shifted in a systematic manner relative to the target star velocity. The weaker narrow absorption is therefore fundamentally different from the saturated absorption. The ionization mechanism is unlikely due directly to these stars or their direct surroundings, especially since $\ratio >1$ in many of these components. Instead they likely trace the more diffuse matter at the boundaries between hot and cool/warm gas that have radiatively cooled or where the ionization is from  high energy radiation from the interfaces or other hot cooling plasma. The weak narrow components are therefore tracers of  past or present hot plasmas cooling from $T>10^6$ K (see below).\footnote{Following one of the referee suggestions, we also note an additional outcome from our study is that the strong dependence between the saturation level of \siiv\ and the type of $R$-value along the sightline to the target star implies that very few energetic photons actually escape the dense \hii\ regions around massive hot stars. It is therefore not suprising that the \hei/H$\alpha$ line intensity ratio in the WIM is significantly less than that observed in the \hii\ regions \citep{reynolds95,madsen06}. }

We showed that the broad components can be so broad that nonthermal motions are important (perhaps dominant) or that multiple narrow components are present.  However, the broad components are unlikely to simply be unidentified blends of narrow components like the $T<7\times 10^4$ K narrow components discussed above, given the observed distribution of $\ratio$ in the broad and narrow components is different. The fact that no warm gas is observed in the \ovi\ and \nv\ implies that there must be a change in the properties of the highly ionized gas between the maximum value of $E_i($\civ$)=64.5$ eV and minimum value of $E_i($\nv$)=77.5$ eV. This is above the \heii\ absorption edge, which abruptly reduces the number of stellar ionizing photons at $>54$ eV, although one may argue that narrow components seen in \civ\ and \siiv\ could mostly arise owing to photoionization by starlight at $E_i<54$ eV. However, Fig.~\ref{f-compmod} rules out this conclusion because photoionization by stars cannot provide the observed large $\ratio$ ratio, except if this ratio is boosted by EUV radiation  \citep{knauth03}. We also saw that the \aliii\ and \siiv\ profiles have little in common (outside the components associated with the stars themselves along $R>0$ sight lines), also suggesting that \siiv\ is not produced by photoionization by stars as \aliii\ can be \citep{savage90,howk99}, possibly because the \hei\ absorption edge at 24.6 eV in stellar atmospheres strongly affects the ionizing radiation field (see \S\ref{s-match}). Localized high energy photons in the vicinity of target stars also cannot be the sole contenders for the sources of ionization because Fig.~\ref{f-compmod} shows that the amount of X-rays along the sight line does not explain solely the large scatter observed in $\ratio$ observed at $T<7 \times 10^4$ K. Therefore both NEI and photoionization with a contribution from cooling radiation must play a role in the origin of the narrow components seen in \siiv\ and \civ. While the photoionization model utilizing emission from a diffuse cooling hot plasma fails to produce large amounts of \ovi\ or \nv\ relative to \civ\ \citep{knauth03}, NEI calculations (see Fig.~\ref{f-ionicf}) could a priori imply \ovi\ and \nv\ in gas at $T<10^5$ K, and yet dominant or even observable narrow components are not found for these ions as for \siiv\ or \civ, implying that NEI conditions must occur in large part near $T_p$ for \ovi\ and \nv\ in the Galactic disk and the 4 Galactic halo sight lines at $1<|z|<3.6$ kpc.  

The recent simulations of TMLs undertaken by \citet{kwak10} may provide a beginning of an explanation for the absence of strong narrow \ovi\ and \nv\ components. They estimate the column densities of \civ, \nv, and \ovi\ in the actively mixing region where the gas remains around $10^5$ K and in the radiatively cooled region at $T\sim 10^4$ K. They found with the NEI calculations that 30\% of $N($\civ$)$ is found in the cooled region, but only about 15\% for \ovi\ and \nv. If we consider {\it only}\ the $R=0$ sight lines, we find that the column density for the unsaturated narrow components of \civ\ are on average 20\% of the total column density (with a large dispersion), which is comparable to the theoretical results. The narrow \ovi\ and \nv\ components are not detected because they are likely too weak to be found in the dominant broad profiles of these ions. It would be interesting to know if these types of models produce absorption profiles consistent with the observed \nv\ and \ovi\ profiles (i.e., absence of discernible narrow components) using the same spectral resolutions.  It is also interesting to note that \citet{kwak10} do not consider  self-photoionization via cooling radiation in their models; allowing for it would likely further raise the amount of \civ\ in the cooler region. In this case, NEI might be a mechanism sufficient by itself to produce the bulk of the narrow components along $R=0$ sight lines. While it is beyond the scope of this work to present a detailed sight line analysis, it seems evident that the quality of the present data with the information about the types of regions probed along the sight lines (see Table~\ref{t-datao}) combined  with these types of new diagnostics will further help the understanding of the highly ionized plasma in the Milky Way.

Regarding the broad component sample, we also found two interesting correlations. The first correlation links the breadth of the profiles (i.e., $b$) with the column densities, which is perspicuously observed for \nv\ and \ovi, and maybe present for the broad \civ\ sample (see Fig.~\ref{f-bn}). The second correlation shows an unambiguous dependency of the distribution of $b$-values and the mean/median values of $b$ with the ionization energies of ions (see Figs.~\ref{f-bhist} and \ref{f-bei}). Since we argued that \nv\ and \ovi\ probe similar types of ionized gas, the immediate conclusion from Figs.~\ref{f-bhist} and \ref{f-bei} is that temperature and/or $b_{\rm nt}$ of the \ovi\ and \nv\ gas must be different (otherwise we should have $\langle b($\nv$)\rangle \ga \langle b($\ovi$)\rangle$ since $b^2 \propto 1/A$). In \S\ref{s-match}, we noted that collisional ionization models predict different temperatures for different high ions \citep[e.g.,][]{borkowski90,gnat07}, so assuming for now that nonthermal motions are negligible, Eqn.~\ref{e-b} leads to $b({\mbox \ovi})/b({\mbox \civ})  = \sqrt{T({\mbox \ovi})/T({\mbox \civ}) A_{\rm C}/A_{O}} = 0.87 \sqrt{T({\mbox \ovi})/T({\mbox \civ})}$ and $b({\mbox \ovi})/b({\mbox \nv})  = 0.94 \sqrt{T({\mbox \ovi})/T({\mbox \nv})}$. For \nv\ and \ovi, the observations give for the mean and median values  $b({\mbox \ovi})/b({\mbox \nv})  = 1.3$ and 1.1, respectively, while the CI and CIE models imply $b({\mbox \ovi})/b({\mbox \nv})  \sim 1.1$ (see \S\ref{s-match}). Doing the same exercise for \civ\ and \ovi\ does not lead to the same agreement, but we also showed that 40\% of the \civ\ components are narrow, lowering the mean value of $b($\civ$)$.  This supports a scenario in which the thermal motions dominate the broadening of the high ions with an increase of the temperature as the ionization energy of the ions increases. In this case, multiple velocity components are required for the broad components in order to explain their breadth. As the number of components would increase with larger $N$ and $b$, we expect in this case a correlation between $N$ and $b$ as found for the \ovi\ and \nv\ (and broad \civ, see Fig.~\ref{f-bn}). We note two caveats in this interpretation: 1) the dispersion in $b$ is large; 2) the comparison is made between the mean/median $b$-values. Regarding the latter point, we, however, also showed in \S\ref{s-match} and Fig.~\ref{f-compcs} that ion-matched components suggest a similar trend for the broad \civ--\siiv\ and \civ--\nv\ pairs. Another possibility is, of course, that $b_{\rm nt}\gg b_{\rm th}$, i.e.,  $b({\mbox \ovi})/b({\mbox \nv})  = b_{\rm nt}({\mbox \ovi})/b_{\rm nt}({\mbox \nv})$. In this context the result from Fig.~\ref{f-bei} could be simply interpreted as nonthermal motions increase with increasing $E_i$, and the correlation  between $b$ and $N$ could be interpreted as a level of disturbance applied to the gas where the amount of high ions increase with the turbulence, e.g., in TMLs \citep{begelman90}. It is likely that both these scenarios and the intermediate one ($b_{\rm nt} \approx b_{\rm th}$) occur.

\section{Summary}\label{s-sum}

We study the properties of transition temperature plasmas in the ISM traced by \siiv, \civ, \nv, and \ovi.  With ionization potentials for creation of 34, 48, 78 and 114 eV, respectively, these ions trace gas in the unstable temperature range from $10^5$ to $10^6$ K.  The \siiv\ and \civ\ can also trace warm ionized gas with $T \la 7 \times 10^4$ K.  We investigate the properties of this gas along the lines of sight to 38 stars in the Milky Way disk using 1.5--2.7 \km\ resolution spectra from STIS of \siiv, \civ, and \nv\,  and 15 \km\ resolution spectra from FUSE of \ovi.   STIS observations of \aliii\ are also studied to provide direct information about photoionized gas in the WIM along each available line of sight.  The observed absorption lines are studied using both the AOD and component fitting methods. The detailed line profiles and  the fit results are given in the Appendix. The high resolution of the observations allows for the first time a clear separation of narrow and broad absorption tracing different temperature regimes of the multi-phase medium along each line of sight in a large sample. Our main results are summarized as follows:

1. Narrow and broad absorption components are seen in \siiv\ and \civ.   The narrow components with $b($\civ$) < 10$ \km\  imply gas with $T<7\times 10^4$ K.   \nv\ and \ovi\ only trace the gas with broad absorption components with $b \ge 10$ \km.   

2. The narrow components can be divided in two categories, the saturated and unsaturated components. The strong, saturated, and  narrow \siiv\  and \civ\ components are signatures of radiative feedback of O-type stars.  The weaker narrow \siiv\  and \civ\ components trace gas in the more diffuse ISM that has radiatively cooled from the transition temperature phase or is photoionized by the EUV radiation from cooling hot gas. 

3. The broad \siiv, \civ, \nv, and \ovi\ components trace collisionally ionized gas  that is probably  undergoing a cooling transition from the hot ISM to the warm/cool ISM. 

4. \siiv\ and \civ\  have very different component behavior than \aliii\ in most cases. \aliii\ is likely mostly produced in the WIM suggesting that \siiv\ and \civ\ have different sites of origin than does \aliii. Most of the observed \siiv\ and \civ\  in the Galactic disk does not arise in the WIM very likely because the WIM is strongly affected by the presence of the \hei\ absorption edge at 24.6 eV in stellar atmospheres.

5.  The median and average of the Doppler parameter, $b$,  distributions for \aliii, \siiv, \civ, \nv, and \ovi\ increase with the energy, $E_i$, required to produce the ion. The distributions for \aliii, \siiv, and \civ\ are skewed to higher $b$-values and are consistent with a lognormal distribution.  In contrast, the distributions for \nv\ and \ovi\ are normal.  

6. For the broad \siiv\ and \civ\ components with matching velocities,  $ \langle N($\civ$)/N($\siiv$) \rangle = 3.9 \pm 1.9$. The small dispersion in this ratio suggests that a regulation process is operating.  That process could be controlled by the cooling of hot gas. In the narrow unsaturated  \siiv\ and \civ\ components  $N($\civ$)/N($\siiv$)$ ranges from 0.4 to 11.1 with a medium of 4.1 in sight lines unaffected directly by the X-ray emission from OB associations.  The large observed range is certainly  related to origins of these ions in both photoionized  and cooling collisionally ionized gas.

7.  Combining the findings from our analyses of the b-values and ionic ratios of the high ions, we argue that CIE is not a viable model in the Milky Way.      

8. Most of the observed interstellar \siiv, \civ, \nv, and \ovi\ absorption in the galactic disk appears to  provide a direct or an indirect diagnostic of cooling hot gas in the ISM. Only the strong saturated components of \siiv\ and associated \civ\ directly trace the O-type star environments.

\acknowledgments
We appreciate comments from Ed Jenkins and Andrew J. Fox. Support for this research was provided by NASA through grant HST-AR-11265.01-A from the Space Telescope Science Institute, which is operated by the Association of Universities for Research in Astronomy, Incorporated, under NASA contract NAS5-26555. This research has made extensive use of the NASA Astrophysics Data System Abstract Service and the Centre de Donn\'ees de Strasbourg (CDS). We finally thank you the referee for his/her thorough reading of our manuscript and useful suggestions.

\appendix

\section{A. Summary Figure}\label{s-fig}

Figure~\ref{f-sum} consists of several panels that show the profiles, normalized profiles, AOD profiles as well as the fitting results and AOD ionic ratios.  At the lower right corner is the star ID, star name, spectral type/luminosity class, Galactic coordinates, and distance (derived from spectroscopic parallax) of the star as well as the $R$-value indicating the importance of X-ray emission tracing hot gas in OB associations along the path to the target star (see \S\ref{s-description}).  The panel on the left shows the absorption profiles with the fitted continua (blue curves) for the available high ions and \aliii. The next plot shows the normalized profiles with the vertical green dotted lines representing the velocity centroids from the \siiv\ component fit and the vertical blue dashed line showing the position of the stellar velocity. If part of the spectrum is gray near the \ovi\ or near \siiv\ $\lambda$1393, this region highlights contaminated velocities by unrelated lines (see \S\ref{s-cont}).  The sub-panels are ordered by $E_i$ from top (high $E_i$, generally \ovi) to bottom (low $E_i$,  singly-ionized species).  Next, the upper plot shows the $N_a(v)$ apparent optical depth profiles for each doublet and \ovi\ $\lambda$1031 (\ovi\ $\lambda 1037$ is always contaminated and cannot be used).  There is generally a good agreement between the $N_a(v)$ for a given species (except for obvious parts that are saturated). The lower plot shows the column density ratios as a function of velocity for the high ions where the pixel bin has a size of the \fuse\ resolution when \civ\ or \nv\ are compared to \ovi\ or the STIS resolution element when high ions in the STIS bandpass are compared.   The mean and dispersion on the ionic ratios based on the column-density ratio profiles are indicated by the horizontal solid and dotted lines, respectively. Note that there is not generally a simple gradient velocity in the ratio as tentatively suggested by \citet{indebetouw04a}. The last panel to the far right shows the results from the component fitting.  The red curve indicates the global component model, and the blue lines show the individual components.  The vertical green dotted lines represent the velocity centroids for \siiv\ while the red tick-marks show the velocity-centroids for each species. In this panel, we also highlight again in gray the velocity interval over which \siiv\ $1393$ can be contaminated by \Niii\ $\lambda$1393. 

\section{B. Broad components with \MakeLowercase{\it b}\,$\ge 20$ \km}\label{s-abroad}

We argued in \S\ref{s-anal} that components with $b\ge 20$ \km\ may be more prone to uncertainties, and we review here their importance for each ion and possible troublesome cases. 

For \aliii, only 3/56 components have $b\ge 20$ \km. In two of these cases (stars 7 and 17), the profiles are complex and the broad components are blended with several narrower components and do not align with broad components seen in the other species. In star 20, however, the broad \aliii\ component aligns with the broad component seen at the same velocities in the high ions, giving some confidence in this result. 
 
For \siiv, 12/138 have components with $b\ge 20$ \km. Several of the \siiv\ profiles are complex, but for many of those, the broad components align with the broad components seen in other ion profiles. There are two cases (stars 15, 18) where the results may be somewhat uncertain because the broad component does not align with other broad components seen in \civ\ or \nv. In two other cases (stars 27, 29), the profiles are simple but the S/N is quite low, and it could be possible that with better S/N, additional components may provide a better fit. 
 
For \civ, 35/130 have components with $b\ge 20$ \km. As for \siiv, while many \civ\ profiles are complex, the broad components align with the broad components seen in other ion profiles, giving further support to the results of the fit. For three stars (27, 29, 37), the S/N is quite low. For star 7, the broad component could be present to overcome some uncertainties in the continuum placement and does not align so well with the broad component seen in \aliii\ or \siiv. In the spectra of stars 19 and 33, the \civ\ profiles are complex and the broad components may be suspect as they are not evident in the profiles of other species.

For \nv, 19/24 components have $b\ge 20$ \km. The velocity structures of the profiles are not complicated, with often just one component for the given S/N. However, as we noted in \S\ref{s-anal}, the \nv\ absorption is often weak, and the stellar continuum  near this doublet can be complicated, leading the errors to be dominated by the continuum placement. The results regarded uncertain owing to complicated stellar continua in this region of the spectrum (see \S\ref{s-anal}) are followed by colons in Table~\ref{t-fit} instead of formal errors.  

\section{C. Volume Densities}\label{s-vol}

As most of our sight lines are at $|z|\le 1.3$ kpc (36 stars), our sample allows us to  derive the volume density of a given ion that should be a good representation of the midplane density. The average volume density for each sight line is calculated by $\log n =\log(N/d)$ (where $n$ is in cm$^{-3}$ and $N$ in cm$^{-2}$), where $N$ is the total integrated column density, and $d$ is the distance to the star. As the column densities are lognormal distributed, we estimate the average and dispersion on $\log n$. Figure~\ref{f-avgn} shows $N$ versus $d$ for \aliii, \siiv, \civ, \nv, and \ovi\ where the solid and dotted lines are the average volume density for the $R\ge 0$ and $R=0$ cases.

In Table~\ref{t-dens}, we summarize the average volume densities of the various high ions and \aliii\ (see Table~\ref{t-aodf} for the individual estimate of $n$). We consider various samples based on their $R$ values. As \citet{bowen08} pointed out in their \ovi\ survey and which apply here, neither the upper or lower limits are randomly distributed. Hence the Kaplan-Meier estimator generally used in samples where both detection and limits are present is not adequate. We therefore treat limits as real values but for that reason we also show in this table the average values when limits are excluded. For \ovi\ and \civ, there is only 1 upper and 1 lower limits, respectively, and therefore these samples are not really affected by the limits (see rows 1 and 3 of Table~\ref{t-dens}). The \aliii\ sample has 3 lower limits and hence is only marginally affected by the limits. However, the \siiv\ sample has 8 lower limits and the \nv\  sample has 8 upper limits, changing the average values by 0.2 dex. 

There is an evident increase in the volume densities from $R=0$  to $R>0$ to the Carina nebula, implying that the stellar environments affect the production of the high ions. This effect is more notable for \aliii\ and \siiv\ than \civ, \nv, and \ovi.  This is not unforeseen  as \aliii\ and \siiv\ can be more easily photoionized by hot stars. The much higher volume densities near Carina are likely due to the extremely active star formation in that region,  without doubt the most active region in our survey. 

The average densities found for \aliii, \siiv, \civ, and \ovi\ for the $R = 0$ line-of-sight sample are smaller  by 0.29, 0.16, 0.14 and 0.15 dex, respectively,  than the mid-plane densities found by \citet{savage09} in their study of the extension of the high ions into the lower Galactic halo.   The nature of the stellar sight-line sample can strongly affect the derived values of average density.  For example, \citet{bowen08} estimate $\langle \log n($\ovi$)\rangle = -7.78 \pm 0.33$ (stdev) their $R = 0$ sample of 42 stars with distances from 200 to 2000 pc and $\langle \log n($\ovi$)\rangle = -7.92 \pm 0.33$ (stdev)  for  their $R = 0$ sample of  56 stars with distances $>2000$ pc.  A combination of sample bias and the use of very different techniques to estimate the average density in the different studies could explain the differences among the various average density determinations found in the \citet{bowen08}, \citet{savage09}, and Table~\ref{t-dens} of this paper; we note that in all these cases the dispersions are quite large. The new results for \nv\ in this paper could be combined with measures of \nv\ column densities toward extragalactic objects to obtain an improved scale height estimate for Galactic \nv.

\clearpage

\begin{figure*}[tbp]
\epsscale{1.2} 
\plotone{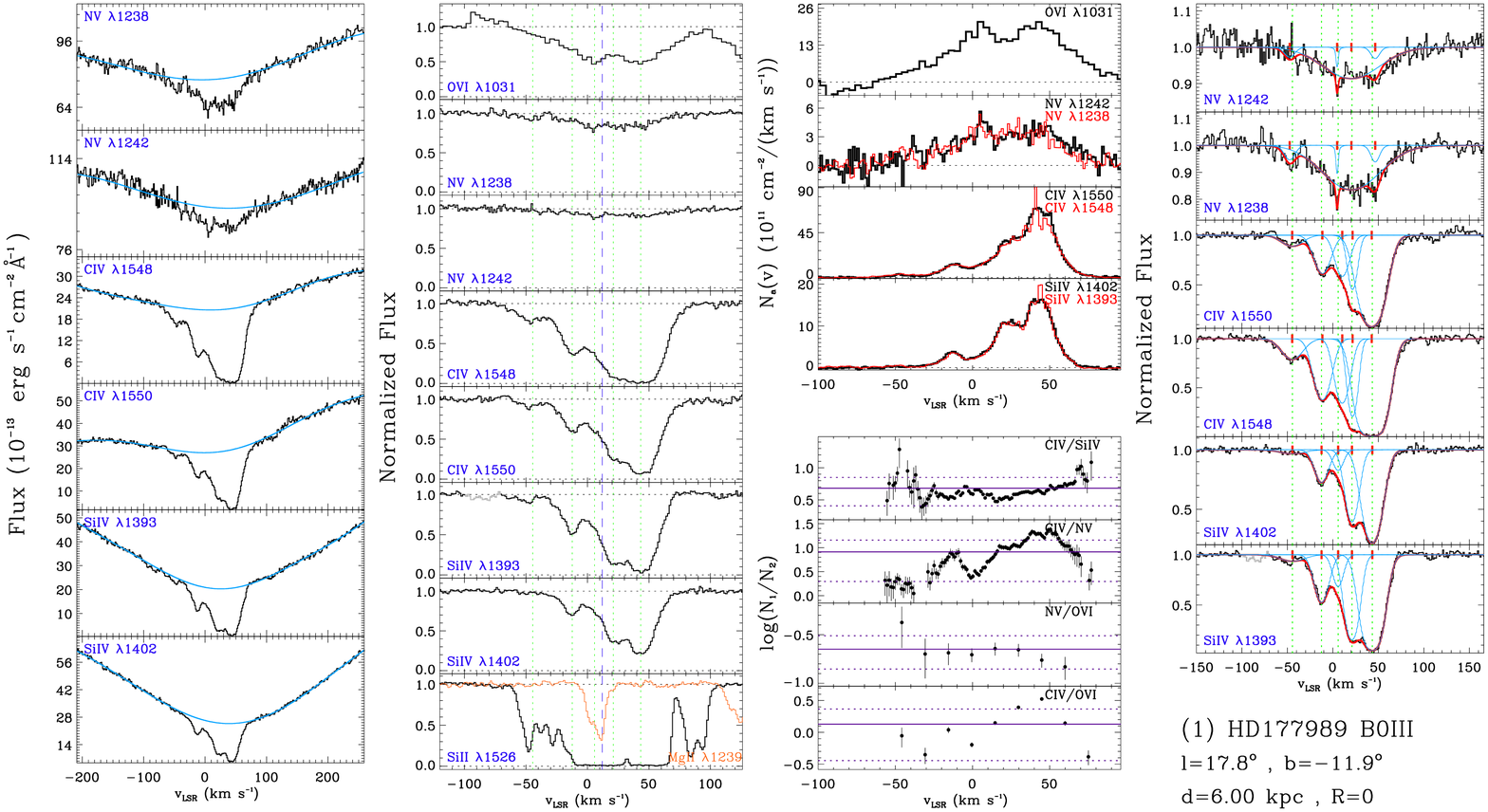}
\end{figure*}

\begin{figure*}[tbp]
\epsscale{1.2} 
\plotone{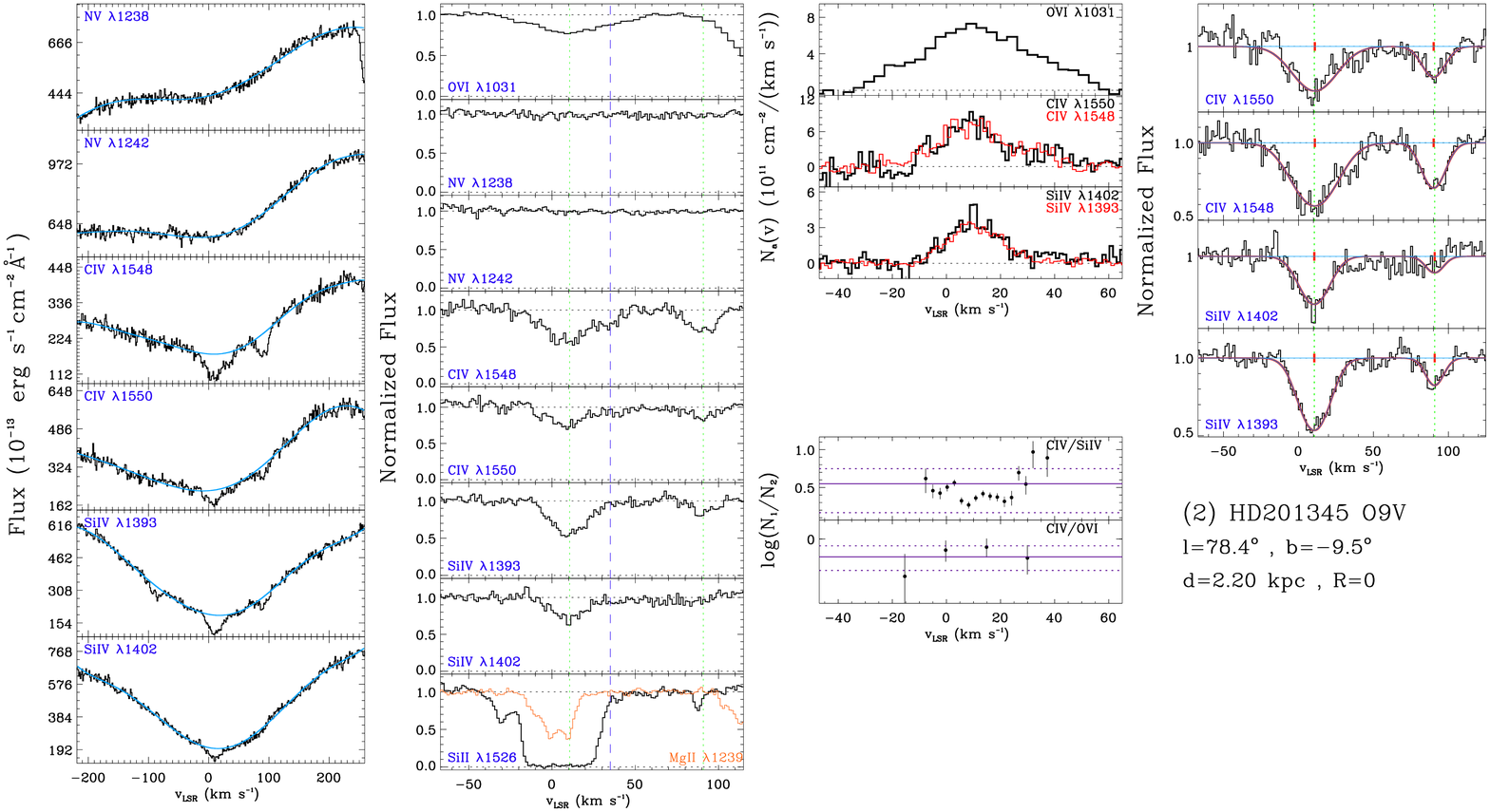}
\caption{Summary figures for each line of sight. See text in Appendix~\ref{s-fig} for full details. \label{f-sum}}
\end{figure*}

\clearpage

\begin{figure*}[tbp]
\epsscale{1.2} 
\plotone{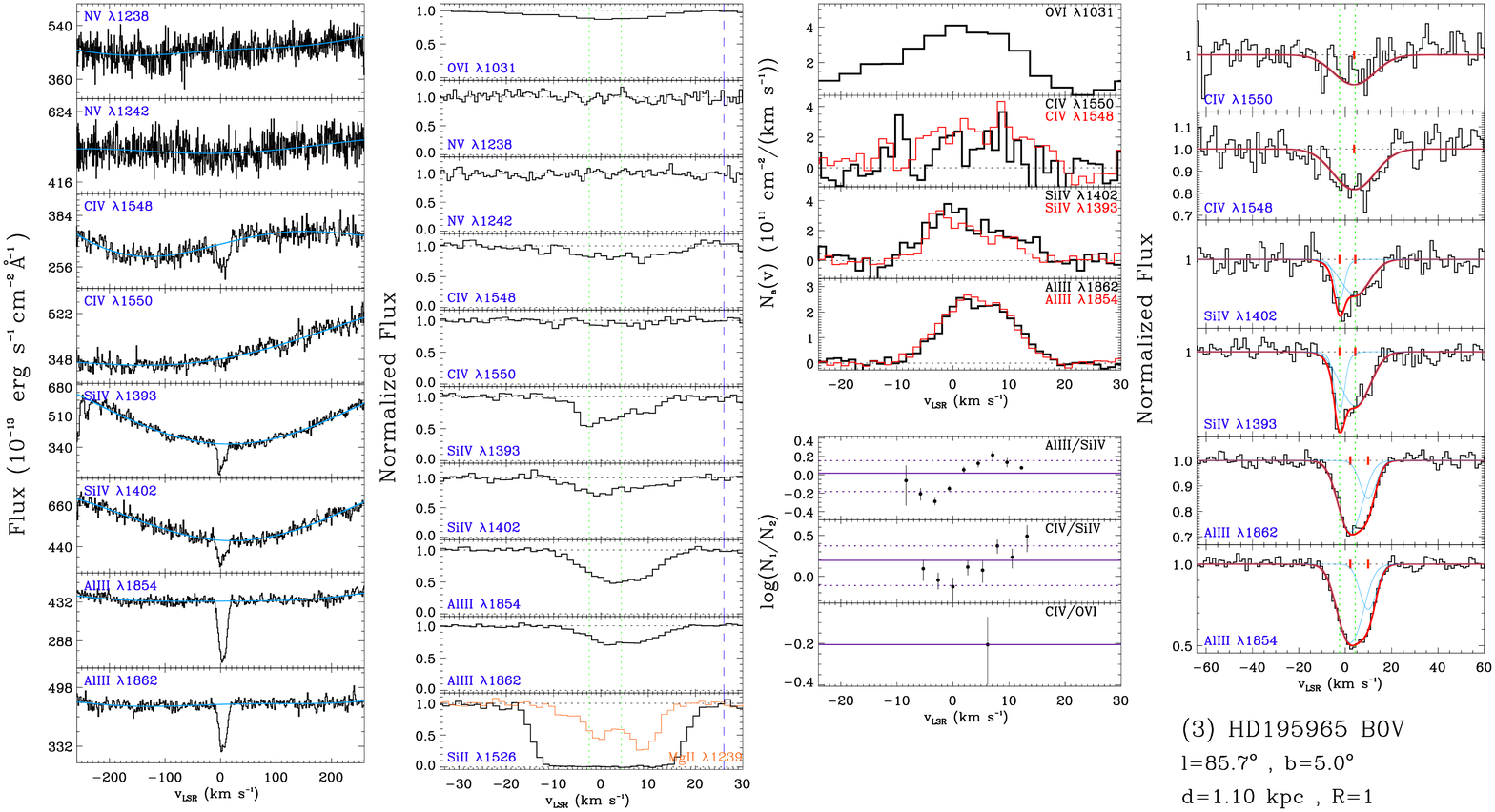}
\end{figure*}

\begin{figure*}[tbp]
\epsscale{1.2} 
\plotone{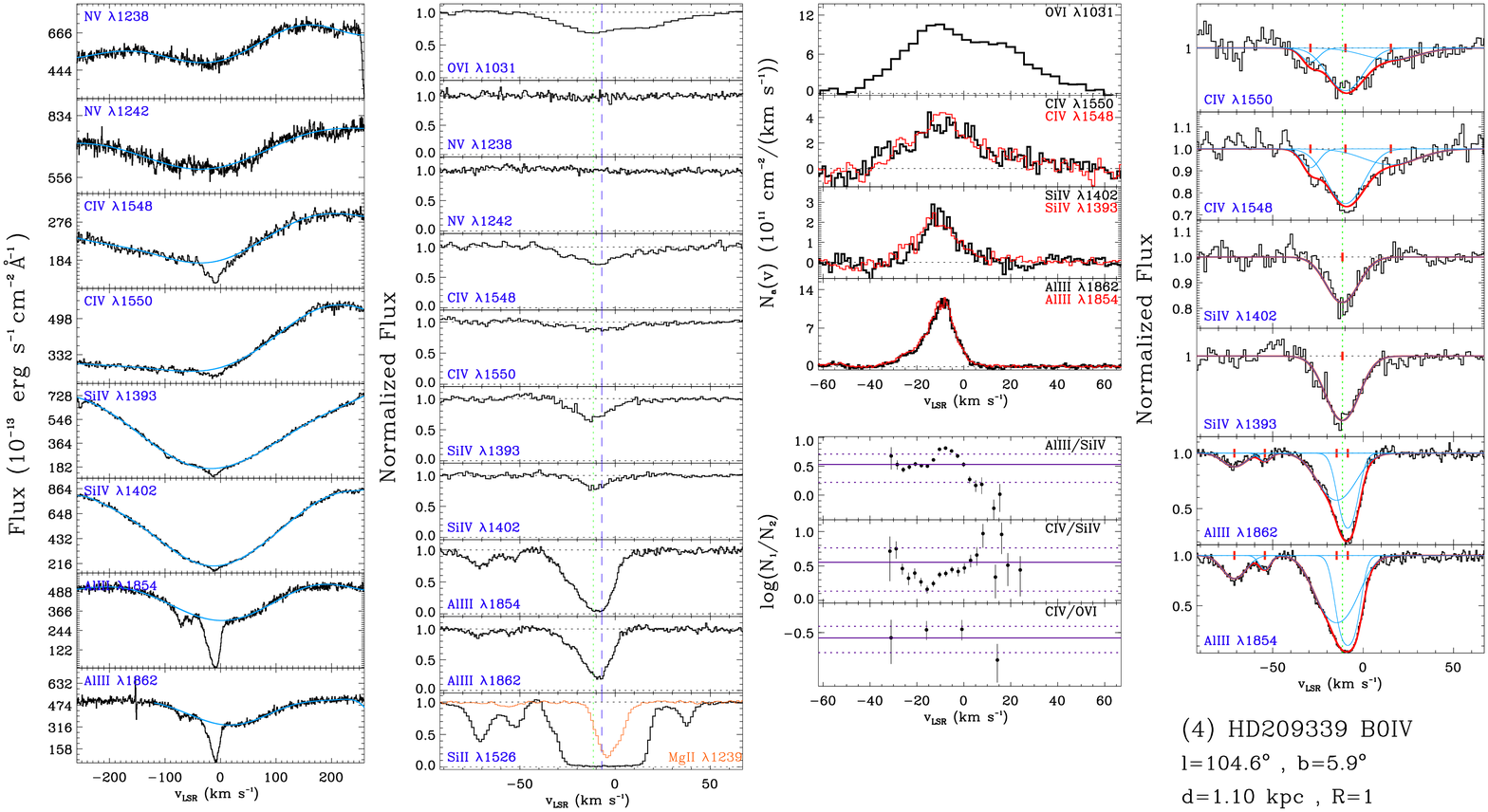}
\figurenum{\ref{f-sum}}
\caption{{\it Continued}}
\end{figure*}

\clearpage

\begin{figure*}[tbp]
\epsscale{1.2} 
\plotone{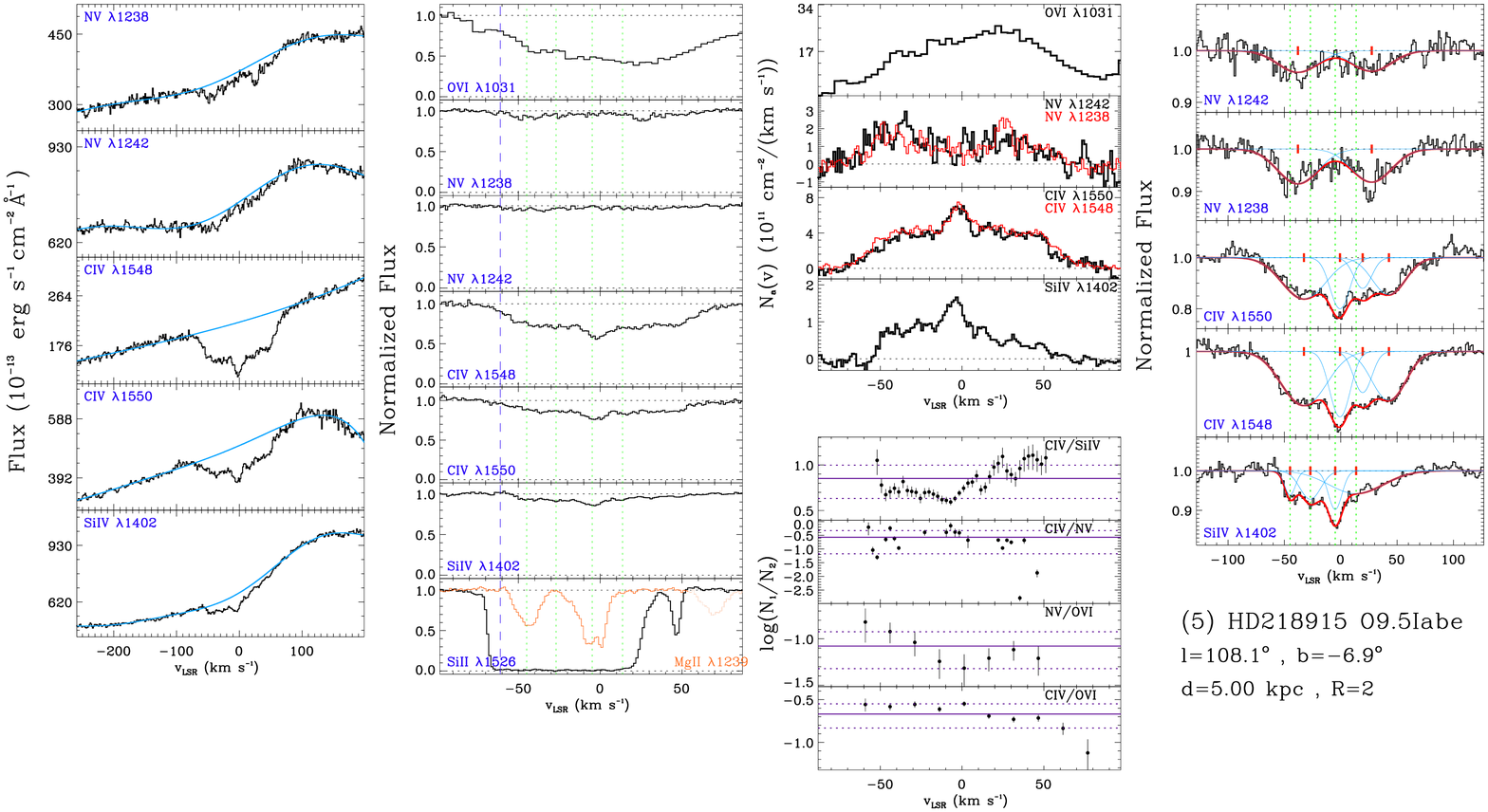}
\end{figure*}

\begin{figure*}[tbp]
\epsscale{1.2} 
\plotone{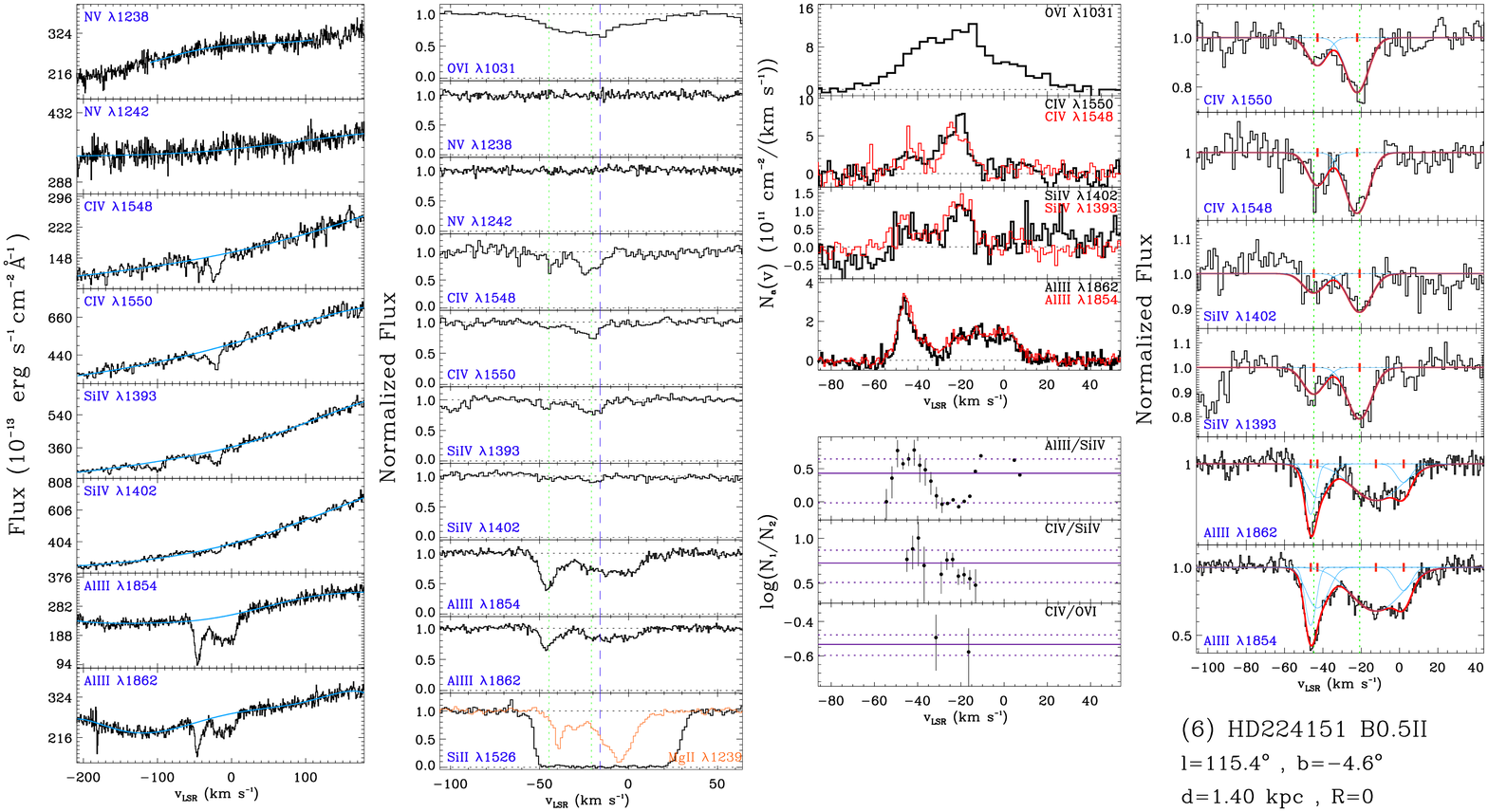}
\figurenum{\ref{f-sum}}
\caption{{\it Continued}}
\end{figure*}

\clearpage

\begin{figure*}[tbp]
\epsscale{1.2} 
\plotone{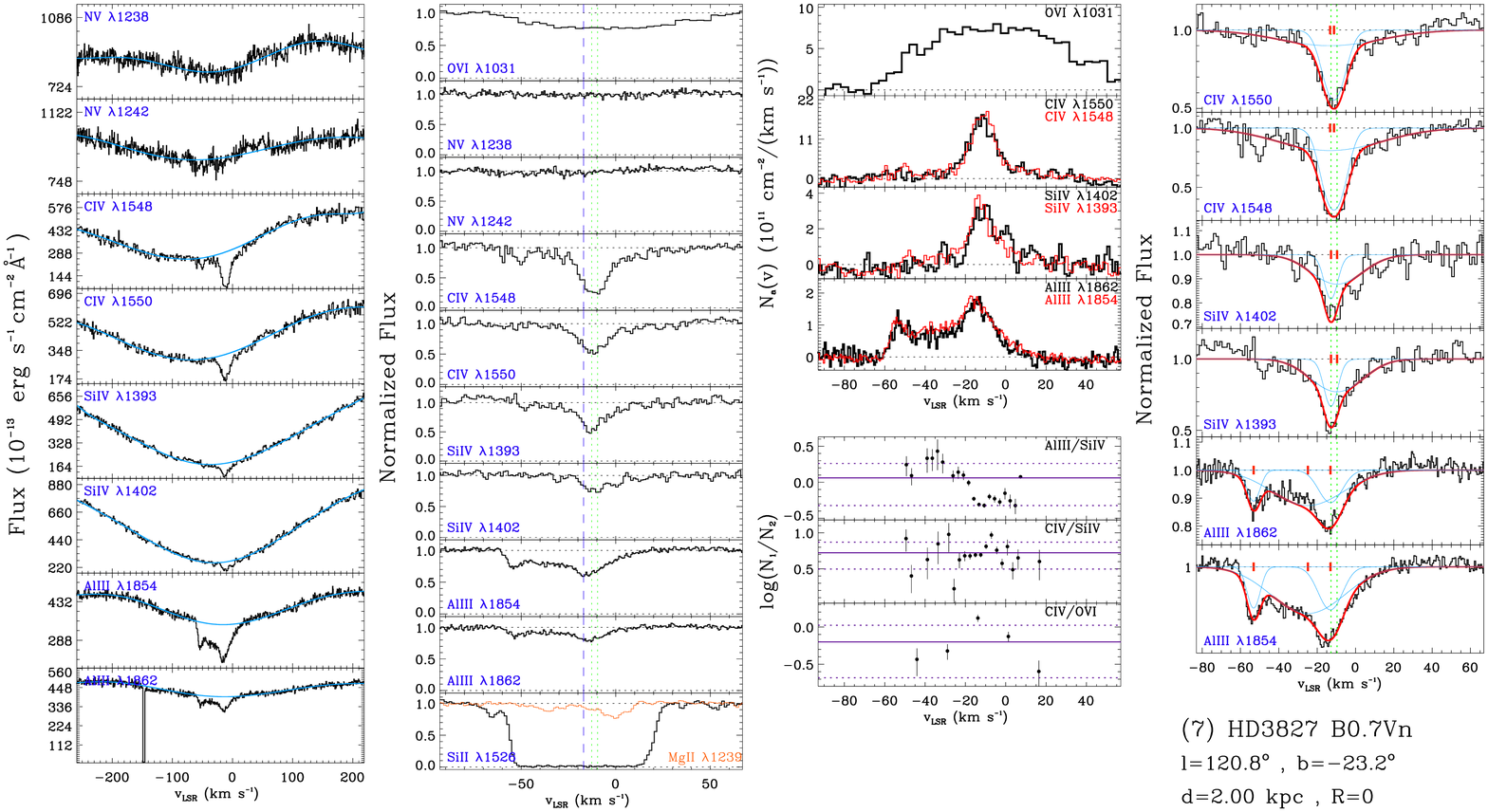}
\end{figure*}

\begin{figure*}[tbp]
\epsscale{1.2} 
\plotone{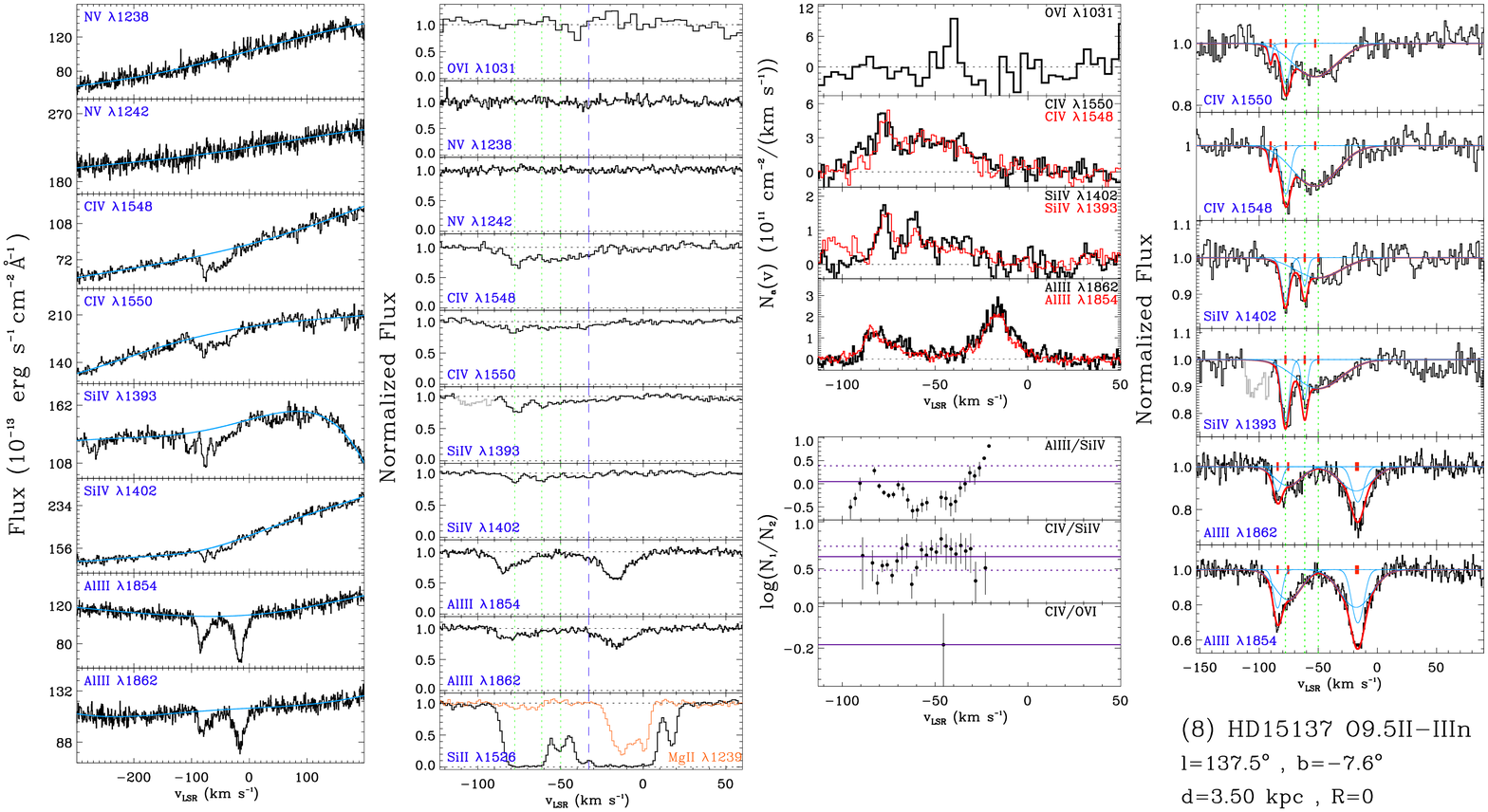}
\figurenum{\ref{f-sum}}
\caption{{\it Continued}}
\end{figure*}

\clearpage

\begin{figure*}[tbp]
\epsscale{1.2} 
\plotone{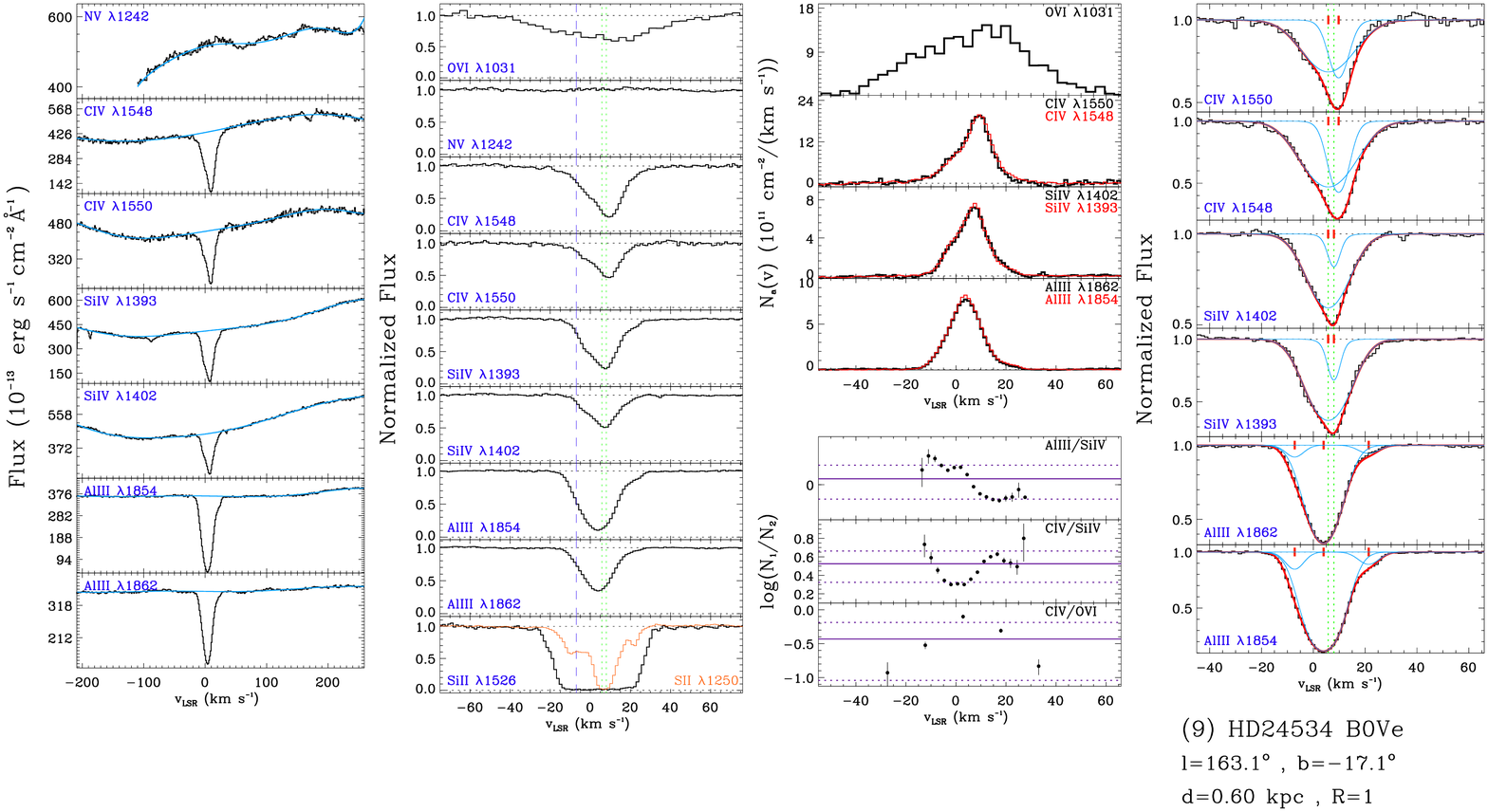}
\end{figure*}

\begin{figure*}[tbp]
\epsscale{1.2} 
\plotone{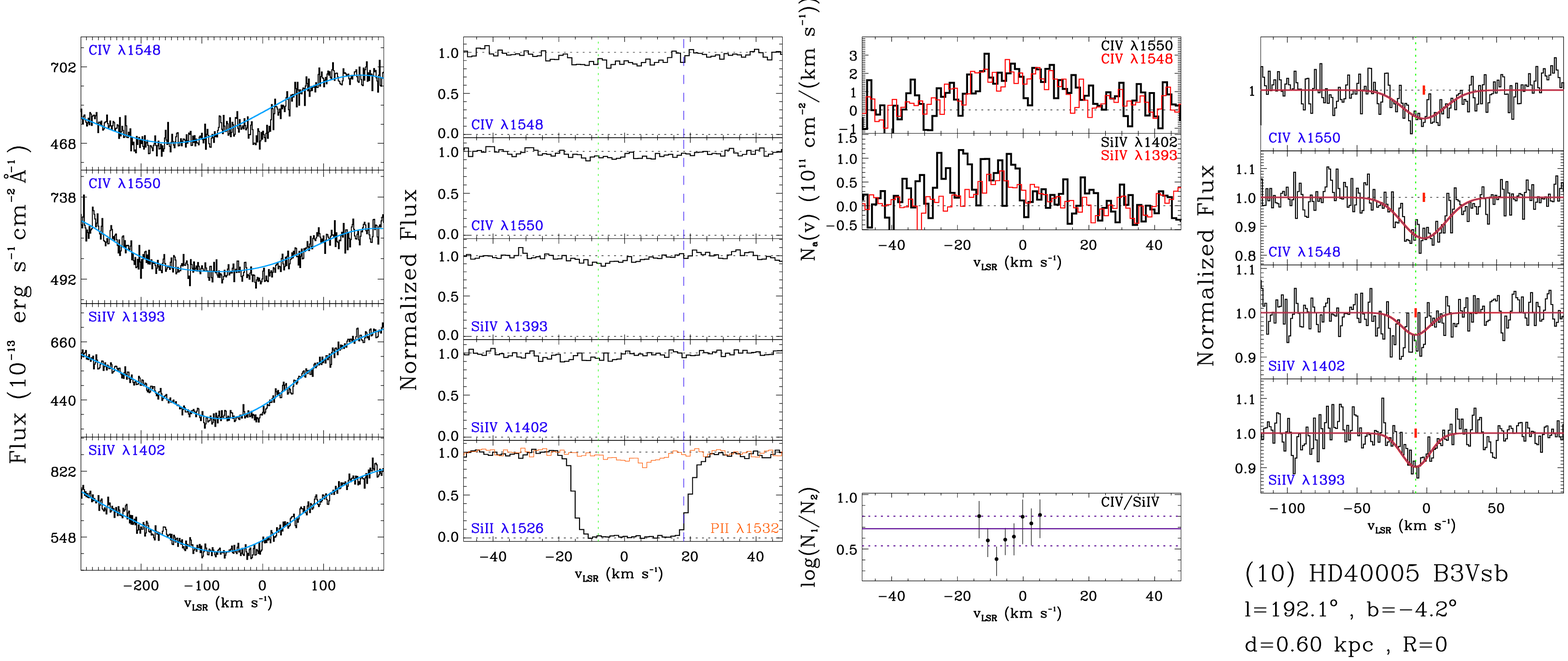}
\figurenum{\ref{f-sum}}
\caption{{\it Continued}}
\end{figure*}

\clearpage

\begin{figure*}[tbp]
\epsscale{1.2} 
\plotone{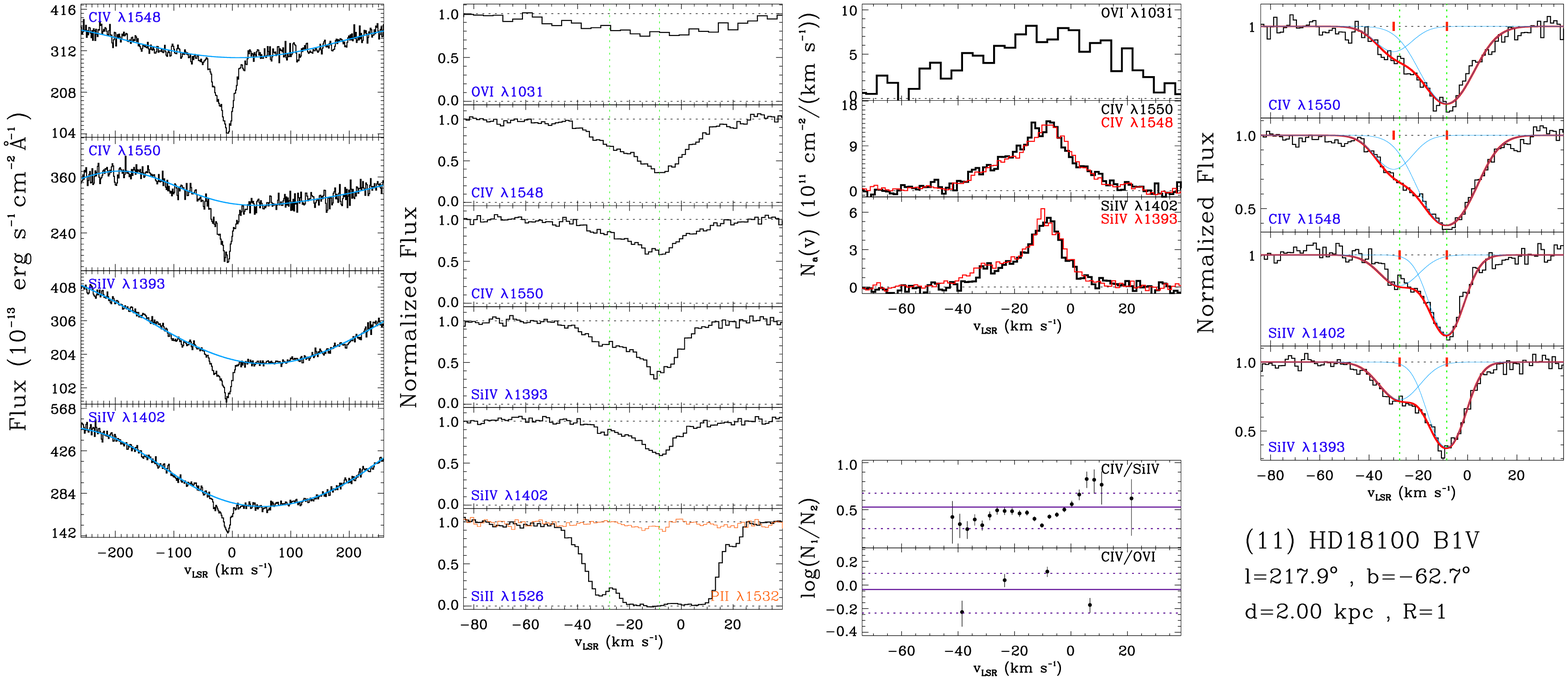}
\end{figure*}

\begin{figure*}[tbp]
\epsscale{1.2} 
\plotone{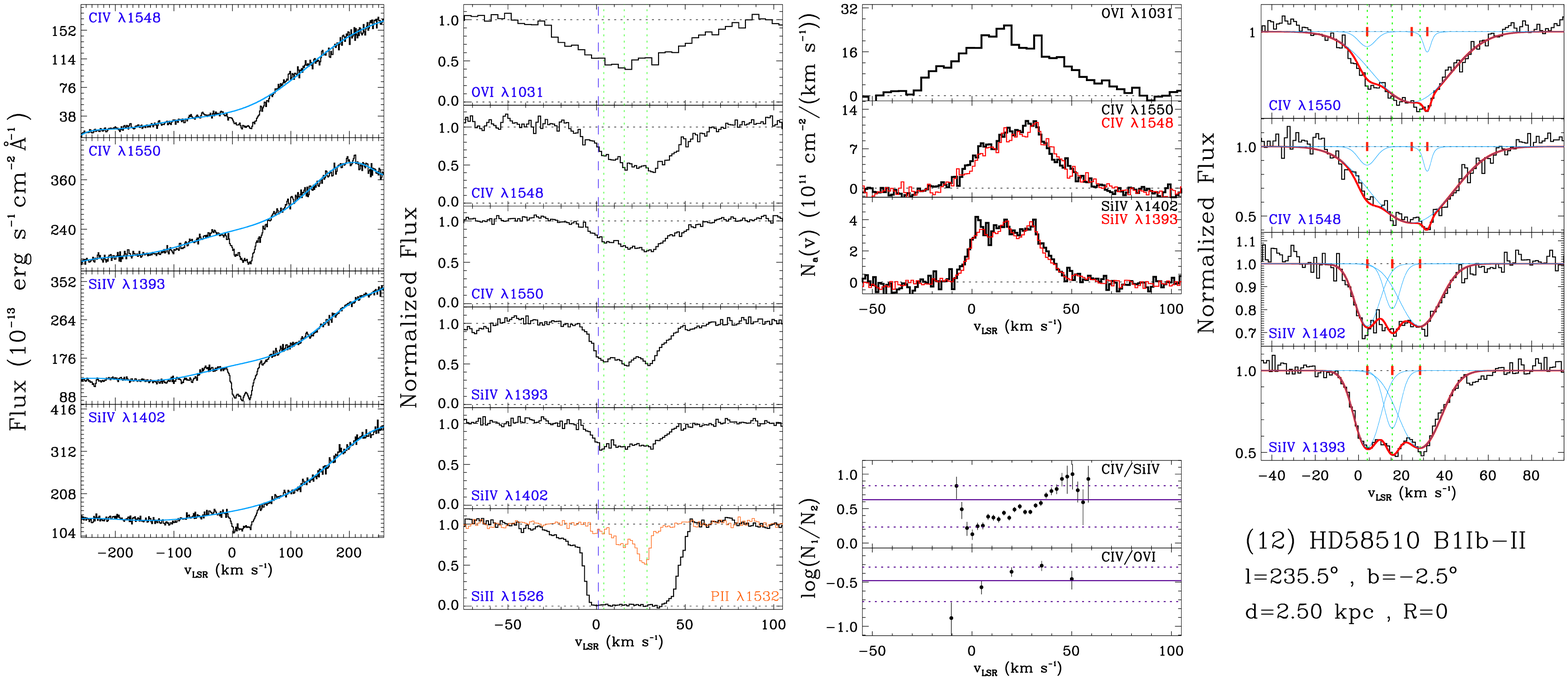}
\figurenum{\ref{f-sum}}
\caption{{\it Continued}}
\end{figure*}

\clearpage

\begin{figure*}[tbp]
\epsscale{1.2} 
\plotone{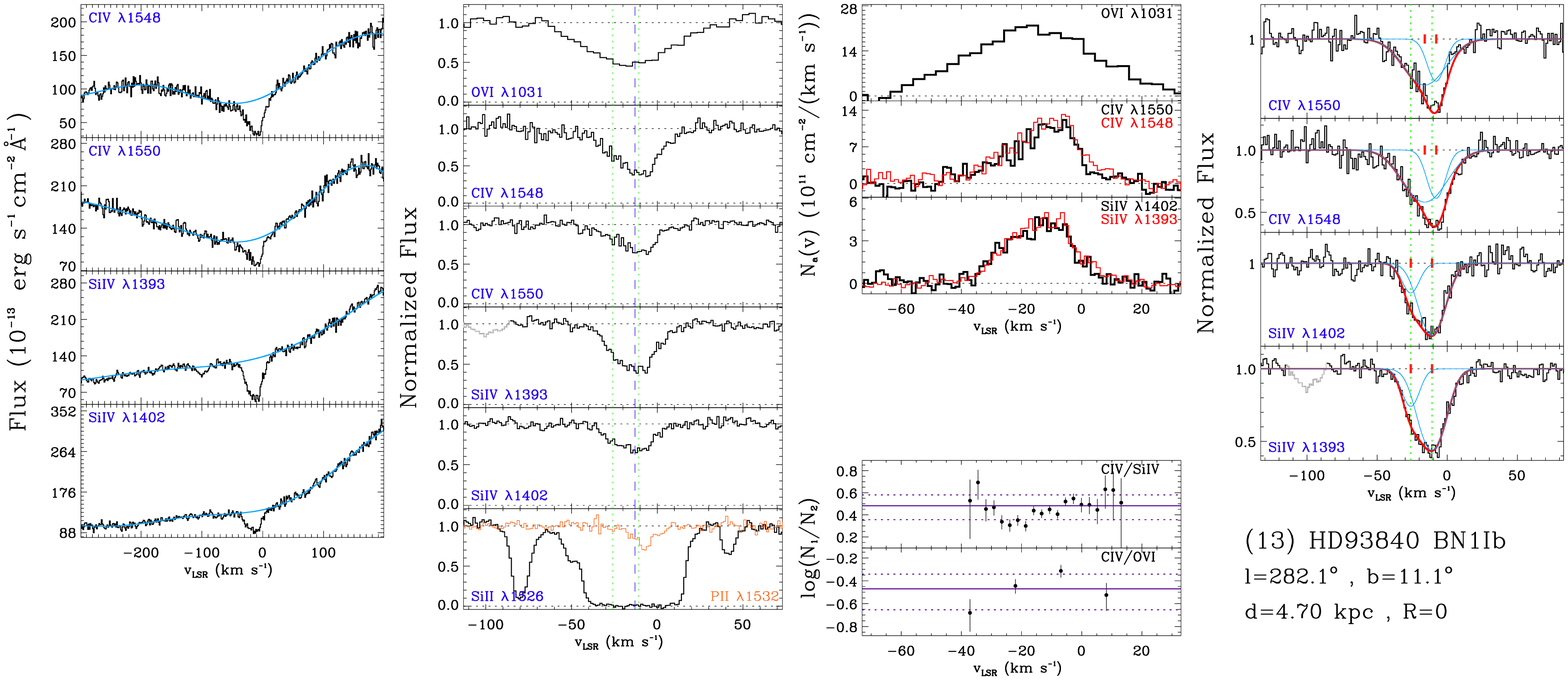}
\end{figure*}

\begin{figure*}[tbp]
\epsscale{1.2} 
\plotone{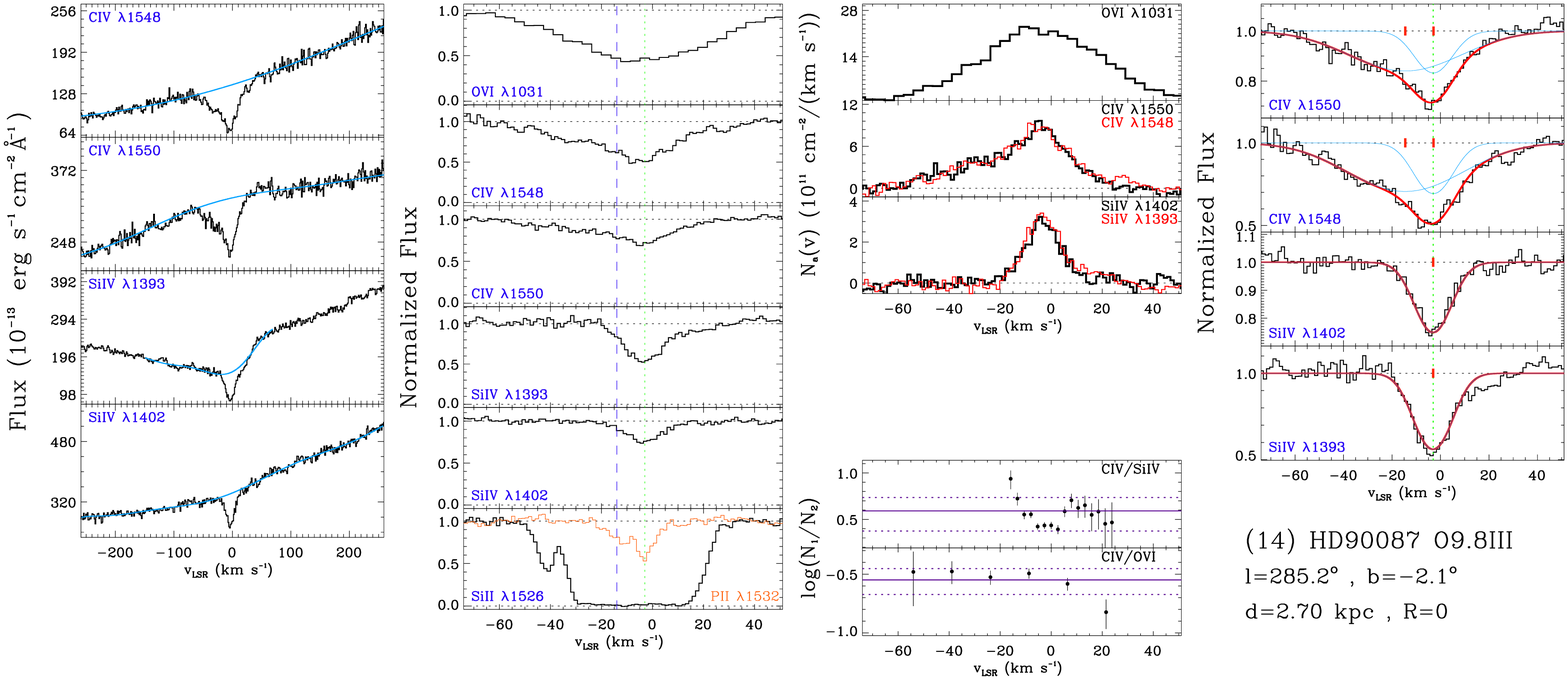}
\figurenum{\ref{f-sum}}
\caption{{\it Continued}}
\end{figure*}

\clearpage 

\begin{figure*}[tbp]
\epsscale{1.2} 
\plotone{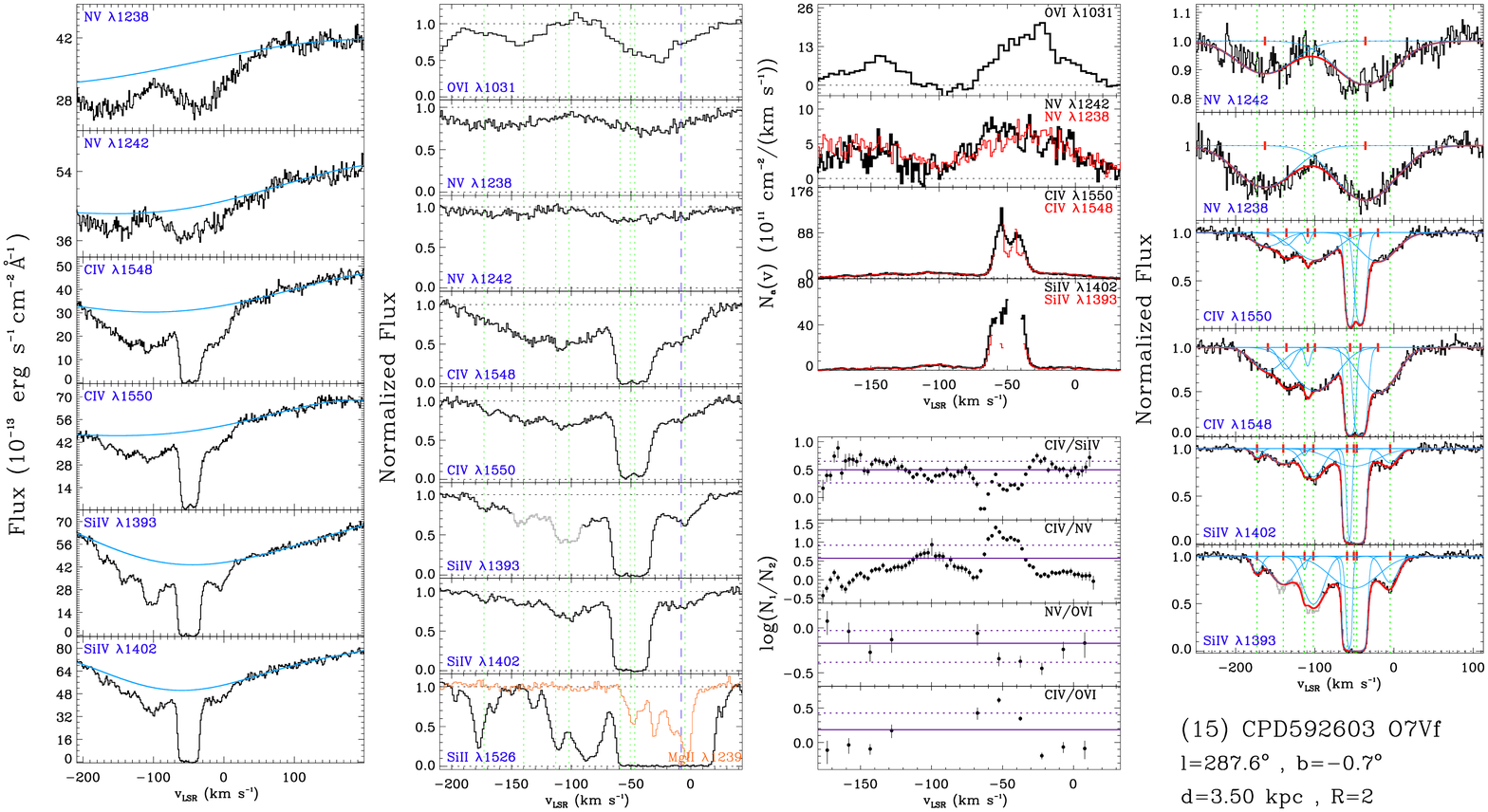}
\end{figure*}

\begin{figure*}[tbp]
\epsscale{1.2} 
\plotone{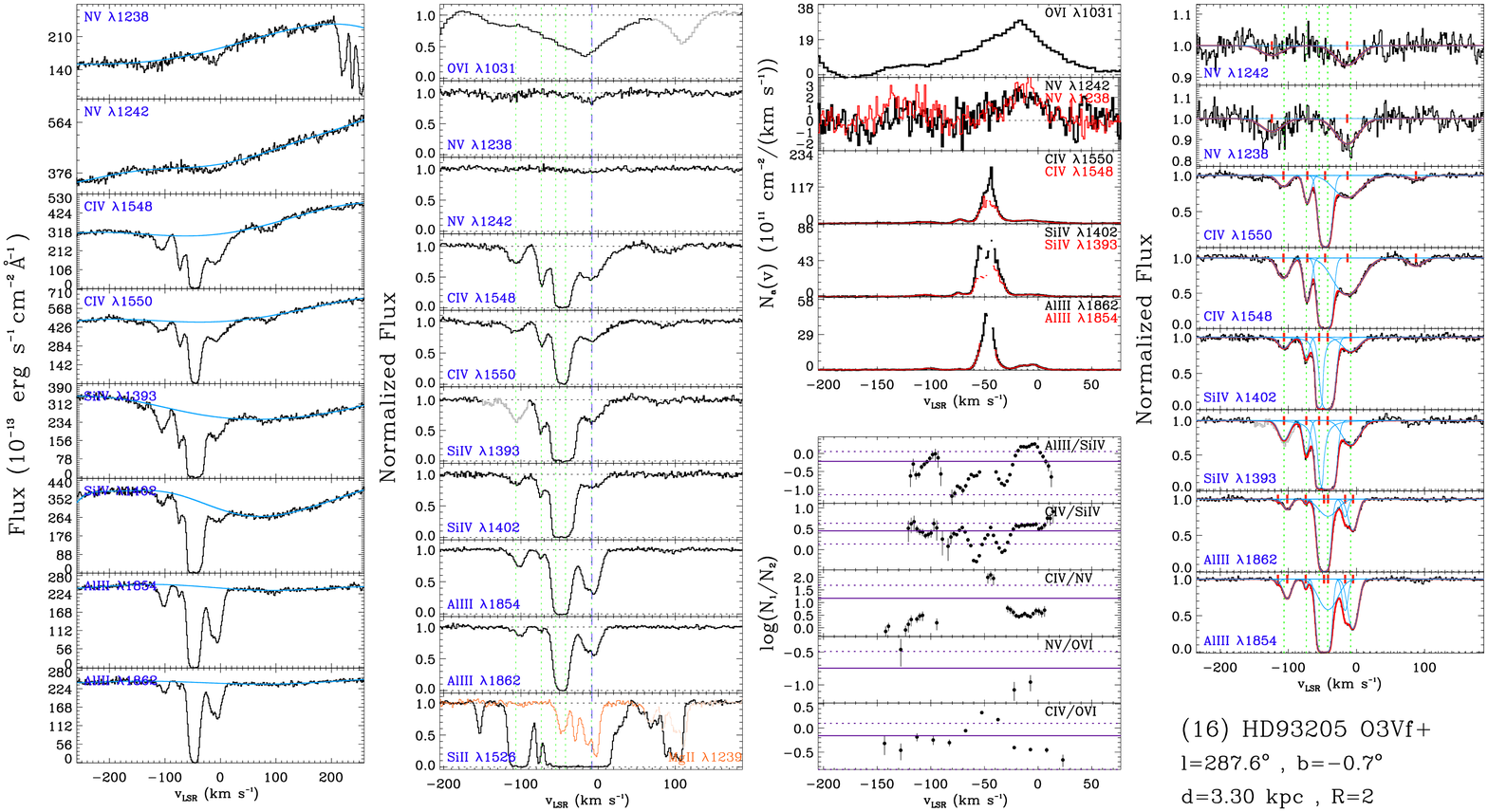}
\figurenum{\ref{f-sum}}
\caption{{\it Continued}}
\end{figure*}

\clearpage

\begin{figure*}[tbp]
\epsscale{1.2} 
\plotone{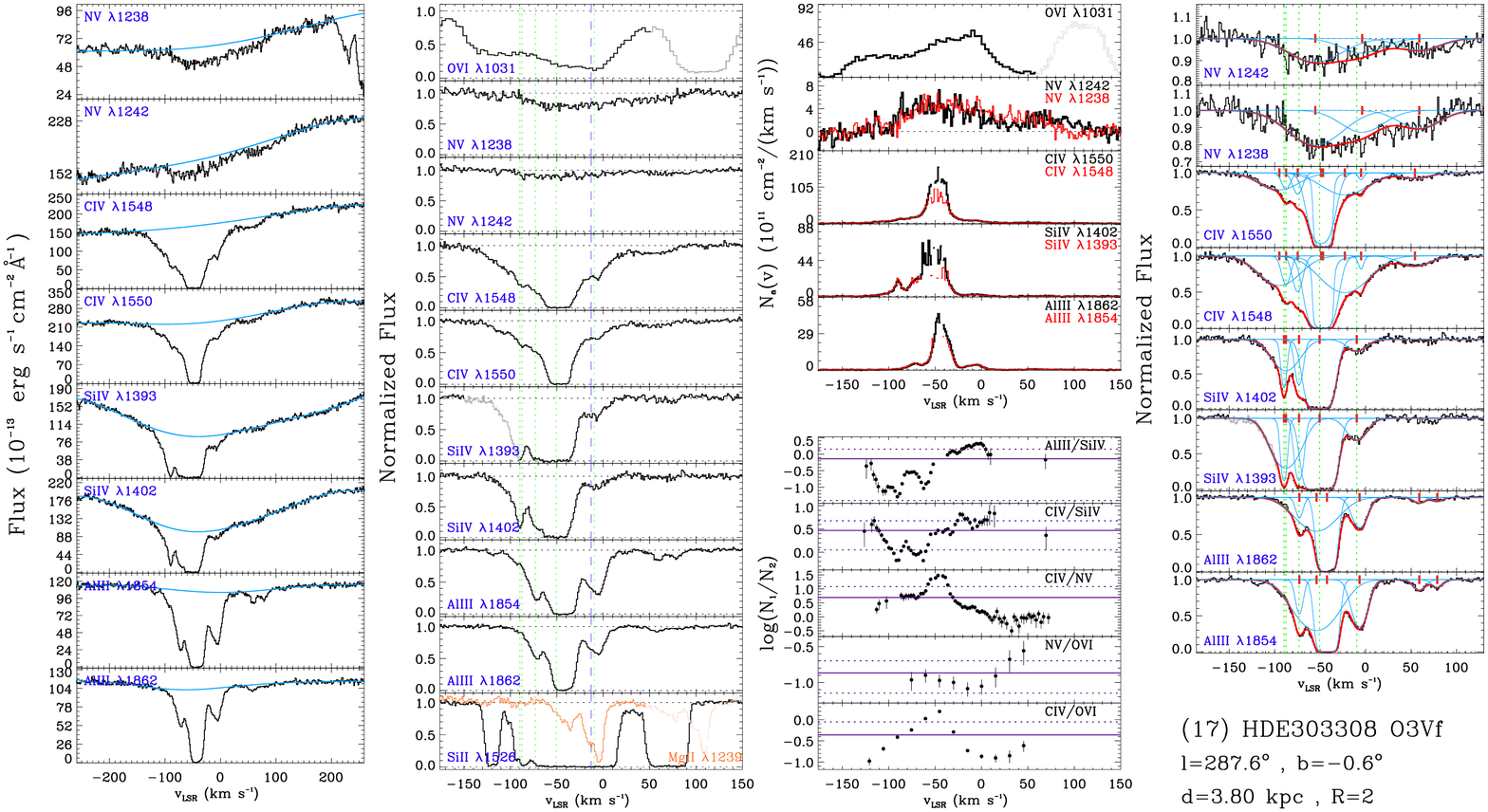}
\end{figure*}

\begin{figure*}[tbp]
\epsscale{1.2} 
\plotone{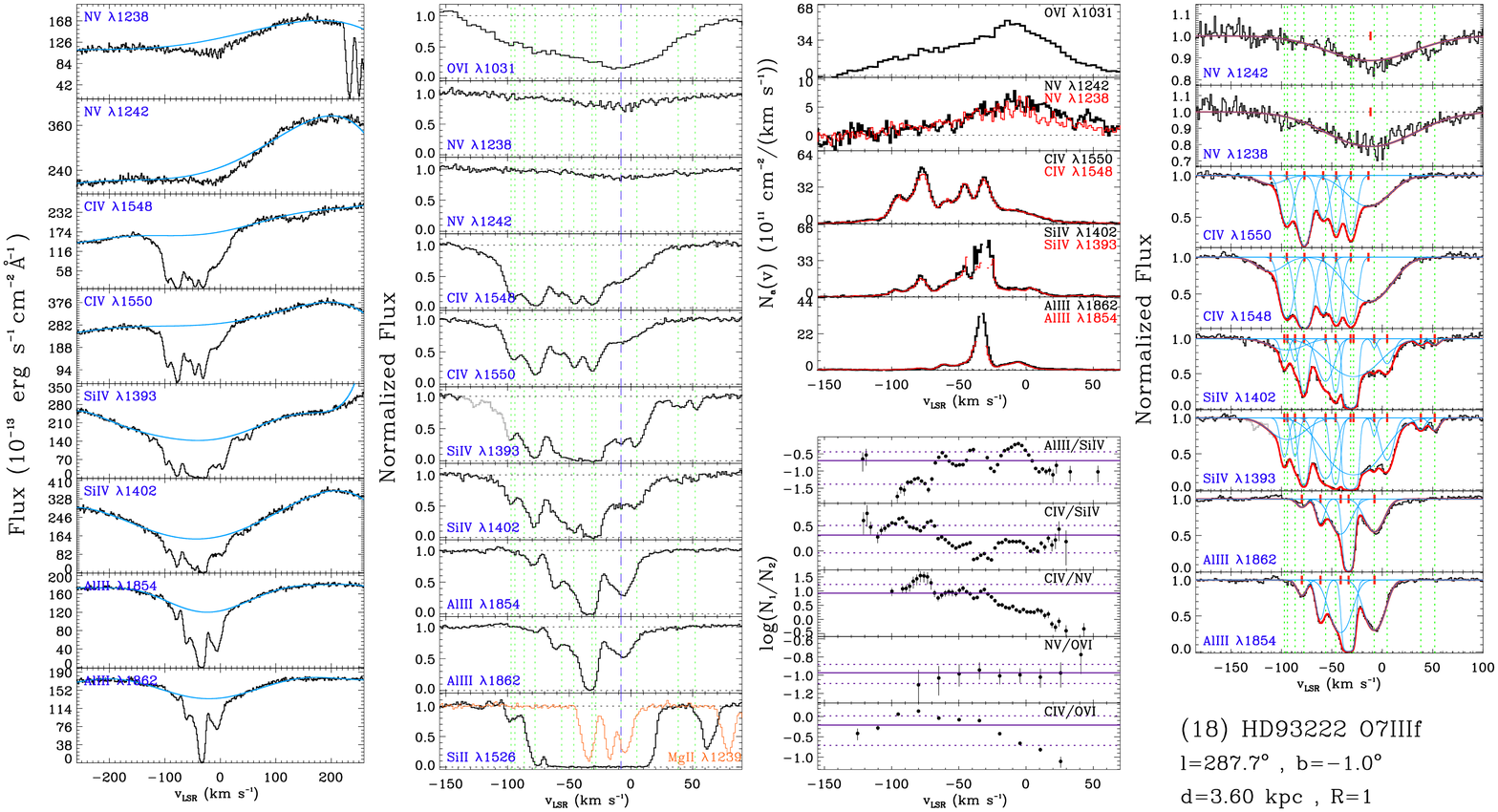}
\figurenum{\ref{f-sum}}
\caption{{\it Continued}}
\end{figure*}

\clearpage 

\begin{figure*}[tbp]
\epsscale{1.2} 
\plotone{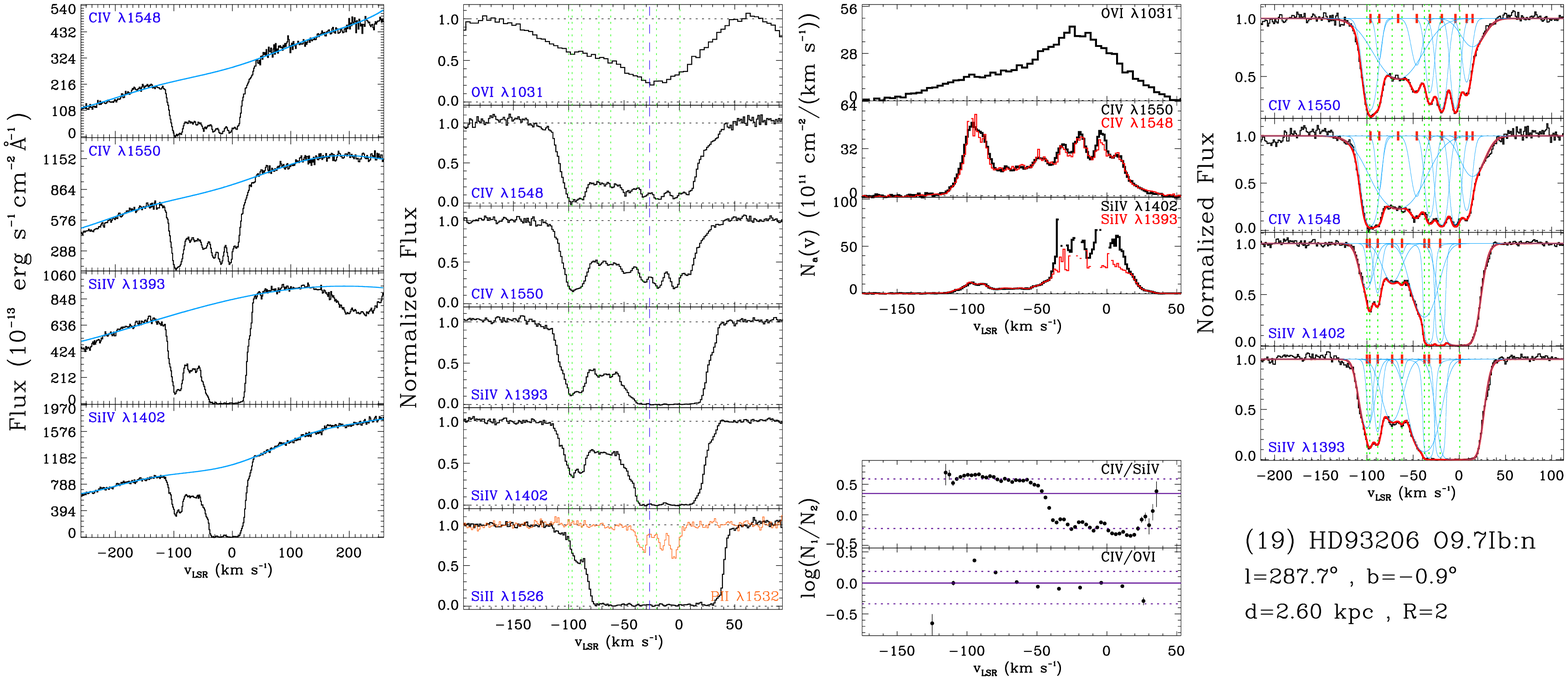}
\end{figure*}

\begin{figure*}[tbp]
\epsscale{1.2} 
\plotone{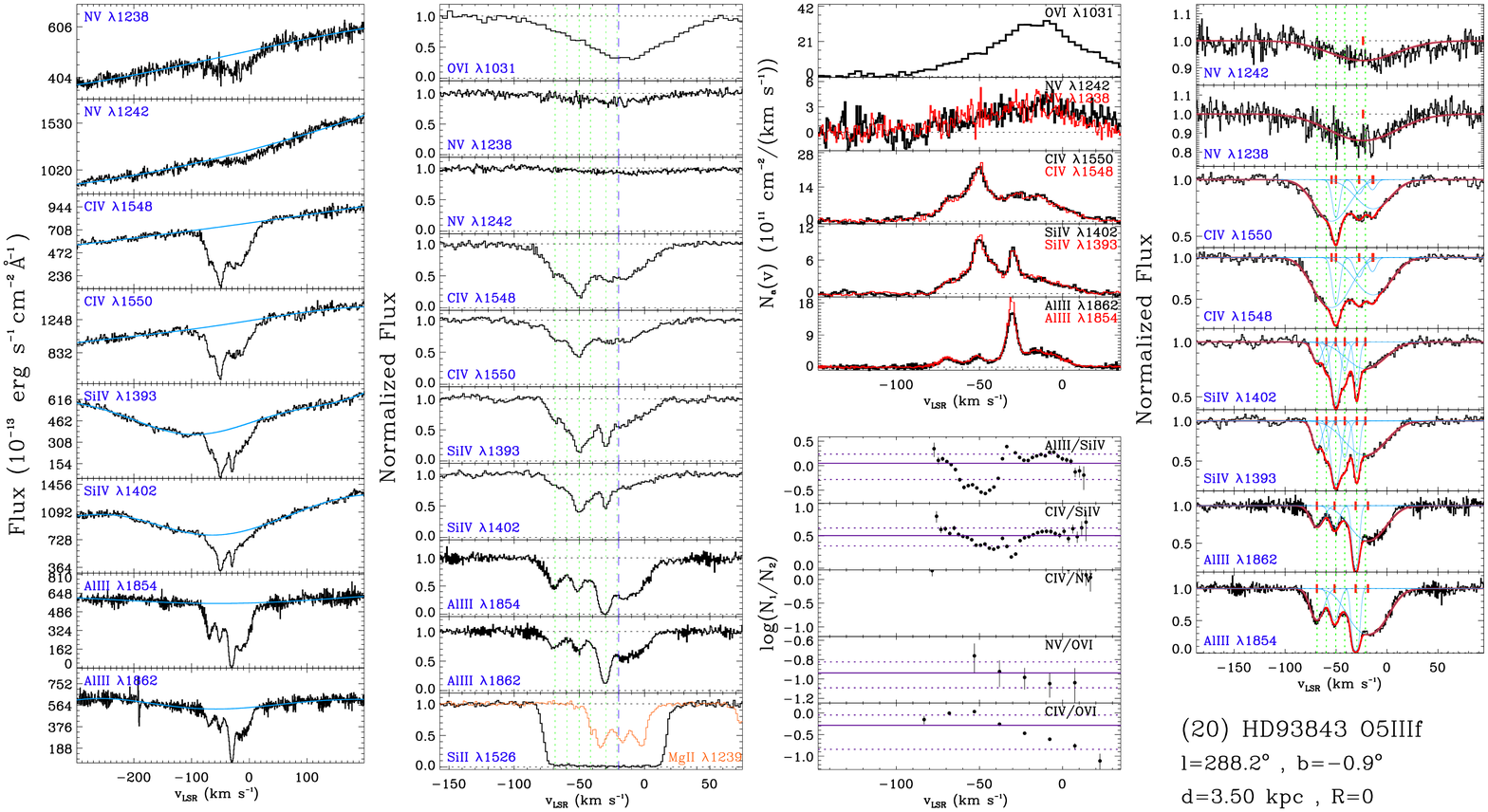}
\figurenum{\ref{f-sum}}
\caption{{\it Continued}}
\end{figure*}

\clearpage 

\begin{figure*}[tbp]
\epsscale{1.2} 
\plotone{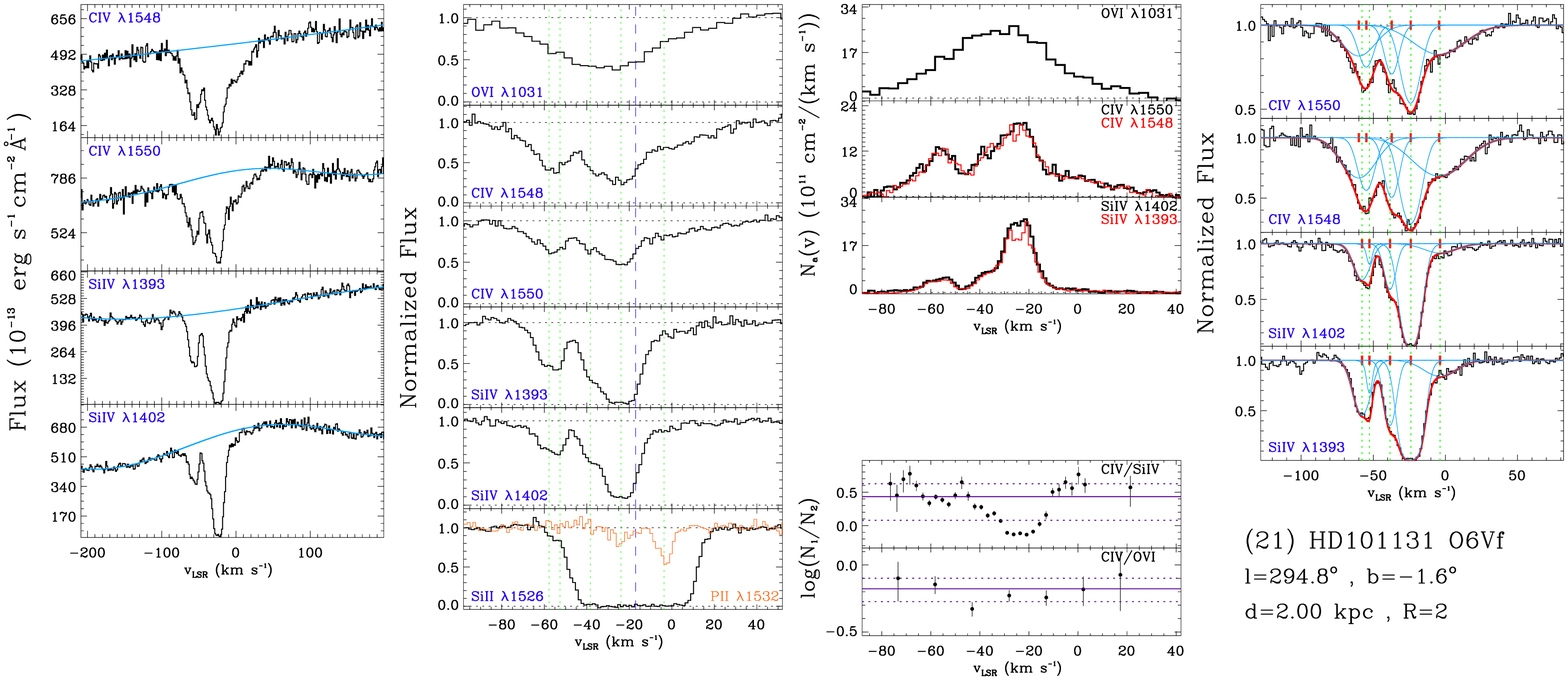}
\end{figure*}

\begin{figure*}[tbp]
\epsscale{1.2} 
\plotone{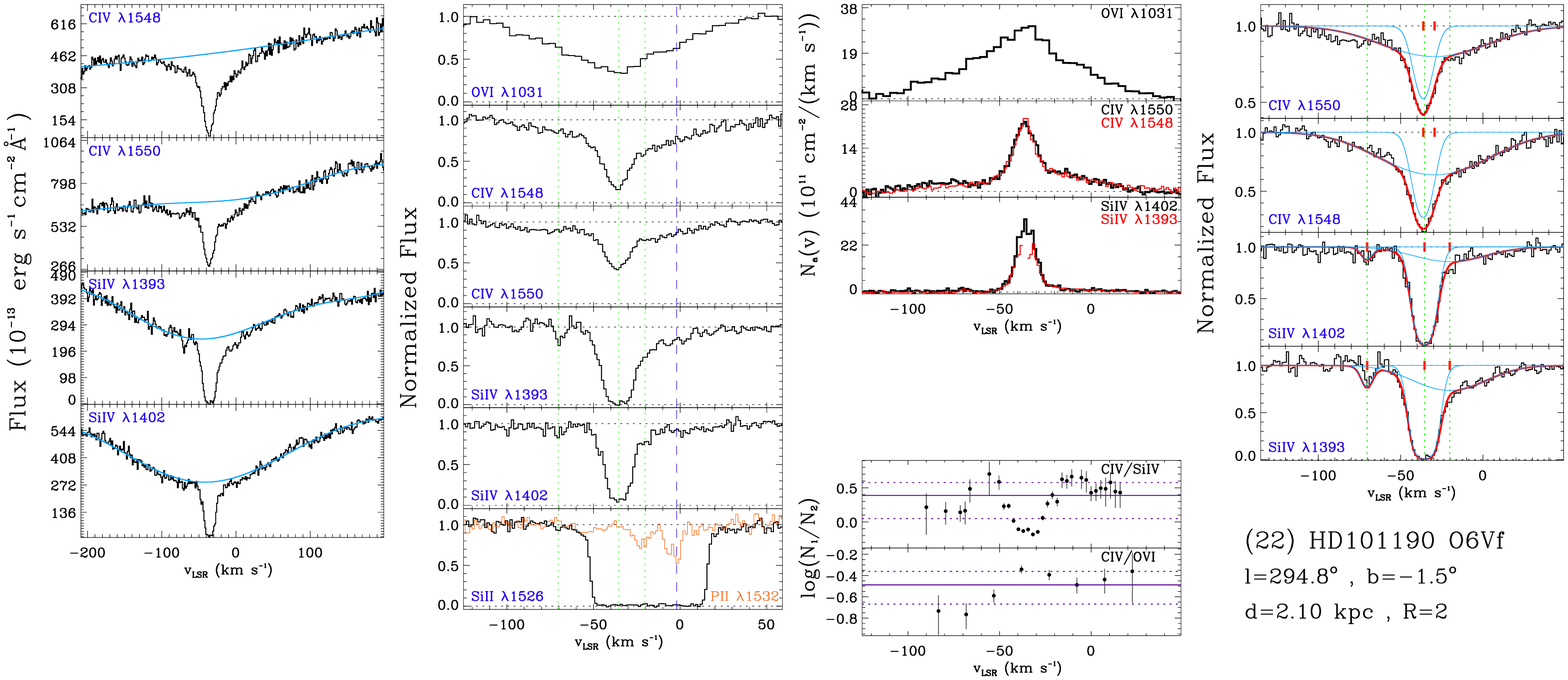}
\figurenum{\ref{f-sum}}
\caption{{\it Continued}}
\end{figure*}

\clearpage 

\begin{figure*}[tbp]
\epsscale{1.2} 
\plotone{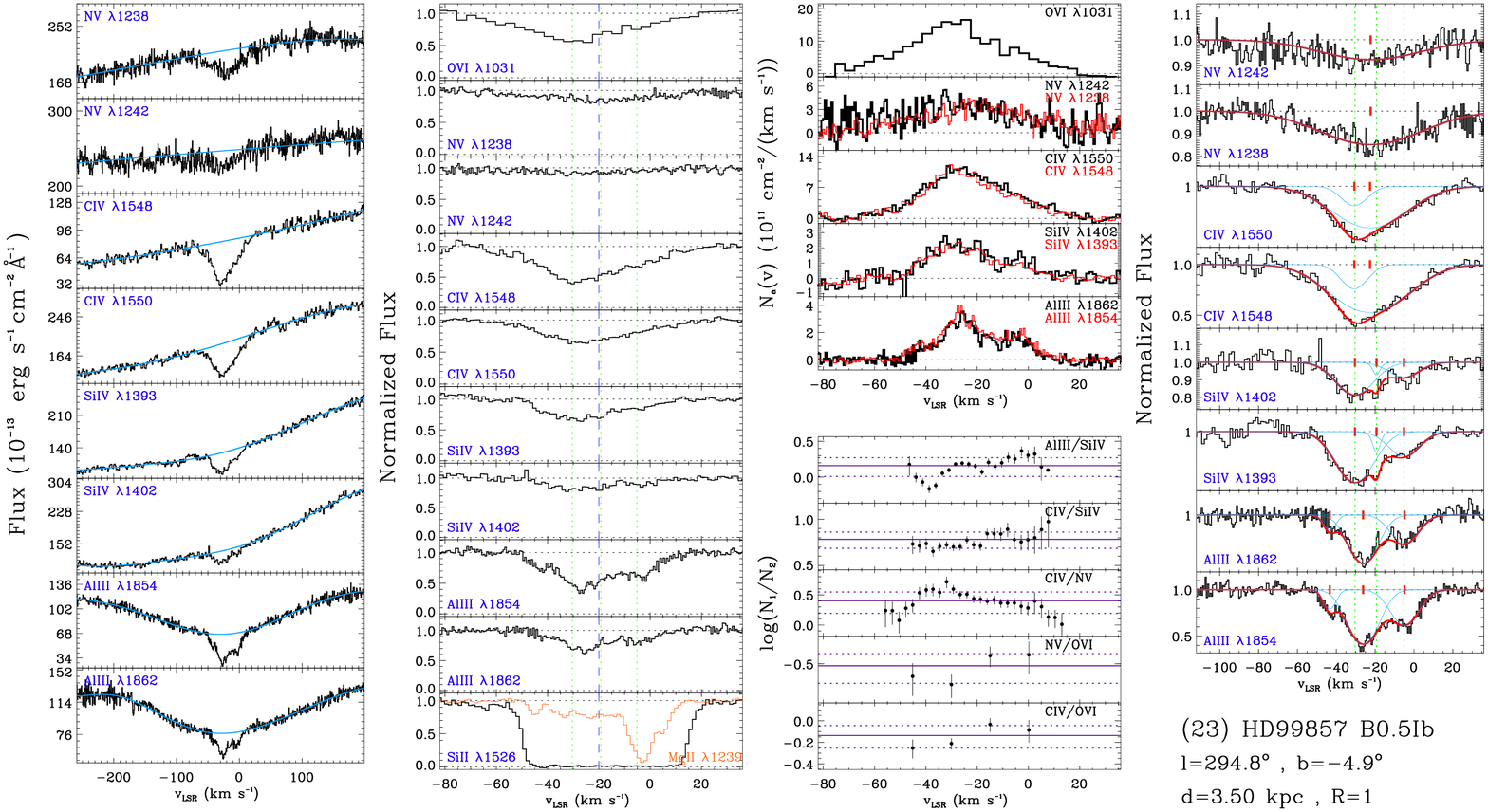}
\end{figure*}

\begin{figure*}[tbp]
\epsscale{1.2} 
\plotone{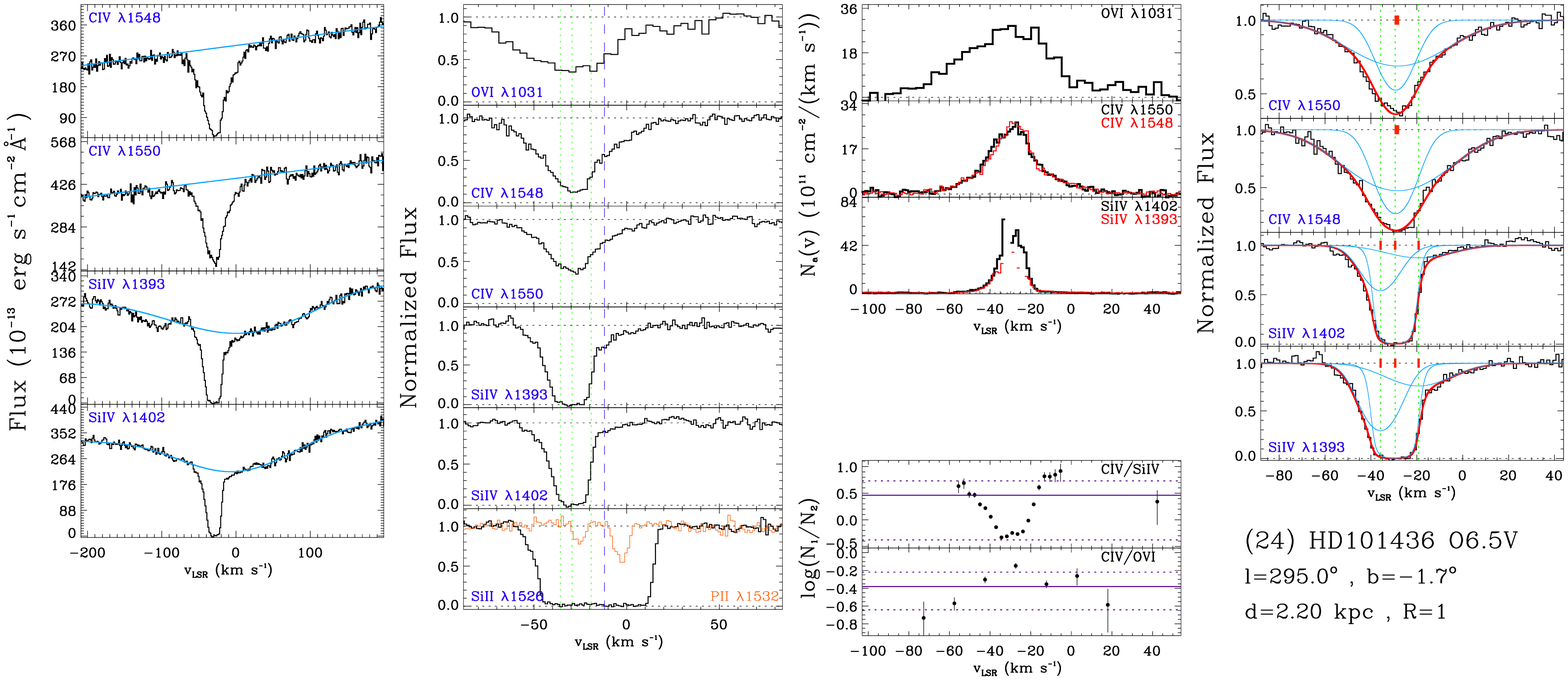}
\figurenum{\ref{f-sum}}
\caption{{\it Continued}}
\end{figure*}

\clearpage 

\begin{figure*}[tbp]
\epsscale{1.2} 
\plotone{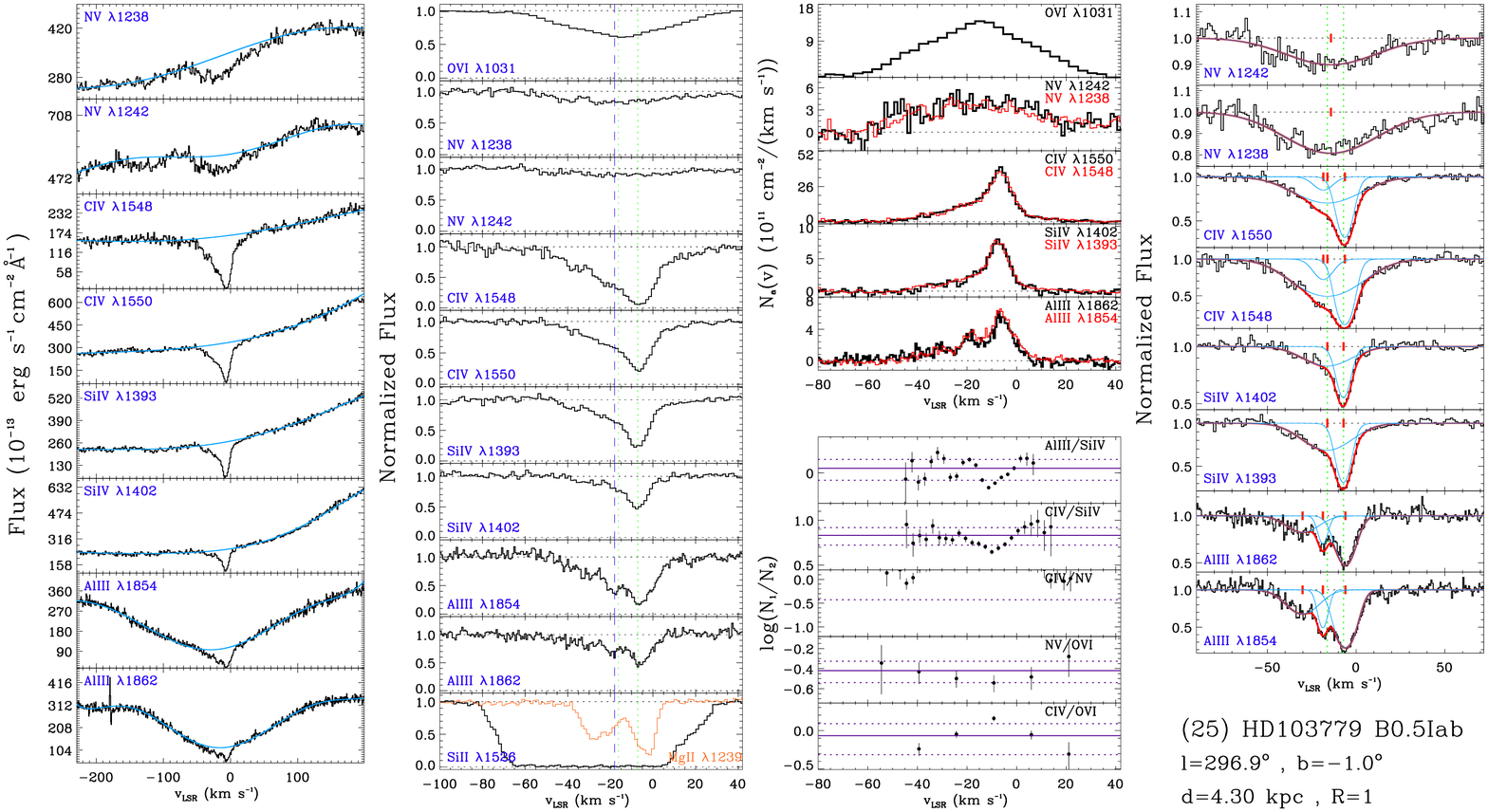}
\end{figure*}

\begin{figure*}[tbp]
\epsscale{1.2} 
\plotone{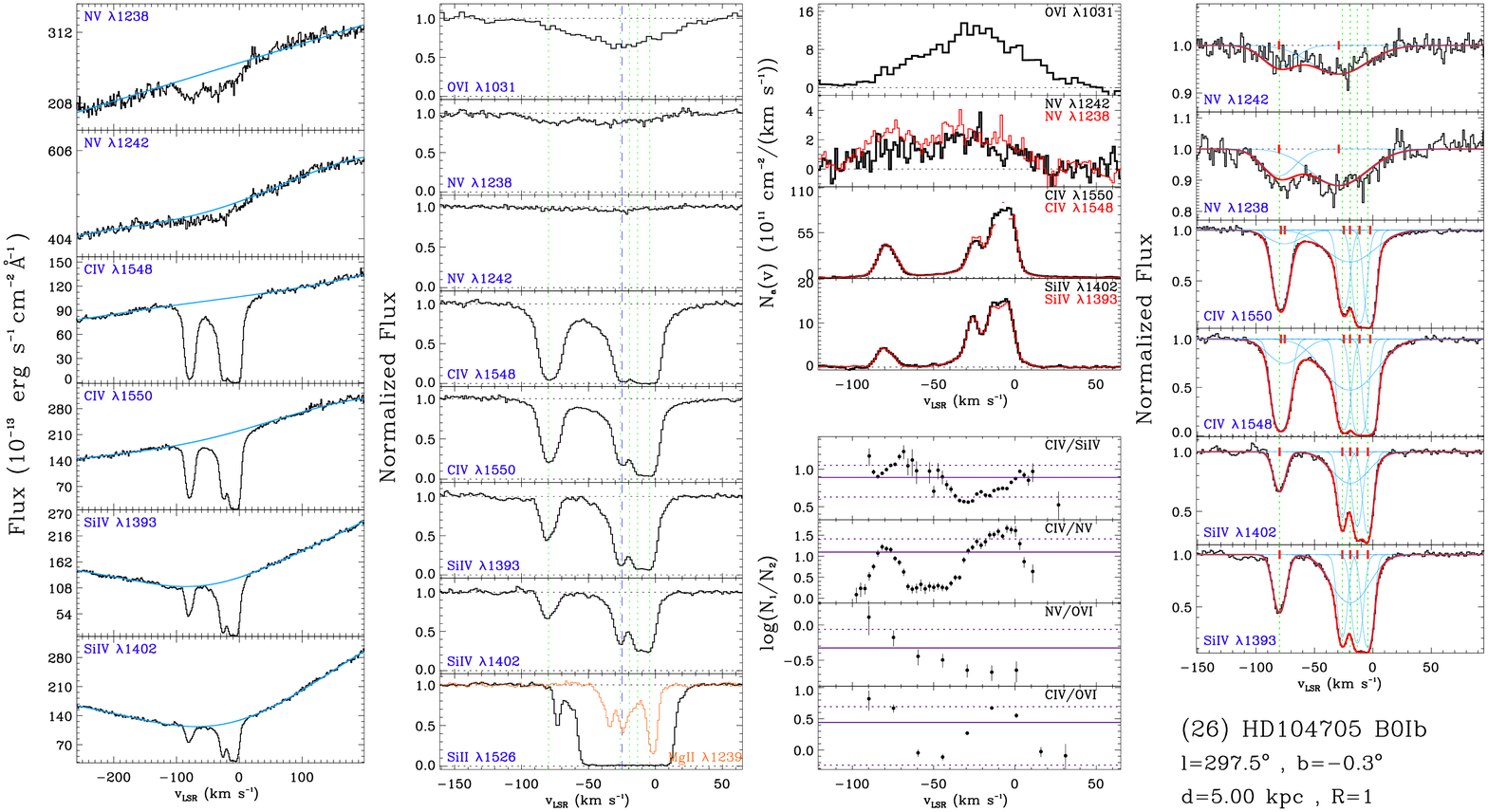}
\figurenum{\ref{f-sum}}
\caption{{\it Continued}}
\end{figure*}

\clearpage 

\begin{figure*}[tbp]
\epsscale{1.2} 
\plotone{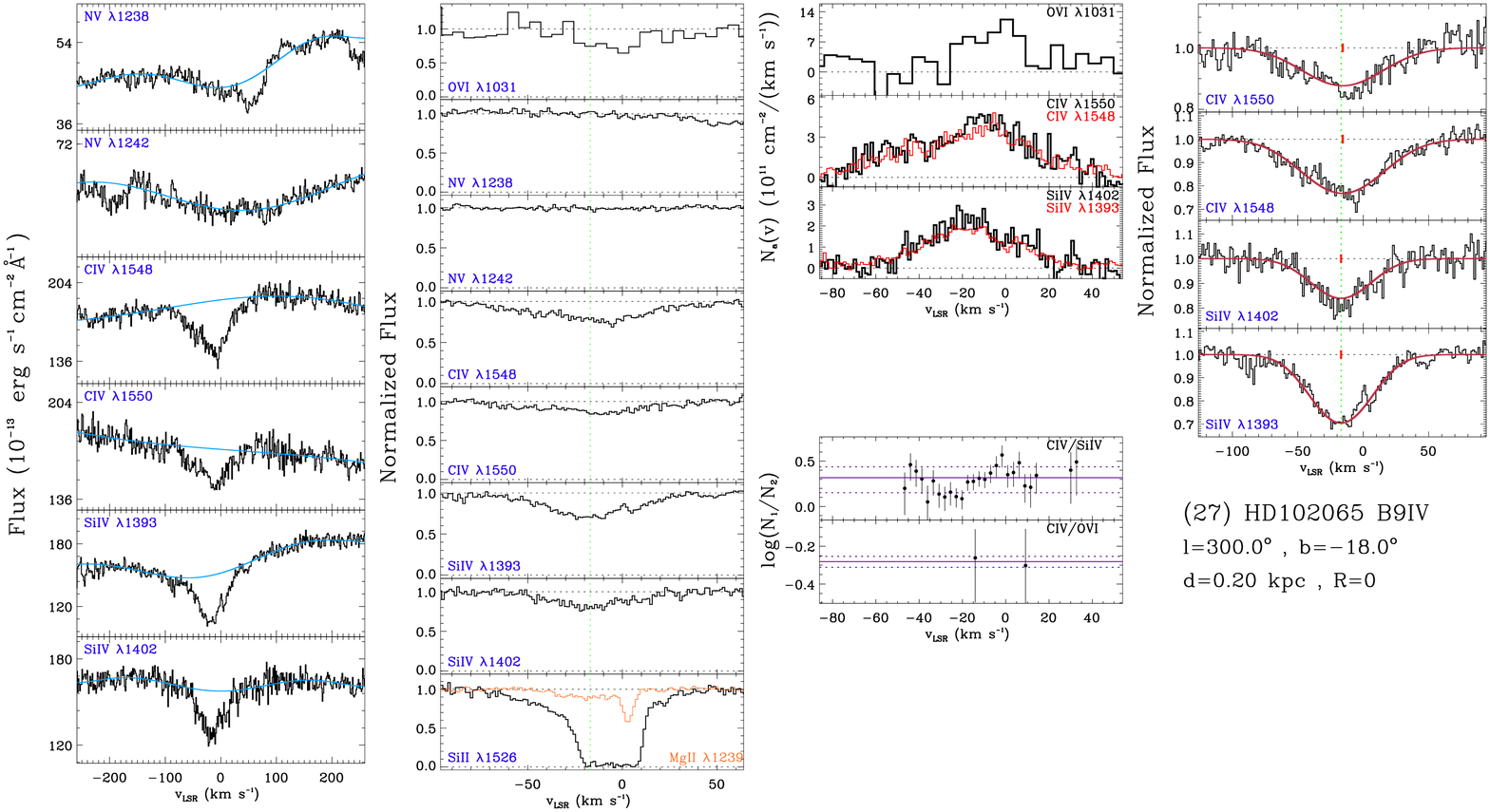}
\end{figure*}

\begin{figure*}[tbp]
\epsscale{1.2} 
\plotone{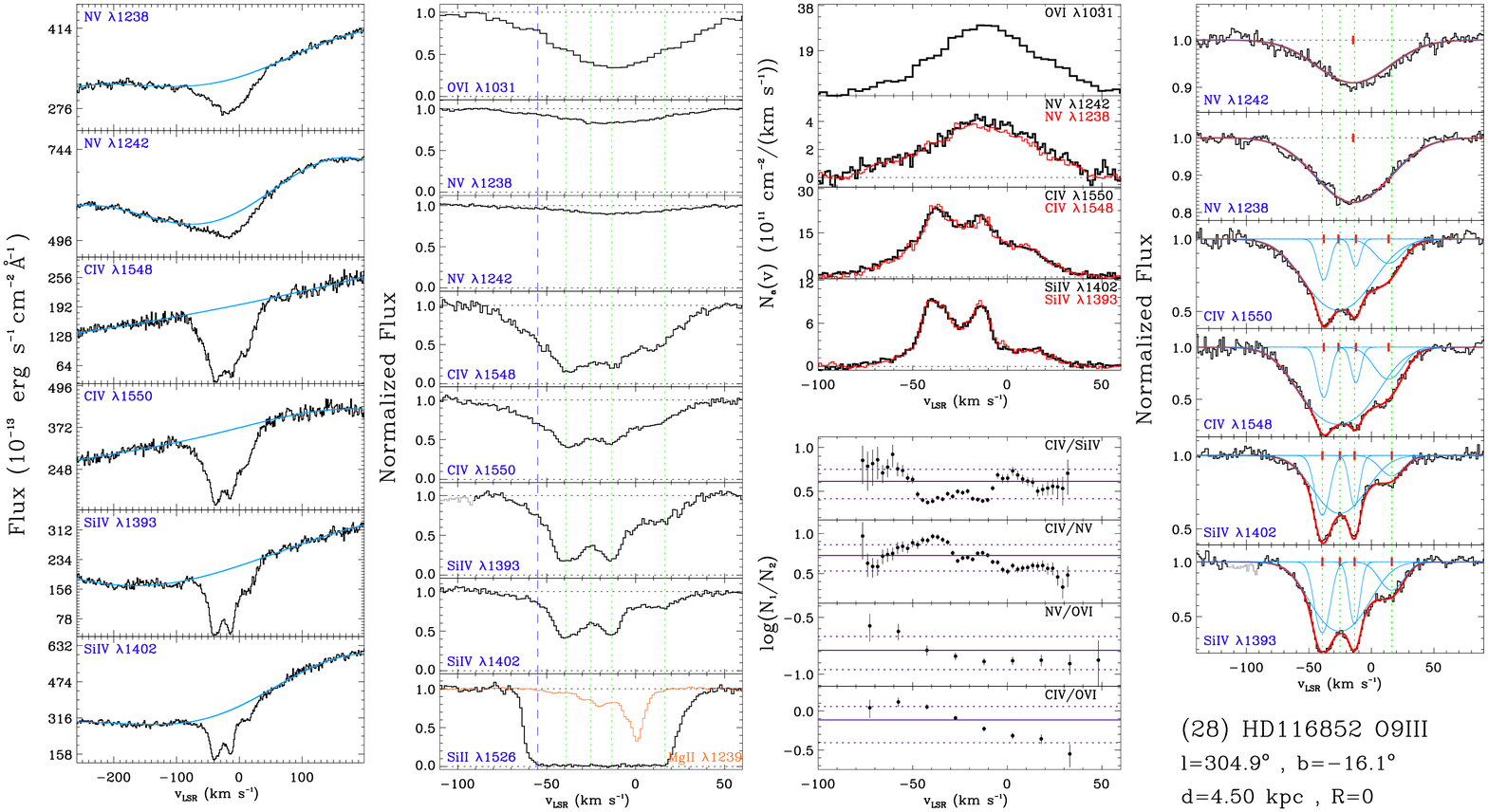}
\figurenum{\ref{f-sum}}
\caption{{\it Continued}}
\end{figure*}

\clearpage 

\begin{figure*}[tbp]
\epsscale{1.2} 
\plotone{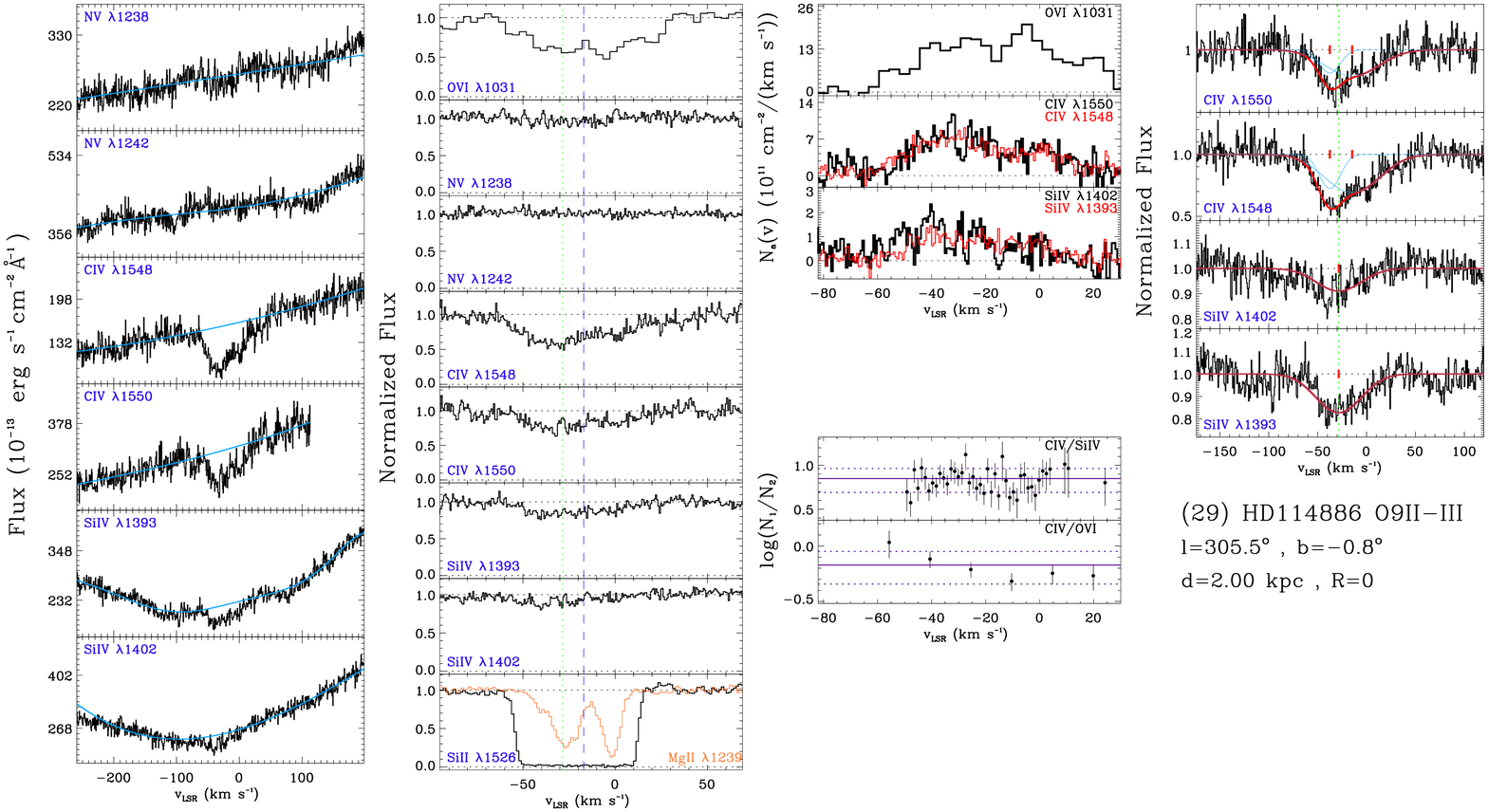}
\end{figure*}

\begin{figure*}[tbp]
\epsscale{1.2} 
\plotone{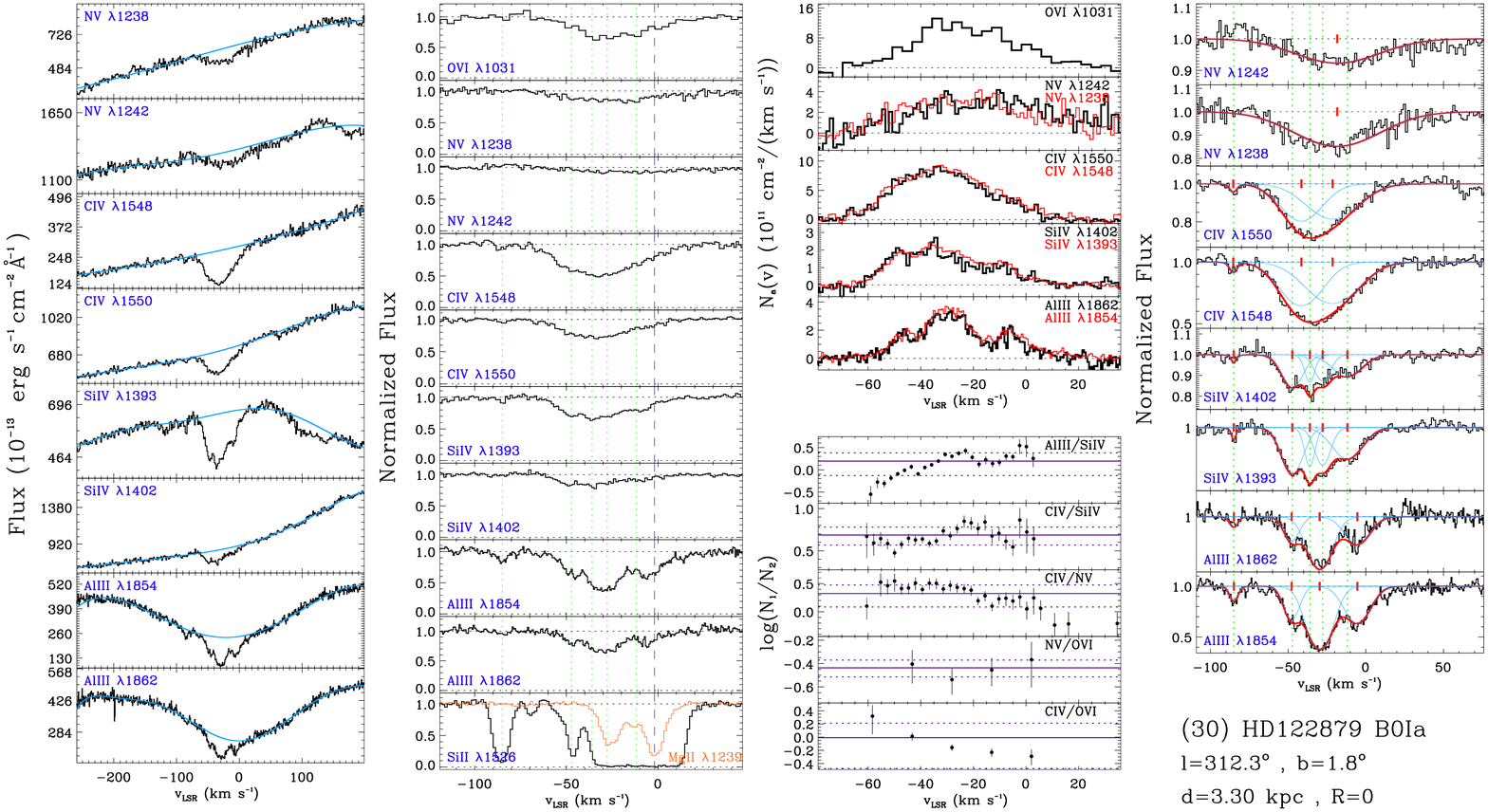}
\figurenum{\ref{f-sum}}
\caption{{\it Continued}}
\end{figure*}

\clearpage 

\begin{figure*}[tbp]
\epsscale{1.2} 
\plotone{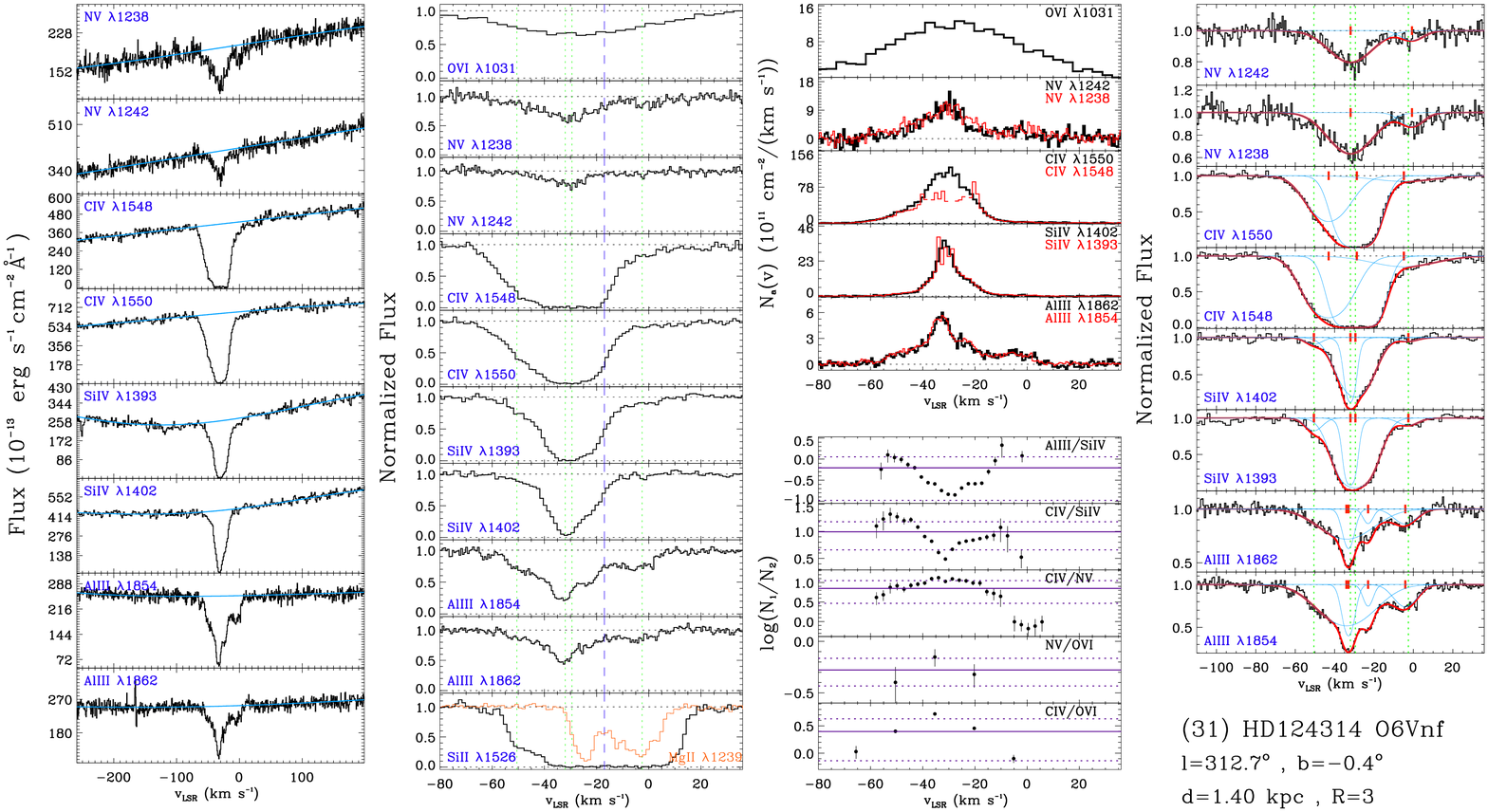}
\end{figure*}

\begin{figure*}[tbp]
\epsscale{1.2} 
\plotone{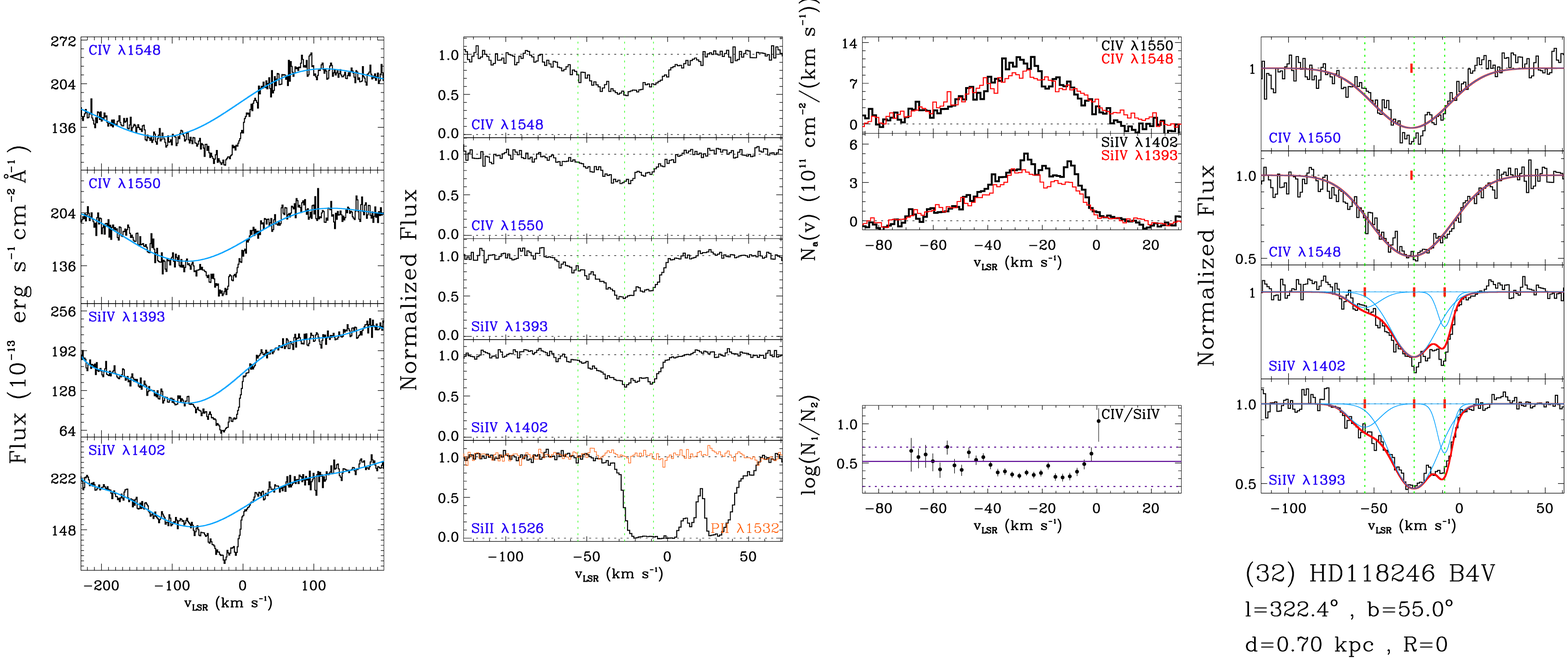}
\figurenum{\ref{f-sum}}
\caption{{\it Continued}}
\end{figure*}

\clearpage 

\begin{figure*}[tbp]
\epsscale{1.2} 
\plotone{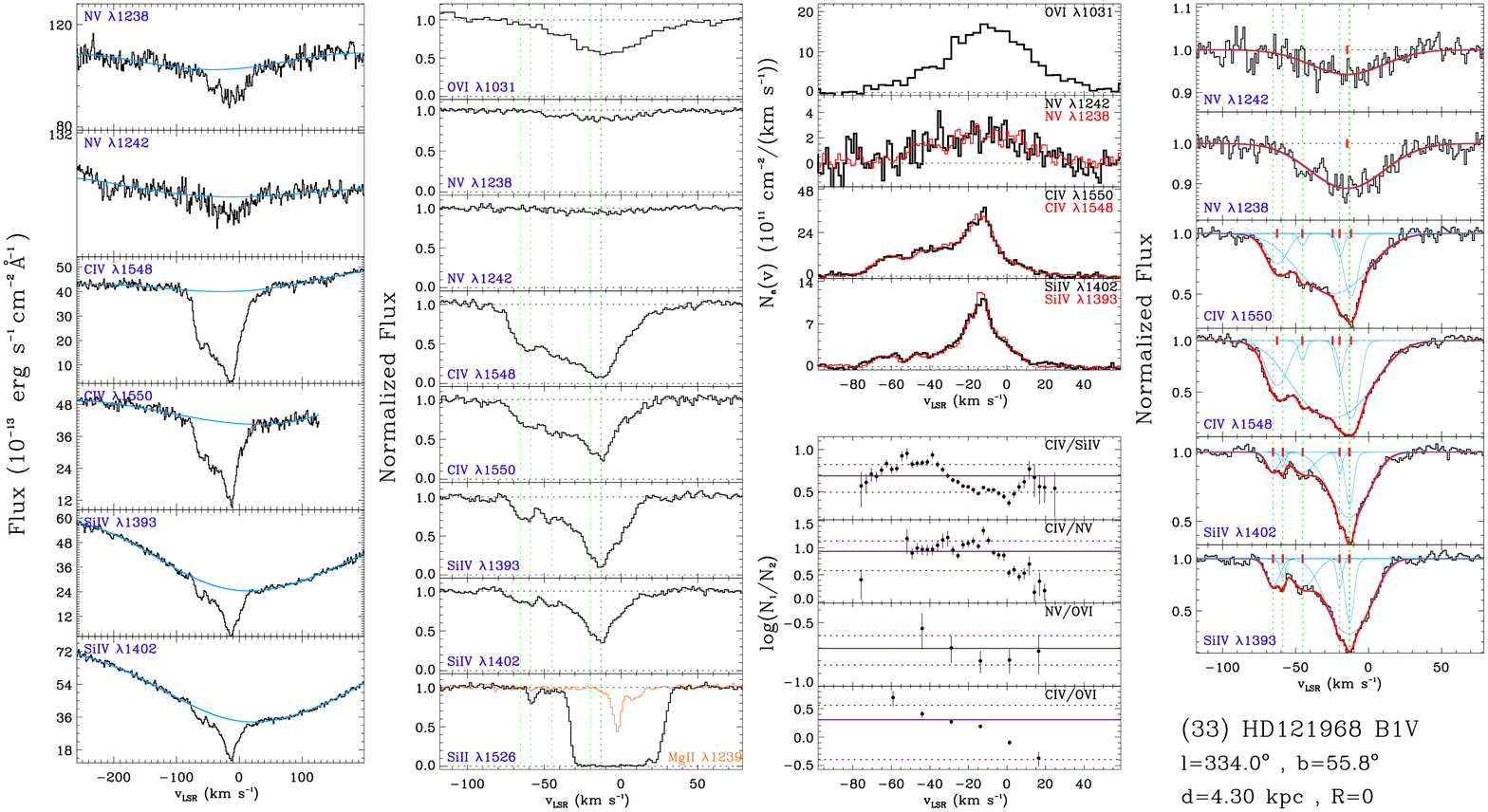}
\end{figure*}

\begin{figure*}[tbp]
\epsscale{1.2} 
\plotone{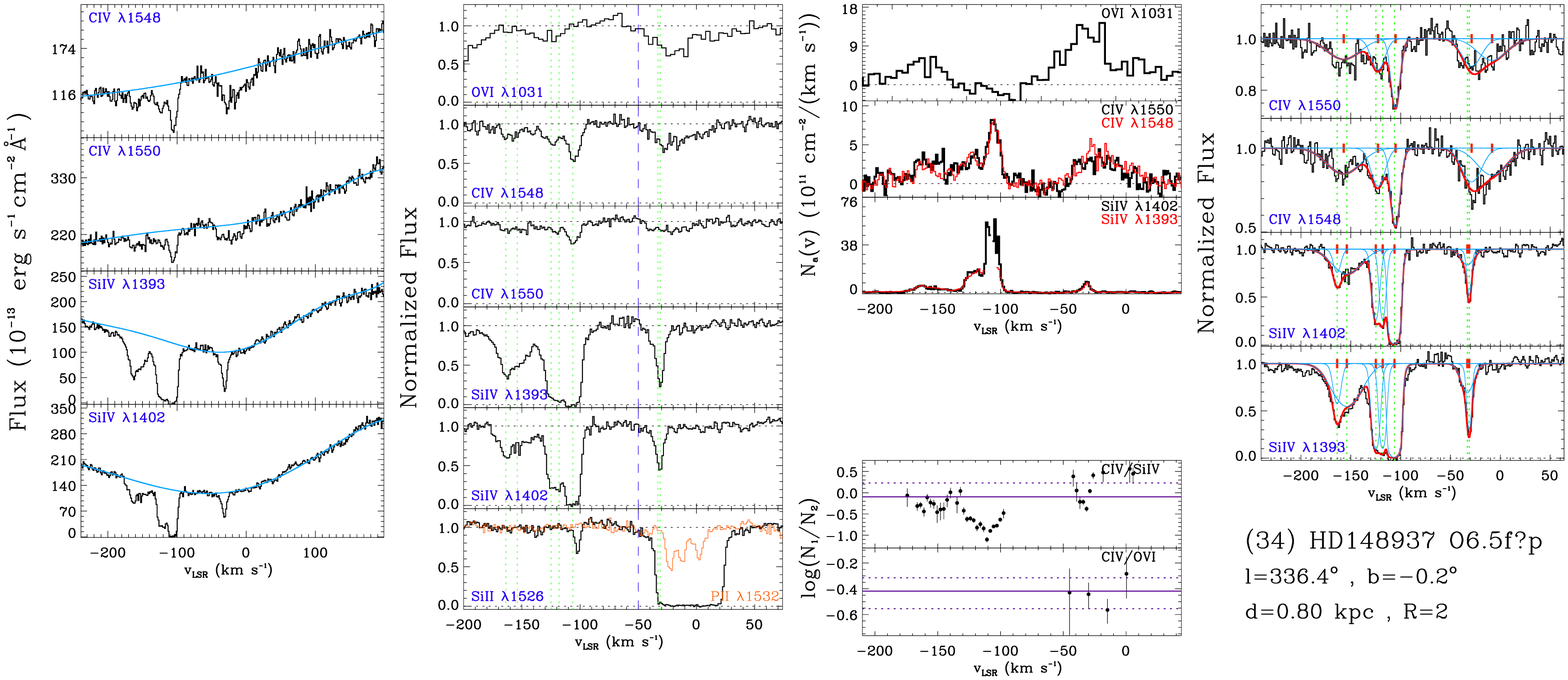}
\figurenum{\ref{f-sum}}
\caption{{\it Continued}}
\end{figure*}

\clearpage

\begin{figure*}[tbp]
\epsscale{1.2} 
\plotone{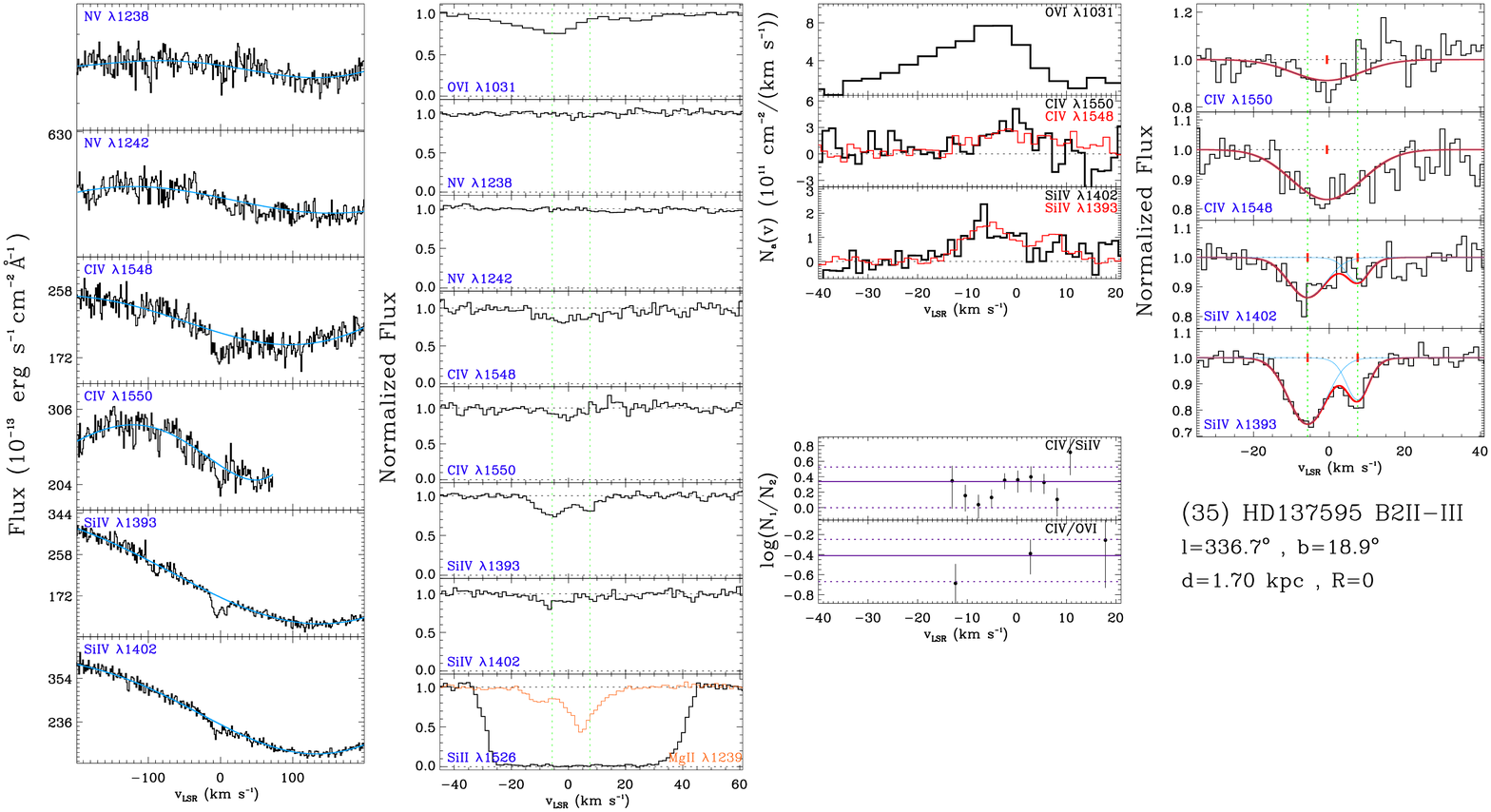}
\end{figure*}

\begin{figure*}[tbp]
\epsscale{1.2} 
\plotone{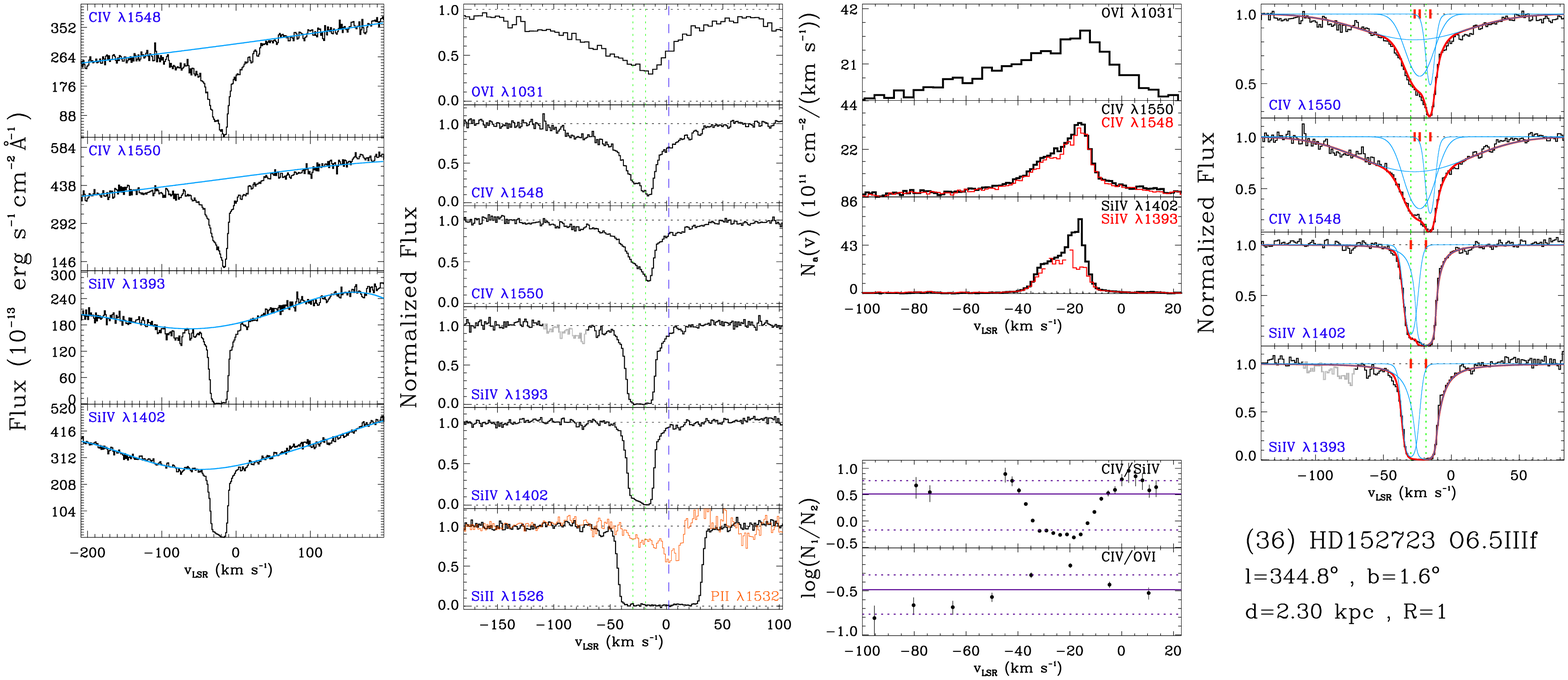}
\figurenum{\ref{f-sum}}
\caption{{\it Continued}}
\end{figure*}

\clearpage 

\begin{figure*}[tbp]
\epsscale{1.2} 
\plotone{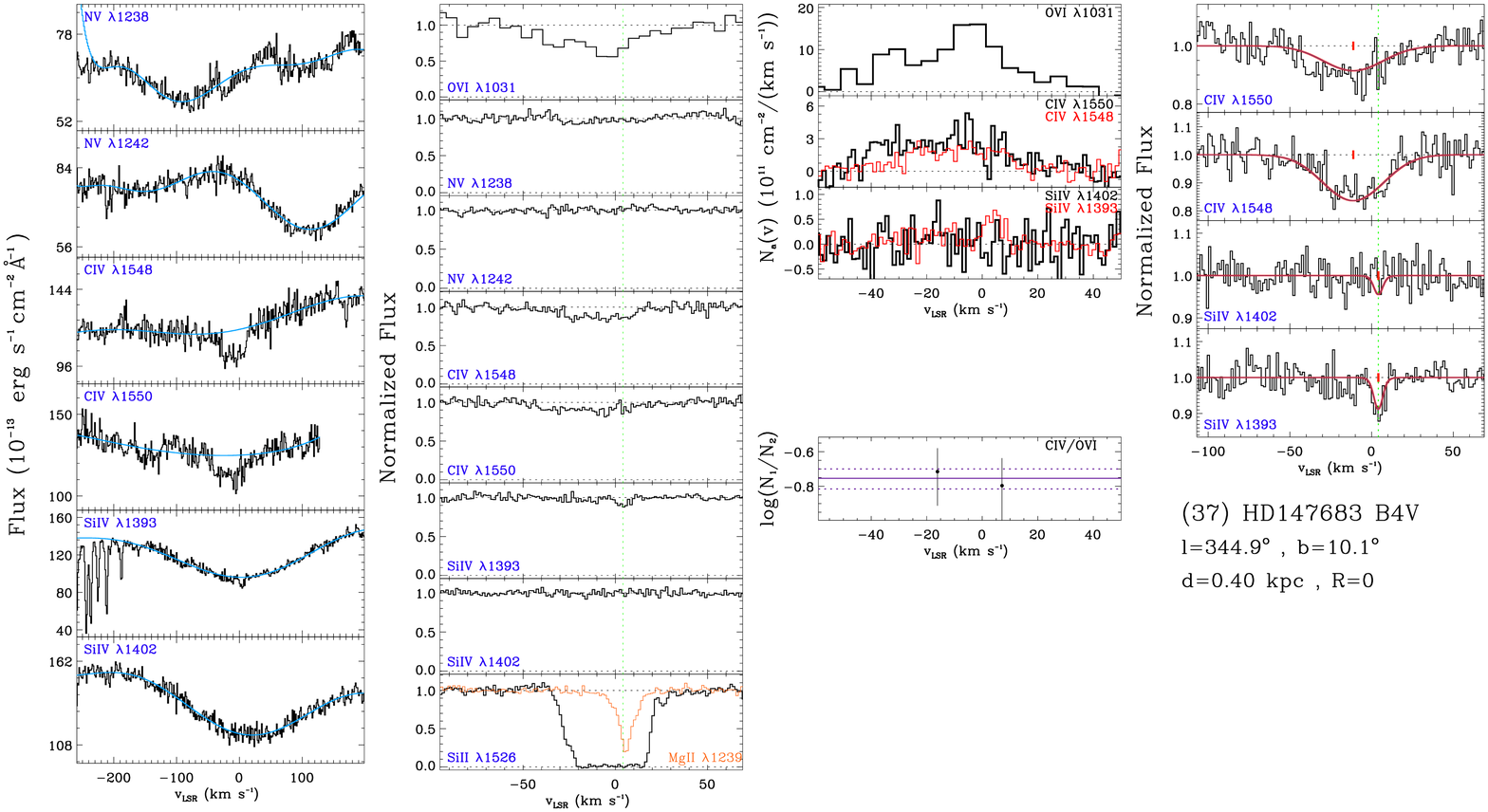}
\end{figure*}

\begin{figure*}[tbp]
\epsscale{1.2} 
\plotone{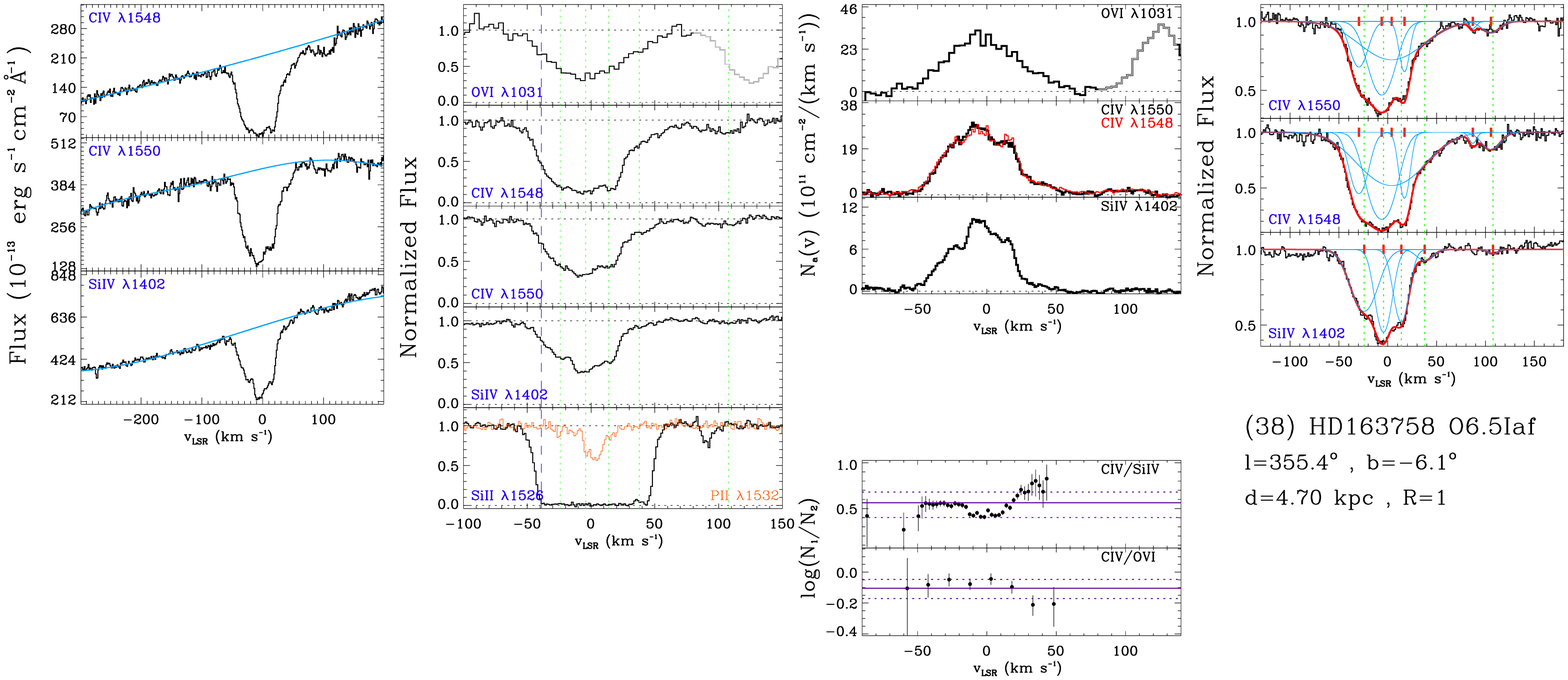}
\figurenum{\ref{f-sum}}
\caption{{\it Continued}}
\end{figure*}

\begin{figure}[tbp]
\epsscale{0.6} 
\plotone{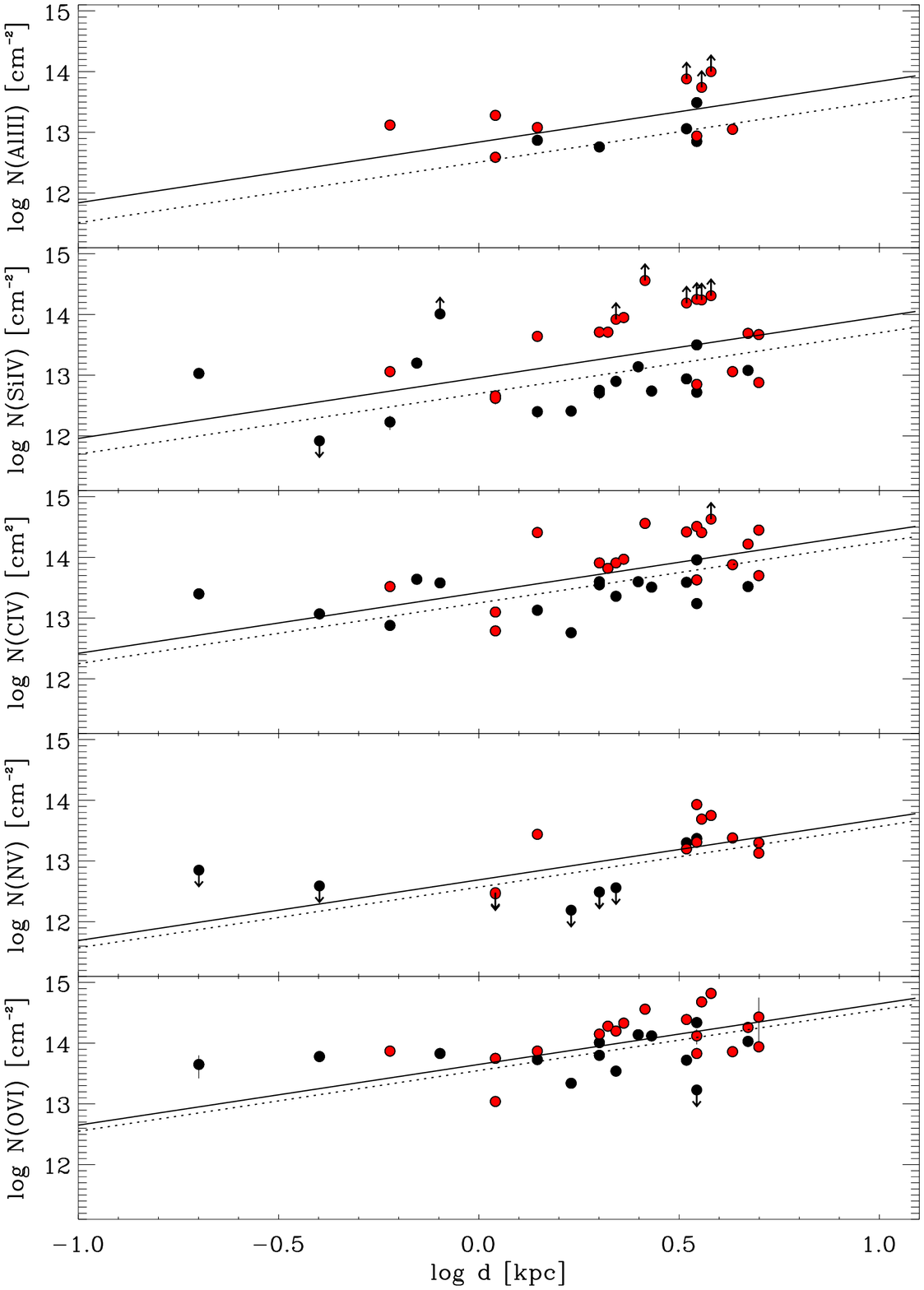}
\caption{The total column density against the star distance for stars at $|z| \le 1.2$ kpc.  The black filled circles are sight lines with $R=0$ and the red  filled circles are for $R>0$.  The dotted and solid lines are the mean $\log n$ for the $R = 0$ (black circles) and $R\ge 0$ samples (red plus black symbols), respectively. 
\label{f-avgn}}
\end{figure}

\clearpage

{\LongTables
\begin{deluxetable}{lcclcccl}
\tabletypesize{\scriptsize}
\tablewidth{0pt}
\tablecaption{Data Summary \label{t-data}}
\tablehead{
\colhead{Star} & 
\colhead{ID} & 
\colhead{Dataset} & 
\colhead{PI/Proposal ID} &
\colhead{Exp. Time [ks]} &
\colhead{Grating} &
\colhead{Wavelength Range [\AA]} &
\colhead{Aperture}
}
\startdata
CPD592603 & 15&  O40P01030 & Walborn/7137     & 1750 &  E140H & 1316--1518 & $0\farcs2\times0\farcs09$  \\
          &   &  O4QX03010 & Danks/7301       & 5160 &  E140H & 1141--1335 & $0\farcs2\times0\farcs09$  \\
          &   &  O4QX03020 &		      & 3180 &  E140H & 1316--1518 & $0\farcs2\times0\farcs09$  \\
          &   &  O4QX03030 &		      & 4800 &  E140H & 1494--1688 & $0\farcs2\times0\farcs09$  \\
HD3827    & 7 &  O6LZ02010 & Lauroesch/9434   & 300  &  E140H & 1390--1586 & $0\farcs2\times0\farcs2$	\\
          &   &  O54359010 & Jenkins/8043     & 1362 &  E140H & 1160--1357 & $0\farcs1\times0\farcs03$  \\
	  &   &  O54359030 &		      & 1296 &  E230H & 1624--1896 & $0\farcs1\times0\farcs03$  \\
	  &   &  O54359020 &		      & 1296 &  E230H & 1624--1896 & $0\farcs1\times0\farcs03$  \\
\enddata
\tablecomments{The resolving powers ($\lambda/\Delta\lambda$) of the STIS E140H/E230H gratings are 
roughly 200,000 (167,000 for E230H), 114,000, and 114,000 for the $0\farcs1\times0\farcs03$, $0\farcs2\times0\farcs09$ 
(or $0\farcs1\times0\farcs09$), and $0\farcs2\times0\farcs2$ apertures, respectively.}
\end{deluxetable}
}

\clearpage
\begin{landscape}
\begin{deluxetable}{llllcccccccl}
\tablewidth{0pc}
\tabcolsep=2pt
\tablecaption{Sight Line Overview \label{t-datao}}
\tabletypesize{\scriptsize}
\tablehead{
\colhead{ID} &
\colhead{Star} &
\colhead{Sp/L$^{\rm{(ref)}}$} &
\colhead{$l$} &
\colhead{$b$} &
\colhead{$V$} &
\colhead{$d$} &
\colhead{$z$} &
\colhead{$v_{\rm c}$} &
\colhead{$v_\star$} &
\colhead{$R$} &
\colhead{Physical Region Along the Line of Sight}
\\
\colhead{} &
\colhead{} &
\colhead{} &
\colhead{(\degr)} &
\colhead{(\degr)} &
\colhead{(mag)} &
\colhead{(kpc)} &
\colhead{(kpc)} &
\colhead{(\km)}	&
\colhead{(\km)}	&
\colhead{}	&
\colhead{}
}
\startdata
 (1) & HD177989    & B0III$^{(10)}$         &	    017.81   &    $ -11.88  $  & 9.34  & 6.0 & $-1.2$	&$+12.7$   &  $ +12$ & 0   & Scutum Supershell, Scutum \& Carina arms Loop I\\
 (2) & HD201345    & O9V$^{(10)}$           &	    078.44   &    $ -09.54  $  & 7.75  & 2.2 & $-0.4$	&$+15.7$   &  $ +35$ & 0   & Loop II\\
 (3) & HD195965    & B0V$^{(10)}$           &	    085.71   &    $ +05.00  $  & 6.98  & 1.1 & $+0.1$	&$+16.7$   &  $ +26$ & 1   & Cyg OB7 Assoc., Loop II\\
 (4) & HD209339    & B0IV$^{(6)}$           &	    104.58   &    $ +05.87  $  & 6.69  & 1.1 & $+0.1$	&$+13.0$   &  $  -7$ & 1   & Cep OB2 Assoc., Loop II/III\\
 (5) & HD218915    & O9.5Iabe$^{(10)}$      &	    108.06   &    $ -06.89  $  & 7.22  & 5.0 & $-0.6$	&$-10.4$   &  $ -61$ & 2   & Loop II/III, Perseus arm\\
 (6) & HD224151    & B0.5II$^{(10)}$        &	    115.44   &    $ -04.64  $  & 6.04  & 1.4 & $-0.1$	&$ +8.8$   &  $ -16$ & 0   & Loop II/III\\
 (7) & HD3827	   & B0.7Vn$^{(10)}$        &	    120.79   &    $ -23.23  $  & 7.95  & 2.0 & $-0.8$	&$ +4.3$   &  $ -17$ & 0   & Low Galactic halo, Loop II\\
 (8) & HD15137     & O9.5II-IIIn$^{(10)}$   &	    137.46   &    $ -07.58  $  & 7.87  & 3.5 & $-0.5$	&$ +1.7$   &  $ -33$ & 0   & Perseus arm, Loop II/III\\
\enddata
\tablecomments{$V$ and distance, $d$, taken from \citet{bowen08} or, if the star is not in by Bowen et al., 
we calculated it following their method.
$v_{\rm c}$ is the velocity correction to move the velocities from the heliocentric frame to the LSR frame ($v_{\rm LSR} = v_{\rm helio} + v_{\rm c}$). 
$v_\star$ is the radial stellar velocity in the LSR frame (helioncentric velocities are taken from the Simbad database). 
\\\
Sco OB2\_2: Upper Scorpius; Sco OB2\_3: Upper Centaurus Lupus; Sco OB2\_4: Lower Centaurus Crux following \citet{dezeeuw99}.
\\
Spectral type/luminosity class references: 
(1) \citet{burgh07};    
(2) \citet{hiltner69}; 
(3) \citet{houck82}; 
(4) \citet{houck75}; 
(6) \citet{belczynski09}; 
(7) \citet{lyubimkov02}; 
(8) \citet{maiz04};
(9) \citet{mathys88} 
(10) \citet{savage01a}; 
(11) \citet{walborn72} ; 
 }
\end{deluxetable}
\clearpage
\end{landscape}

{\LongTables
\begin{deluxetable}{lcc}
\tabcolsep=3pt
\tablecolumns{3}
\tablewidth{0pt} 
\tabletypesize{\scriptsize}
\tablecaption{Average LSR Velocities and Apparent Column densities \label{t-aod}} 
\tablehead{\colhead{Ion}    &  \colhead{$v_a$}&\colhead{$N_a$} \\
	\colhead{}   &\colhead{(\km)} &  \colhead{[cm$^{-2}$]}}
\startdata
  \cutinhead{  (1)      HD177989 -- $[ -100, +   96]$ }
\siiv\ $\lambda$1393    & $   28.1  \pm    0.5  $ & $ 13.82 \,^{+0.01}_{-0.01} $ \\
\siiv\ $\lambda$1402    & $   29.7  \pm    0.5  $ & $ 13.82 \,^{+0.01}_{-0.01} $ \\
\civ\ $\lambda$1548     & $   29.4  \pm    0.2  $ & $ 14.43 \,^{+0.01}_{-0.01} $ \\
\civ\ $\lambda$1550     & $   31.2  \pm    0.5  $ & $ 14.44 \,^{+0.01}_{-0.01} $ \\
\nv\ $\lambda$1238      & $   17.4  \pm    2.7  $ & $ 13.47 :			 $ \\
\nv\ $\lambda$1242      & $   22.7  \pm    4.2  $ & $ 13.50 :			 $ \\
\ovi\ $\lambda$1031     & $   29.2  \pm    5.6  $ & $ 14.20 \,^{+0.04}_{-0.05} $ \\
  \cutinhead{  (2)      HD201345 -- $[  -47, +   65]$ }
\siiv\ $\lambda$1393    & $   12.4  \pm    2.1  $ & $ 12.91 \,^{+0.03}_{-0.03} $ \\
\enddata		
\tablecomments{The symbol ``$>$" indicates that the flux reaches zero and hence the apparent column density value is a lower limit. 
 The symbol ``$(>)$" means that the flux reaches zero in the strong line of the doublet but not in the weak line; the weak line may 
 suffer from saturation in these cases.  The symbol ``$<$" indicates that there is no detection; the value quoted is a 3$\sigma$ upper limit. 
 The $^*$  implies that \ovi\ is partially contaminated by H$_2$ and hence a smaller velocity interval was used to estimate $v_a$ and $N_a$.
 The colon indicates that the column density is uncertain owing to uncertainties in the continuum placement. 
 }
\end{deluxetable}
}

\clearpage

\begin{landscape}
\begin{deluxetable}{clcccccccccc}
\tabcolsep=3pt
\tablewidth{0pt} 
\tabletypesize{\scriptsize}
\tablecaption{Adopted Apparent Column Densities and Average Densities ($N_a/d$) \label{t-aodf}} 
\tablehead{\colhead{ID}    & \colhead{Star}    & \colhead{$\log N($\aliii$)$}    &  \colhead{$\log N($\siiv$)$}&\colhead{$\log N($\civ$)$} &  \colhead{$\log N($\nv$)$} &  \colhead{$\log N($\ovi$)$}
					    & \colhead{$\log n($\aliii$)$}    &  \colhead{$\log n($\siiv$)$}&\colhead{$\log n($\civ$)$} &  \colhead{$\log n($\nv$)$} &  \colhead{$\log n($\ovi$)$} \\
\colhead{}    & \colhead{}    & \colhead{[cm$^{-2}$]}    &  \colhead{[cm$^{-2}$]}&\colhead{[cm$^{-2}$]} &  \colhead{[cm$^{-2}$]} &  \colhead{[cm$^{-2}$]}
					  & \colhead{[cm$^{-3}$]}    &  \colhead{[cm$^{-3}$]}&\colhead{[cm$^{-3}$]} &  \colhead{[cm$^{-3}$]} &  \colhead{[cm$^{-3}$]}					
}
\startdata
   (1) &      HD177989   &   \nodata                     & $ 13.82 \,^{+0.01}_{-0.01} $ & $ 14.43 \,^{+0.01}_{-0.01} $ & $ 13.49 :		   $ & $ 14.20 \,^{+0.04}_{-0.05} $ &   \nodata	& $ -8.45  $ & $ -7.84  $ & $ -8.78  $ & $ -8.07  $ \\
   (2) &      HD201345   &   \nodata                     & $ 12.96 \,^{+0.05}_{-0.06} $ & $ 13.36 \,^{+0.06}_{-0.07} $ & $ < 12.56                 $ & $ 13.54 \,^{+0.03}_{-0.04} $  &   \nodata	& $ -8.88  $ & $ -8.48  $ & $ < -9.28$ & $ -8.30  $ \\
   (3) &      HD195965   &  $  12.59 \,^{+0.02}_{-0.02} $ & $ 12.66 \,^{+0.03}_{-0.04} $ & $ 12.79 \,^{+0.05}_{-0.06} $ & $ < 12.46                 $ & $ 13.04 \,^{+0.03}_{-0.04} $ &  $  -8.94  $ 	& $ -8.87  $ 	& $ -8.74  $ & $ < -9.08$ & $ -8.49  $ \\
   (4) &      HD209339   &  $  13.28 \,^{+0.01}_{-0.01} $ & $ 12.62 \,^{+0.08}_{-0.10} $ & $ 13.10 \,^{+0.05}_{-0.06} $ & $ < 12.48                 $ & $ 13.75 \,^{+0.02}_{-0.02} $ &  $  -8.25  $ 	& $ -8.91  $ 	& $ -8.43  $ & $ < -9.05$ & $ -7.78  $ \\
   (5) &      HD218915   &   \nodata                     & $ 12.88 \,^{+0.03}_{-0.03} $ & $ 13.70 \,^{+0.02}_{-0.02} $ & $ 13.13 :		   $ & $ 14.43 \,^{+0.32}_{-0.42} $ &   \nodata	& $ -9.31  $ & $ -8.49  $ & $ -9.06  $ & $ -7.76  $ \\
   (6) &      HD224151   &  $ 12.87 \,^{+0.03}_{-0.03} $ & $ 12.40 \,^{+0.09}_{-0.11} $ & $ 13.13 \,^{+0.06}_{-0.07} $ & \nodata                      & $ 13.73 \,^{+0.04}_{-0.04} $ &  $ -8.77  $ 	& $ -9.24  $ 	& $ -8.51  $ & \nodata  & $ -7.91  $ \\
   (7) &        HD3827   &  $  12.76 \,^{+0.02}_{-0.02} $ & $ 12.71 \,^{+0.09}_{-0.11} $ & $ 13.55 \,^{+0.03}_{-0.04} $ & $ < 12.49                 $ & $ 13.80 \,^{+0.02}_{-0.02} $ &  $  -9.03  $ 	& $ -9.08  $ 	& $ -8.25  $ & $ < -9.30$ & $ -7.99  $ \\
   (8) &       HD15137   &  $ 12.85 \,^{+0.02}_{-0.02} $ & $ 12.72 \,^{+0.05}_{-0.06} $ & $ 13.24 \,^{+0.04}_{-0.04} $ & \nodata                      & $ < 13.23                  $ &  $ -9.18  $ 	& $ -9.31  $ 	& $ -8.80  $ & \nodata  & $ < -8.80 		 $ \\
\enddata		
\tablecomments{The symbol ``$>$" indicates that the flux reaches zero and hence the apparent column density value is a lower limit. 
The symbol ``$<$" indicates that there is no detection; the value quoted is a 3$\sigma$ upper limit. 
The colon indicates that the column density is uncertain owing to uncertainties in the continuum placement. 
$a$: The stronger line of the doublet is saturated so that the column density is based only on the weaker transition; 
in this case some saturation correction may be needed \citep[see][]{savage91}}
\end{deluxetable}
\clearpage
\end{landscape}


{\LongTables
\begin{deluxetable}{clccccc}
\tabcolsep=3pt
\tablecolumns{7}
\tablewidth{0pt} 
\tabletypesize{\scriptsize}
\tablecaption{Ionic Ratios from the Total Apparent Column Densities and Apparent Column Density Profiles \label{t-aodr}} 
\tablehead{\colhead{ID}    & \colhead{Star}    & \colhead{\aliii/\siiv}    &  \colhead{\civ/\siiv}&\colhead{\civ/\nv} &  \colhead{\nv/\ovi} &  \colhead{\civ/\ovi}}
\startdata
   (1) &      HD177989   &  \nodata                      & $  0.62 \,^{+0.01}_{-0.01} $ & $  0.95 :		     $ & $ -0.72 :		    $ & $  0.23 \,^{+0.04}_{-0.05} $ \\
       &                 &  \nodata                      & $  0.68 \,^{+0.17}_{-0.27} $ & $  0.91 \,^{+0.25}_{-0.62} $ & $ -0.65 \,^{+0.14}_{-0.20} $ & $  0.13 \,^{+0.24}_{-0.57} $ \\
   (2) &      HD201345   &  \nodata                      & $  0.40 \,^{+0.08}_{-0.09} $ & $ >  0.80                  $ & $ < -0.98                  $ & $ -0.18 \,^{+0.07}_{-0.08} $ \\
       &                 &  \nodata                      & $  0.55 \,^{+0.20}_{-0.38} $ & \nodata                      & \nodata                      & $ -0.13 \,^{+0.08}_{-0.10} $ \\
   (3) &      HD195965   &  $ -0.07 \,^{+0.04}_{-0.04} $ & $  0.13 \,^{+0.06}_{-0.07} $ & $ >  0.34                  $ & $ < -0.58                  $ & $ -0.25 \,^{+0.06}_{-0.07} $ \\
       &                 &  $  0.02 \,^{+0.14}_{-0.20} $ & $  0.19 \,^{+0.18}_{-0.31} $ & \nodata                      & \nodata                      & $ -0.21 \,^{+0.36}_{-0.56} $ \\
   (4) &      HD209339   &  $  0.66 \,^{+0.08}_{-0.10} $ & $  0.48 \,^{+0.09}_{-0.12} $ & $ >  0.62                  $ & $ < -1.27                  $ & $ -0.65 \,^{+0.06}_{-0.07} $ \\
       &                 &  $  0.53 \,^{+0.18}_{-0.31} $ & $  0.55 \,^{+0.21}_{-0.42} $ & \nodata                      & \nodata                      & $ -0.54 \,^{+0.09}_{-0.11} $ \\
\enddata		
\tablecomments{Column density Ratios on the same row that of the ID-Star line are estimated from the total column density summarized in Table~\ref{t-aodf}. 
The immediate next row summarizes the average results based on the pixe-by-pixel analysis of the AOD ratio profiles. 
Values preceded by ``$<$'' in the \civ/\siiv\ column are upper limit (\siiv\ absorption is saturated), while in the \nv/\ovi\ column those 
are 3$\sigma$ upper limits (\nv\ is not detected). Values preceded by ``$>$'' in the \civ/\nv\ column are 3$\sigma$ lower limit (\nv\ is not detected).}
\end{deluxetable}
}

\clearpage

{\LongTables
\begin{deluxetable}{lccc}
\tabcolsep=3pt
\tablecolumns{4}
\tablewidth{0pt} 
\tabletypesize{\scriptsize}
\tablecaption{Results from the Profile Fits \label{t-fit}} 
\tablehead{\colhead{No.}    &  \colhead{$v$}&\colhead{$b$} &  \colhead{$\log N$}\\
	\colhead{}    &  \colhead{(\km)}&\colhead{(\km)} &  \colhead{[cm$^{-2}$]}}
\startdata
  \cutinhead{ (1) --    HD177989   -- $ R = 0 $ }
  \cutinhead{--   \civ\ --}
   1 & $        -44.7  \pm          0.4  $ & $         15.8  \pm          0.6 $ & $       12.93  \pm         0.01  $ \\
   2 & $        -11.5  \pm          0.3  $ & $         12.0  \pm          0.4 $ & $       13.43  \pm         0.02  $ \\
   3 & $         10.3  \pm          1.2  $ & $         10.3  \pm          1.3 $ & $       13.41  \pm         0.09  $ \\
   4 & $         21.6  \pm          0.3  $ & $          6.9  \pm          0.4 $ & $       13.40  \pm         0.08  $ \\
   5 & $         42.7  \pm          0.1  $ & $         16.0  \pm          0.1 $ & $       14.27  \pm         0.01  $ \\
  \cutinhead{--    \nv\ --}
   1 & $        -47.5  \pm          1.2  $ & $          7.8  :  		$ & $	    12.19  :   			$ \\
   2 & $          4.7  \pm          0.3  $ & $          1.4  :  		$ & $	    11.91  :   			$ \\
   3 & $         20.3  \pm          1.1  $ & $         43.1  :  		$ & $	    13.43  :   			$ \\
\enddata		
\tablecomments{Each doublet of a given ion was simultaneously fitted, but each species was independently fitted. 
A value followed by ``:'' mean that the flux in this component is zero in both lines of the doublet, except for \nv\ where it indicates 
that the values are uncertain owing to uncertainties in the continuum placement. Velocities are in the LSR frame.}
\end{deluxetable}
}

\clearpage

\begin{deluxetable}{lcccc}
\tabcolsep=1pt
\tablecolumns{5}
\tablewidth{0pt} 
\tabletypesize{\footnotesize}
\tablecaption{Column densities in Broad and Narrow Components \label{t-statn}} 
\tablehead{\colhead{Sample}    &\colhead{$\langle \log N \rangle$}    &  \colhead{$\log N_{1/2}$} &\colhead{$\langle \log N \rangle$}    &  \colhead{$\log N_{1/2}$} \\
\colhead{}    &\colhead{$[{\rm cm^{-2}}]$}    &  \colhead{$[{\rm cm^{-2}}]$} &\colhead{$[{\rm cm^{-2}}]$} &  \colhead{$[{\rm cm^{-2}}]$}
}
\startdata
 &	\multicolumn{2}{c}{Broad } & \multicolumn{2}{c}{Narrow } \\
  \cutinhead{--  \aliii\ -- }
All 	 & $ 12.59\pm 0.39 $ & $12.56$	&  $ 12.04 \pm 0.44  $  & $12.01$  \\ 
$R=0$ 	 & $ 12.50\pm 0.32 $ & $12.51$	&  $ 12.09 \pm 0.46  $  & $11.89$  \\ 
$R>0$ 	 & $ 12.64\pm 0.42 $ & $12.80$	&  $ 12.01 \pm 0.45  $  & $12.01$  \\
  \cutinhead{--  \siiv\ -- }
All 	 & $ 12.79\pm 0.47  $ & $12.80$	&  $ 12.40 \pm 0.50  $  & $12.32$  \\
$R=0$ 	 & $ 12.68\pm 0.42  $ & $12.67$	&  $ 12.18 \pm 0.40  $  & $12.28$  \\
$R>0$ 	 & $ 12.86\pm 0.50  $ & $12.84$	&  $ 12.54 \pm 0.51  $  & $12.57$  \\
  \cutinhead{--  \civ\ --}
All 	 & $ 13.33\pm 0.40  $ & $13.36$ &  $ 13.09 \pm 0.60  $   & $13.05$  \\
$R=0$ 	 & $ 13.35\pm 0.37  $ & $13.40$ &  $ 12.67 \pm 0.48  $   & $12.59$  \\
$R>0$ 	 & $ 13.32\pm 0.41  $ & $13.35$ &  $ 13.29 \pm 0.55  $   & $13.38$  \\
  \cutinhead{--  \nv\ --}
All 	 & $ 13.25\pm 0.30  $ & $13.32$ &  $ 12.21 \pm 0.27  $   & $12.19$  \\
$R=0$ 	 & $ 13.34\pm 0.11  $ & $13.37$ &  $ 12.09 \pm 0.16  $   & $12.17$  \\
$R>0$ 	 & $ 13.22\pm 0.34  $ & $13.18$ &  \nodata$^a$ 		 & \nodata$^a$ 
\enddata		
\tablecomments{Saturated components are excluded from the samples. 
For \aliii, \siiv, \civ, and \nv, the cutoff  between narrow and broad components is 6.5, 6.5, 10, and 9 \km, respectively. 
$N_{1/2}$ stands for the median value. $a$: only a single component is available with $\log N = 12.57$}
\end{deluxetable}

\begin{deluxetable}{lccccc}
\tabcolsep=1pt
\tablecolumns{6}
\tablewidth{0pt} 
\tabletypesize{\scriptsize}
\tablecaption{Volume Densities in the Galactic Plane for the Different Ions$^a$   \label{t-dens}} 
\tablehead{  \colhead{Sample} & \colhead{$\langle \log n \rangle_{\mbox{\aliii}}$}&  
\colhead{$\langle \log n \rangle_{\mbox{\siiv}}$}&  \colhead{$\langle \log n \rangle_{\mbox{\civ}}$}&  \colhead{$\langle \log n \rangle_{\mbox{\nv}}$}&  \colhead{$\langle \log n \rangle_{\mbox{\ovi}}$}}
\startdata
$R\ge 0$	&     $ -8.65 \pm  0.11 $ $(14)$ & $  -8.53 \pm 0.10 $ $(36)$ & $ -8.07 \pm 0.07 $ $(36)$ & $ -8.80 \pm 0.09 $ $(21)$ & $ -7.84 \pm 0.07 $ $(34)$ \\ 
$R\ge 0^b$ 	&     $ -8.80 \pm  0.11 $ $(10)$ & $  -8.69 \pm 0.10 $ $(29)$ & $ -8.17 \pm 0.08 $ $(29)$ & $ -8.86 \pm 0.11 $ $(16)$ & $ -7.91 \pm 0.07 $ $(27)$  \\
$R\ge 0^c$ 	&     $ -8.78 \pm  0.11 $ $(11)$ & $  -8.72 \pm 0.09 $ $(28)$ & $ -8.08 \pm 0.07 $ $(35)$ & $ -8.68 \pm 0.09 $ $(13)$ & $ -7.81 \pm 0.06 $ $(33)$  \\
$R=0^b$		&     $ -8.98 \pm  0.09 $ $ (4)$ & $  -8.79 \pm 0.14 $ $(17)$ & $ -8.24 \pm 0.10 $ $(17)$ & $ -8.92 \pm 0.15 $ $ (9)$ & $ -7.94 \pm 0.11 $ $(15)$ \\ 
$R>0^b$   	&     $ -8.68 \pm  0.17 $ $ (6)$ & $  -8.54 \pm 0.15 $ $(12)$ & $ -8.06 \pm 0.12 $ $(12)$ & $ -8.82 \pm 0.12 $ $ (7)$ & $ -7.88 \pm 0.08 $ $(12)$ \\
\enddata		
\tablecomments{$a$: Mean Volume density $\langle \log n \rangle = \langle \log(N/d)\rangle$  with its 
standard deviation of the mean. The value between parentheses is the number
of data points used to estimate $\langle \log(N/d)\rangle$. Only sight line with 
$|z_{\rm h}| \le 1.3$ kpc were considered and note that only 3 sight lines have
$0.9 \le |z_{\rm h}| \le 1.3$ kpc. $b$:  Stars 15 to 21 from the Carina region were excluded. $c$: Upper or lower limits were excluded. 
$d$: Only stars 15 to 21 from the Carina region (lower limits are included in the \aliii, \siiv, and \civ\ samples).}
\end{deluxetable}

\end{document}